\documentclass[useAMS,usenatbib]{mnras}
\usepackage{journals}
\usepackage{graphicx}
\usepackage{color}
\usepackage{times}
\usepackage{hyperref}
\usepackage{amsmath}
\usepackage{bm}
\usepackage{multirow}
\usepackage{rotating}
\usepackage{caption}

\title[Supernovae and their host galaxies -- VII.]{Supernovae and their host galaxies -- VII.
The diversity of Type~Ia supernova progenitors}
\author[A.~A.~Hakobyan~et~al.]{A.~A.~Hakobyan,$^{1}$\thanks{\fontsize{7.3}{8.8}\selectfont{E-mail:
\href{mailto:artur.hakobyan@yerphi.am}{artur.hakobyan@yerphi.am} (AAH);
\href{mailto:l.barkhudaryan@yerphi.am}{l.barkhudaryan@yerphi.am} (LVB)}}
L.~V.~Barkhudaryan,$^{1}$\textcolor[rgb]{0,0,1}{\footnotemark[1]}
A.~G.~Karapetyan,$^{1}$
M.~H.~Gevorgyan,$^{1}$
\newauthor
G.~A.~Mamon,$^{2}$
D.~Kunth,$^{2}$
V.~Adibekyan$^{3}$
and M.~Turatto$^{4}$
\\
$^{1}$Center for Cosmology and Astrophysics, Alikhanian National Science Laboratory, 2 Alikhanian Brothers Str., 0036 Yerevan, Armenia\\
$^{2}$Institut d'Astrophysique de Paris (UMR 7095: CNRS \& Sorbonne Universit\'{e}), 98 bis bd Arago, F-75014 Paris, France\\
$^{3}$Instituto de Astrof\'{i}sica e Ci\^{e}ncia do Espa\c{c}o, Universidade do Porto, CAUP, Rua das Estrelas, P-4150-762 Porto, Portugal\\
$^{4}$INAF -- Osservatorio Astronomico di Padova, Vicolo dell'Osservatorio 5, I-35122 Padova, Italy}
\begin{document}

\date{Accepted 2020 September 21. Received 2020 September 15; in original form 2020 July 20}

\pagerange{\pageref{firstpage}--\pageref{lastpage}} \pubyear{2020}

\maketitle

\label{firstpage}

\begin{abstract}
  We present an analysis of the light curve (LC) decline rates $(\Delta m_{15})$ of 407 normal and peculiar
  supernovae (SNe) Ia and global parameters of their host galaxies.
  As previously known, there is a significant correlation
  between the $\Delta m_{15}$ of normal SNe~Ia and global ages (morphologies, colours, masses)
  of their hosts.
  On average, those normal SNe~Ia that are in galaxies from the Red Sequence
  (early-type, massive, old hosts)
  have faster declining LCs in comparison with those from the Blue Cloud
  (late-type, less massive, younger hosts) of the colour--mass diagram.
  The observed correlations between the $\Delta m_{15}$ of
  normal SNe~Ia and hosts' parameters appear to be due to
  the superposition of at least two distinct populations of faster and slower declining
  normal SNe~Ia from older and younger stellar components.
  We show, for the first time, that the $\Delta m_{15}$ of
  91bg- and 91T-like SNe is independent of host morphology and colour.
  The distribution of hosts on the colour--mass diagram confirms the known tendency for
  91bg-like SNe to occur in globally red/old galaxies while 91T-like events prefer blue/younger hosts.
  On average, the youngest global ages of 02cx-like SNe hosts and their
  positions in the colour--mass diagram hint that these events likely originate from
  young population, but they differ from 91T-like events in the LC decline rate.
  Finally, we discuss the possible explosion channels and present our favoured SN~Ia models
  that have the potential to explain the observed SN-host relations.
\end{abstract}


\section{Introduction}
\label{intro}

In the realm of stellar explosions, Type~Ia supernovae (SNe~Ia) are believed to be
the evolutionary endpoint, accompanied by the thermonuclear explosion,
of carbon-oxygen (CO) white dwarf (WD) stars in interacting close binary systems
\citep[e.g.][]{2016IJMPD..2530024M}.
These events play a key role in understanding the evolution of binary stellar systems
\citep*[e.g.][]{2009ApJ...699.2026R},
the chemical enrichment of galaxies \citep[e.g.][]{1986A&A...154..279M},
and the nature of accelerating expansion of the Universe
\citep[e.g.][]{1998AJ....116.1009R,1999ApJ...517..565P}.

It is now clear that Type Ia SNe are not a homogeneous population
of WD explosions, as they display a variety of photometric
and spectroscopic properties \citep[see reviews by][]{2017hsn..book..317T,2017hsn..book..375J}.
In the local Universe, in comparison with moderately uniform normal SNe~Ia \citep*{1993AJ....106.2383B},
about one third of all SNe~Ia events have peculiar
characteristics \citep[e.g.][]{2011MNRAS.412.1441L,2017ApJ...837..121G}.
The most common subclasses of peculiar SNe~Ia are:
\emph{i)} 91T-like events \citep{1992ApJ...384L..15F,1992AJ....103.1632P},
which are overluminous at the $B$-band maximum ($\sim0.6$~mag more luminous than normal SNe~Ia)
and have slow-declining light curves (LCs),
with distinctive pre-peak spectra dominated by \mbox{Fe\,{\sc iii}} lines,
\emph{ii)} 91bg-like SNe \citep{1992AJ....104.1543F,1993AJ....105..301L,1996MNRAS.283....1T},
which are subluminous events ($\sim2$~mag less luminous than normal ones) and have fast-declining LCs,
with post-maximum spectra dominated by unusually strong
\mbox{O\,{\sc i}} and \mbox{Ti\,{\sc ii}} lines,
and \emph{iii)} low luminosity (more than 2~mag fainter than normal SNe~Ia) and faster declining 02cx-like SNe
\citep[also called SNe~Iax,][]{2003PASP..115..453L,2013ApJ...767...57F},
with early spectra resembling those of 91T-like events.
There is also a tiny percentage of other peculiar SNe~Ia, including the faint but
slowly declining 02es-like SNe, so called Ca-rich transients, the extremely luminous 06gz-like
(also called super-Chandrasekhar) SNe, and other SNe~Ia with spectra showing evidence of interaction
with the circumstellar medium
\citep[see e.g.][for a review on the extremes of SNe~Ia]{2017hsn..book..317T}.

In general, the behaviour of the LCs of SN~Ia depends on the mass of synthesized radioactive $^{56}{\rm Ni}$,
the kinetic energy of the explosion, and the opacity of the ejecta
\citep[e.g.][]{1982ApJ...253..785A,2007Sci...315..825M}.
Importantly, Type Ia SNe show a key relation between their luminosity at the $B$-band maximum
and their LC decline rate $\Delta m_{15}$
\citep[i.e. magnitude difference between the maximum and 15 days after,][]{1993ApJ...413L.105P}.
This is known as the width-luminosity relation (faster declining SNe~Ia are fainter)
that played an enormous role in standardization of SNe~Ia and
their use in cosmology as the best distance indicators.
However, the width-luminosity relation is well established for normal SNe~Ia,
while peculiar events deviate, sometimes very strongly, from that relation
\citep[see e.g.][]{2016MNRAS.460.3529A,2017hsn..book..317T}.
Note that the $\Delta m_{15}$ and colours of SNe~Ia are also related:
the faster declining events correspond to the intrinsically redder SNe
\citep[e.g.][]{1999AJ....118.1766P}.

Theoretically, there are many possibilities in the proposed progenitor channels
for Type Ia SNe that are still under debate
\citep[see e.g.][]{2018PhR...736....1L,2020arXiv200102947R},
however they are generally categorized into the following main classes;
the \emph{single-degenerate} (SD) and \emph{double-degenerate} (DD) channels,
both of which probably occur in nature \citep[e.g.][]{2013FrPhy...8..116H}.
In the SD channel \citep[e.g.][]{1973ApJ...186.1007W},
a degenerate WD grows in mass through accretion from a non-degenerate companion,
consequently causing an explosion when the WD mass reaches the Chandrasekhar mass limit
($\approx 1.4 M_{\odot}$).
The non-degenerate companion can be a main-sequence/subgiant star, or a red giant,
or a helium (He) star \citep*[e.g.][]{1997Sci...276.1378N}.
In the DD channel \citep[e.g.][]{1984ApJS...54..335I,1985ApJS...58..661I},
an explosion occurs when two degenerate WDs
coalescence or interact through accretion with each other in a binary system,
after having been brought together due to the loss of
orbital angular momentum via the emission of gravitational waves.

Many studies attempted to constrain the nature of SNe~Ia progenitors by exploring
the relations between the spectroscopic/photometric properties of SNe~Ia and
the global characteristics of their host galaxies (or of the site of explosion),
such as morphology, colour, star formation rate (SFR), mass, metallicity,
and age of the stellar population \citep[e.g.][]{2000AJ....120.1479H,2001ApJ...554L.193H,
2005ApJ...634..210G,2009ApJ...691..661H,2009ApJ...707.1449N,2010MNRAS.406..782S,
2011ApJ...727..107G,2011ApJ...740...92G,2013MNRAS.435.1680J,2014MNRAS.438.1391P,
2016MNRAS.460.3529A,2018A&A...615A..68R,2019ApJ...886...58M,2020arXiv200609433P}.
In short, these studies showed that more luminous and slower declining SNe~Ia explode,
on average, in galaxies with later morphological type, lower mass, higher specific SFR,
and younger stellar population age (for SN local environment as well).
\citet[][]{2020ApJ...889....8K} recently claimed a significant correlation between the
SN~Ia luminosity (or LC decline rate) and the stellar population age of its host,
at a 99.5 per cent confidence level.
They suggested that the previously reported correlations with host morphology, mass,
and SFR are originated from the difference in population age
\citep[see also][]{2013A&A...560A..66R,2014MNRAS.445.1898C,2019ApJ...874...32R}.
However, SN~Ia samples in these studies consist only of spectroscopically normal events
with known LC properties, or sometimes include only a tiny portion of peculiar SNe~Ia.
Therefore, the relations between LC properties of peculiar SNe~Ia and
the characteristics of their host galaxies have not been explored in such detail
as it was done for normal SNe~Ia.

In our recent paper \citep{2019MNRAS.490..718B},
we comparatively studied elliptical host galaxies
(distances ${\leq {\rm 150~Mpc}}$) of 66 normal SNe~Ia and 41 subluminous/91bg-like events
from the footprint of the Sloan Digital Sky Survey (SDSS), without considering SN~Ia LC properties.
Our results supported the earlier suggestion \citep[e.g.][]{2005ApJ...634..210G,2008ApJ...685..752G}
that the characteristics of normal SNe~Ia and 91bg-like events depend more on age than on
mass or metallicity of the elliptical host galaxies.
We showed, that the majority of the elliptical hosts of 91bg-like events are very old ($>8$~Gyr)
in comparison with those of normal SNe~Ia, which are on average bluer and
might have more residual star formation that gives rise to younger/prompt SNe~Ia progenitors.
In other words, we showed that the age distribution of 91bg-like SNe hosts does not extend down to
the stellar ages that produce significant $u$-band fluxes in early-type hosts of normal SNe~Ia,
thus younger stars in these hosts do not produce 91bg-like SNe.
Therefore, we concluded that the \emph{delay time}\footnote{The delay time is the time interval between
the SN~Ia progenitor formation and the later thermonuclear explosion of WD.}
\emph{distribution} (DTD) of 91bg-like SNe is likely weighted toward
long delay times, larger than several Gyr \citep[see also][]{2017NatAs...1E.135C,2019PASA...36...31P}.
These results led us to favour SN~Ia progenitor models such as
He-ignited violent mergers \citep[e.g.][]{2013ApJ...770L...8P}
that have the potential to explain the different DTDs of normal SNe~Ia and 91bg-like events.

In the present study, we significantly improved the inclusion of various spectroscopic subclasses
of Type Ia SNe (normal SN~Ia and peculiar 91T-, 91bg- and 02cx-like events)
from host galaxies with almost all Hubble morphological types, including objects not only from the SDSS,
but also from the Panoramic Survey Telescope and Rapid Response System (Pan-STARRS)
and the SkyMapper Southern Sky Survey, thus covering the entire sky.
In addition, instead of relying only on the discrete spectroscopic classifications,
SNe~Ia in our sample now have the available continuous and extinction-independent
LC decline rate $(\Delta m_{15})$ values in the $B$-band.
The goals of this paper are to properly identify the diversity of SNe~Ia
and better constrain the nature (i.e. the progenitor channel, through the DTD) of their different subclasses
through a comprehensive study of SN~Ia LC decline rates and global properties of their hosts
(e.g. morphology, stellar mass, colour, and age of stellar population)
in a well-defined sample of more than 400 SNe~Ia from relatively nearby galaxies.

This is the seventh paper of the series, the content of which is as follows.
Section~\ref{samplered} presents the sample selection and reduction.
Our results and interpretations are presented in Section~\ref{RESults}.
Finally, Section~\ref{DISconc} summarizes our conclusions.
To conform to data values used in our series of the papers,
a Hubble constant $H_0=73 \,\rm km \,s^{-1} \,Mpc^{-1}$ is adopted.

\section{Sample selection and reduction}
\label{samplered}

\subsection{SN~Ia sample}
\label{samplered1}

We used the Open Supernova Catalog \citep[][]{2017ApJ...835...64G}
to collect spectroscopically classified SNe~Ia
with distances ${\leq {\rm 150~Mpc}}$
($z \lesssim 0.036$),\footnote{The luminosity distances of SNe and/or host galaxies
are calculated using the recession velocities both
corrected to the centroid of the Local Group and for the Virgocentric infall
\citep[see][for more details]{2012A&A...544A..81H}.}
discovered up to 2019 May~1.
All SNe~Ia are required to have a $B$-band LC decline rate ($\Delta m_{15}$).
We compiled the LC decline rates with their errors from various publications where
different LC fitters were applied on the $B$-band photometry
(or measured directly from the LCs) with wide
temporal coverage for individual SNe
\citep[e.g.][]{2009ApJ...700..331H,2010ApJS..190..418G,2013ApJ...773...53F,
2016MNRAS.460.3529A,2019MNRAS.490.3882S}.
Note that the $B$-band is historically the most often used in SN studies,
and therefore the data is best sampled in this band.
It should be noted also that the $\Delta m_{15}$ of an SN~Ia is a weak function of the
line-of-sight dust extinction towards the SN in host galaxy,
which affects the LCs \citep[][]{1999AJ....118.1766P}:
$\Delta m_{15}^{\rm true} = \Delta m_{15}^{\rm obs} + 0.1 \, E(B-V)$,
where the last term in the equation contains colour excess towards the SN.
The $E(B-V)$ values are mostly distributed within 0 to 0.3~mag,
with the mean value of $\sim0.1$~mag \citep[e.g.][]{1999AJ....118.1766P,2016MNRAS.460.3529A}.
Therefore, the term in the equation is between 0 to 0.03~mag,
while the mean error in $\Delta m_{15}$ estimation is
$\sim0.1$~mag.\footnote{In some cases, the values of $E(B-V)$ towards SN can reach up to $\sim0.5$~mag,
and only in unique cases up to around 1~mag \citep[e.g.][]{2016MNRAS.460.3529A,2020arXiv200615164U}.
However, in all these cases the errors in $\Delta m_{15}$ estimations are higher,
reaching up to about 0.2~mag.}
For this reason, we consider the $B$-band $\Delta m_{15}$
as a practically extinction-independent parameter.
In the compilation of SNe~Ia LC decline rates, we avoided using any transformation from
LC stretch\footnote{The stretch parameter, usually used in cosmological studies for the
standardization of SN~Ia, is related to the width of the SN~Ia LC
\citep[see][for more details]{2015EJPh...36a5007C}.}
to $\Delta m_{15}$, because it is unclear how reliable such a transformation is
\citep[e.g.][]{2016MNRAS.460.3529A,2019MNRAS.486.5785S}.

\begin{table}
  \centering
  \begin{minipage}{84mm}
  \caption{Numbers and fractions of SNe~Ia subclasses in our
           (distances ${\leq {\rm 150~Mpc}}$) and LOSS (distances ${\leq {\rm 80~Mpc}}$)
           volume-limited samples.}
  \tabcolsep 10.5pt
  \label{ourNumLOSS}
  \begin{tabular}{lrrrr}
  \hline
    &\multicolumn{2}{c}{Our}&
    \multicolumn{2}{c}{LOSS}\\
    \multicolumn{1}{c}{SN~subclass}&\multicolumn{1}{c}{$N_{\rm SN}$}&
    \multicolumn{1}{c}{$fr$}&\multicolumn{1}{c}{$N_{\rm SN}$}&\multicolumn{1}{c}{$fr$}\\
  \hline
    normal &303&$73.5^{+2.8}_{-3.0}$&52&$70.3^{+7.0}_{-7.8}$\\
    91T &42&$10.2^{+2.2}_{-1.9}$&7&$9.4^{+5.9}_{-4.1}$\\
    91bg &50&$12.1^{+2.3}_{-2.0}$&11&$14.9^{+6.7}_{-5.2}$\\
    02cx &12&\multirow{2}{*}{$4.1^{+1.5}_{-1.2}$}&2&\multirow{2}{*}{$5.4^{+5.1}_{-3.0}$}\\
    02es &5&&2&\\
    06gz &2& \multicolumn{1}{c}{---} &0& \multicolumn{1}{c}{---}\\
    \\
    All &414&&74&\\
  \hline
  \end{tabular}
  \parbox{\hsize}{\emph{Notes.} To be comparable with the LOSS sample,
                  the fractions (in per cent) of SN subclasses in our sample
                  are calculated, using the approach of \citet{2011PASA...28..128C},
                  without two 06gz-like SNe. Despite the different nature of
                  02cx- and 02es-like events \citep[][]{2015ApJ...799...52W},
                  the LOSS merged these SNe in a single bin, and we therefore presented the
                  corresponding merged fraction in our sample as well
                  (however the SN numbers are presented separately).}
\end{minipage}
\end{table}

The collected SNe~Ia are also required to have available subclasses
of their spectroscopy (e.g. normal, 91T-, 91bg-, and 02cx-like).
Following \citet[][]{2019MNRAS.490..718B}, we carried out an extensive literature search
to compile the information on the subclasses.
As an interactive archive of SN spectra and corresponding references,
we mostly used the Weizmann Interactive Supernova data Repository \citep[][]{2012PASP..124..668Y}.
We considered also the websites of Central Bureau for Astronomical
Telegrams\footnote{\href{http://www.cbat.eps.harvard.edu/iau/cbat.html}{http://www.cbat.eps.harvard.edu/iau/cbat.html}},
the Astronomer's Telegram\footnote{\href{http://www.astronomerstelegram.org/}{http://www.astronomerstelegram.org/}},
and other references with information on the SN subclasses
\citep[e.g.][]{2012MNRAS.425.1789S,2013ApJ...773...53F,2020ApJ...892..121K}.
Note that we included in the 91T-like SNe subclass the transitional 99aa-like events.
Very early spectra of these SNe resemble those of 91T-like events.
The difference with 91T-like SNe is that in the spectra of 99aa-like SNe the \mbox{Ca\,{\sc ii}} lines
are always prominent. However, at maximum light the 99aa-like SNe spectra are similar
to those of normal SNe~Ia. In addition, the photometry shows that 99aa-like SNe are
similarly luminous and slow-declining as 91T-like SNe.
At the same time, we included two transitional 86G-like events in the 91bg-like SNe subclass.
Spectroscopically, these events show properties intermediate between normal and 91bg-like SNe.
The \mbox{Si\,{\sc ii}} lines are as strong as in 91bg-like spectra,
while \mbox{Ti\,{\sc ii}} lines are somewhat weaker.
The LCs of 86G-like events peak at magnitudes intermediate between those of normal and 91bg-like SNe,
and decline rapidly similar to the latter.
For more details on the properties of these transitional events,
the reader is referred to \citet{2017hsn..book..317T}.
In total, we succeeded to collect the subclasses for
414 SNe~Ia with available LC decline rates.
The second column of Table~\ref{ourNumLOSS} shows the numbers of
different subclasses of the collected Type Ia SNe.

We compared the fractions of SN subclasses in our sample with those
in the volume-limited sample of the Lick Observatory Supernova Search
\citep[LOSS;][]{2011MNRAS.412.1441L}.
Based on the detection efficiency of LOSS, the authors noted that their
survey did not miss any SNe~Ia that exploded within the sample of targeted
galaxies out to 80~Mpc, achieving a completeness more than 98 per cent (74 events).
Table~\ref{ourNumLOSS} shows that the representation of the SN subclasses
in our sample is in good agreement within errors with that in the LOSS.
We checked this by applying the two-sample Kolmogorov--Smirnov (KS) and
Anderson--Darling (AD) tests on the reconstructed two discrete data sets (distributions)
with sizes of 414 and 74, using the ordered frequencies $\{303, 42, 50, 12, 5, 2\}_{\rm Our}$
and $\{52, 7, 11, 2, 2, 0\}_{\rm LOSS}$ from Table~\ref{ourNumLOSS}, respectively.
The $P$-value of two-sample KS (AD) test represents the probability that the two distributions being compared
are randomly drawn from the same parent distribution.
Following consolidated tradition, in this paper we adopted
the threshold of $P = 0.05$ for significance levels of statistical tests.
The KS and AD tests showed that the frequencies of SN subclasses in
our and LOSS samples are not significantly different ($P_{\rm KS}=0.728$ and $P_{\rm AD}=0.929$).

\begin{figure}
\begin{center}$
\begin{array}{@{\hspace{0mm}}c@{\hspace{0mm}}}
\includegraphics[width=1\hsize]{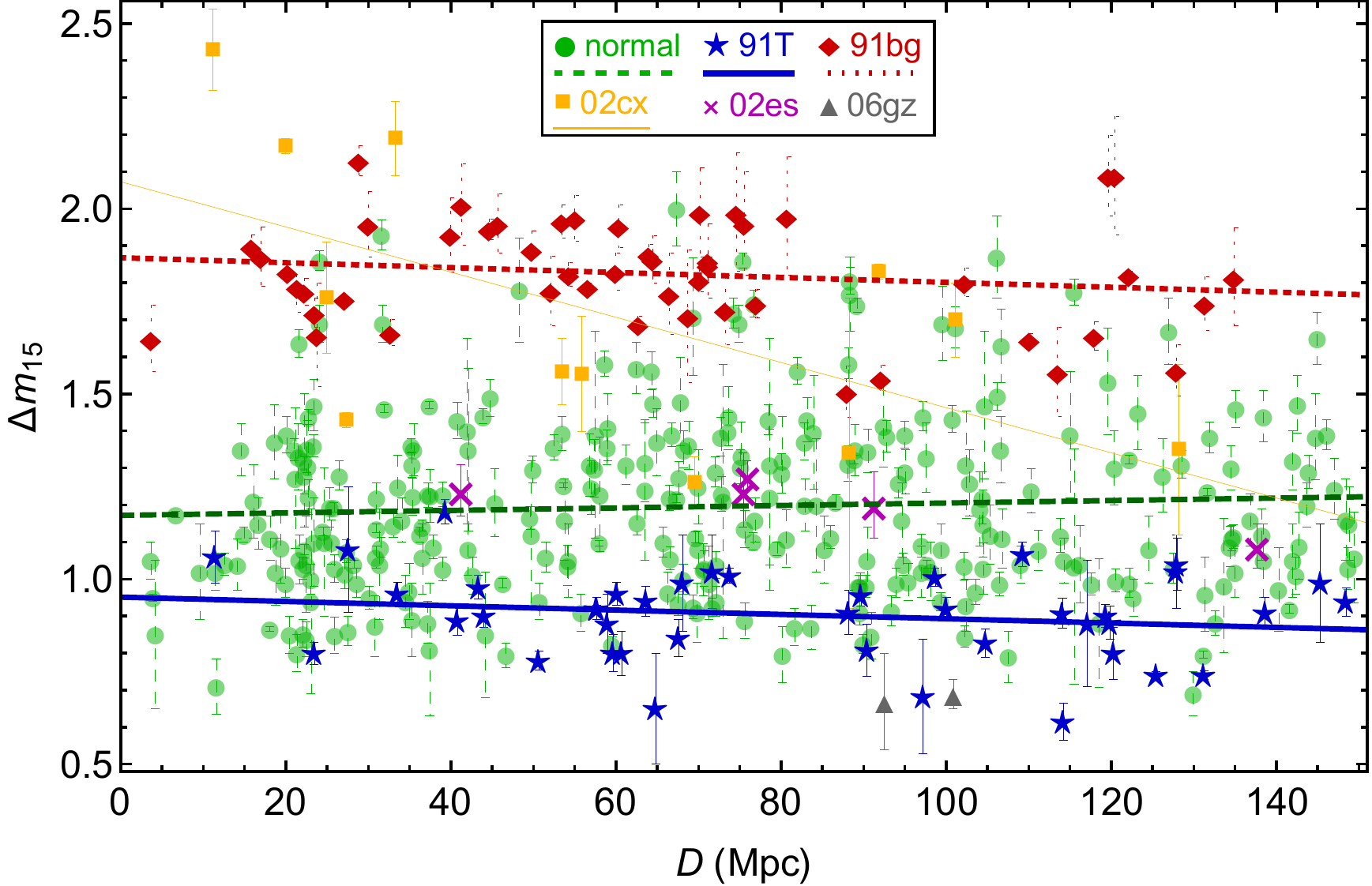}
\end{array}$
\end{center}
\caption{Variation of the $B$-band LC drop, $\Delta m_{15}$, of different subclasses of
         SNe~Ia as a function of the host galaxy distance.
         Because of very few data points for 02es- and 06gz-like SNe,
         their best-fitting lines are useless and not presented.}
\label{Delm15vsDis}
\end{figure}

For small-size samples like in Table~\ref{ourNumLOSS}, in addition to
the original two-sample KS (or AD) test, a
Monte Carlo (MC) simulation is usually used for a better approximation of the $P$-value
\citep[e.g.][]{FeigelsonBabu2012}.
If the original two data samples have $m$ and $n$ members,
the MC simulation randomly generates partitions of joint $m+n$ size sample
into an $m$-set and an $n$-set, computing the two-sample KS (or AD) test statistic
$D(m,n)$ for the generated sets.
The $P^{\rm MC}_{\rm KS}$-value (or $P^{\rm MC}_{\rm AD}$-value)
is the proportion of $D_i(m,n)$ values
that are greater than $d$, where $i$ is the number of iterations
(we used $i = 10^5$) and $d$ is the test statistic
based on the original two data samples.
The described analysis also showed that the frequencies of SN subclasses in our
and LOSS samples are consistent between each other
($P^{\rm MC}_{\rm KS}=0.710$ and $P^{\rm MC}_{\rm AD}=0.650$).
Thus, we believe that the artificial loss/excess of any of the SN subclasses
in our sample is not significant.

It is important to recall that the
91T-like SNe have peak luminosities that are $\sim0.6$~mag
overluminous than do normal SNe~Ia,
while 91bg-like SNe have luminosities that are $\sim2$~mag lower
in comparison with normal ones.
In general, the peak luminosities of 91bg- and 02es-like SNe are comparable.
On the other hand, the peak luminosities of some 06gz-like events are even higher
than those of 91T-like SNe, while the 02cx-like SNe can be less luminous than 91bg-like SNe
\citep[see][for a review on the extremes of SNe~Ia]{2017hsn..book..317T}.
Therefore, the discoveries of the events with lower luminosities
might be complicated at greater distances.

\begin{table}
  \centering
  \begin{minipage}{84mm}
  \caption{Numbers of SNe~Ia subclasses and
           coefficients of the linear best-fits ($\Delta m_{15} = a + b \, D[{\rm Mpc}]$)
           from Fig.~\ref{Delm15vsDis} with results of the Spearman's rank correlation test.}
  \tabcolsep 5.1pt
  \label{Delm15vsDisSpRo}
  \begin{tabular}{lrcrrc}
  \hline
    \multicolumn{1}{c}{SN~subclass}&\multicolumn{1}{c}{$N_{\rm SN}$}&
    \multicolumn{1}{c}{$a$}&
    \multicolumn{1}{c}{$b\times10^{-3}$}&\multicolumn{1}{c}{$r_{\rm s}$}&
    \multicolumn{1}{c}{$P_{\rm s}$}\\
  \hline
    normal &303&$1.17\pm0.03$&$0.3\pm0.3$&0.057&0.320\\
    91T &42&$0.95\pm0.05$&$-0.6\pm0.5$&$-$0.131&0.407\\
    91bg &50&$1.87\pm0.04$&$-0.7\pm0.6$&$-$0.135&0.350\\
    02cx &12&$2.07\pm0.18$&$-6.1\pm2.6$&$-$0.559&0.058\\
  \hline
  \end{tabular}
  \parbox{\hsize}{\emph{Notes.} Spearman's coefficient ($r_{\rm s}\in[-1;1]$)
                  is a nonparametric measure of rank correlation,
                  it assesses how well the relationship between two variables can be described
                  using a monotonic function. Null hypothesis of the test is that the variables
                  are independent, and alternative hypothesis is that they are not ($P_{\rm s} \leq 0.05$).}
\end{minipage}
\end{table}

To check the possible influence of the distance effect, in Fig.~\ref{Delm15vsDis}
we illustrated the dependence of extinction-independent $\Delta m_{15}$,
which is a good proxy for the intrinsic luminosity of a SN~Ia \citep[e.g.][]{1993ApJ...413L.105P},
on distance for the different SN subclasses.
The parameters of best-fitting lines from Fig.~\ref{Delm15vsDis} and results of
the Spearman's rank correlation test for $\Delta m_{15}$ versus distance (in Mpc)
are presented in Table~\ref{Delm15vsDisSpRo}.
The Spearman's rank test shows not significant trends between
the $\Delta m_{15}$ and distances for all the SN subclasses.
Only for 02cx-like SNe, the $P_{\rm s}$-value of negative trend ($r_{\rm s}<0$)
is close to the 0.05 threshold, but still statistically not significant
(see Table~\ref{Delm15vsDisSpRo}).

Note that, in this study we avoided merging 02cx- and 02es-like SNe subsamples,
because they are two distinct subclasses based on their spectroscopic, photometric,
and host galaxy properties \citep[e.g.][]{2015ApJ...799...52W}.
For this reason and due to a small number of 02es-like events,
we simply removed these five SNe from our analysis.
For the same reasons, we also removed two 06gz-like SNe from our statistical/comparative study.
These unusual and rare SNe (02es- and 06gz-like) and their host galaxies
will be the subject of a forthcoming paper in this series.

\begin{figure}
\begin{center}$
\begin{array}{@{\hspace{0mm}}c@{\hspace{0mm}}}
\includegraphics[width=1\hsize]{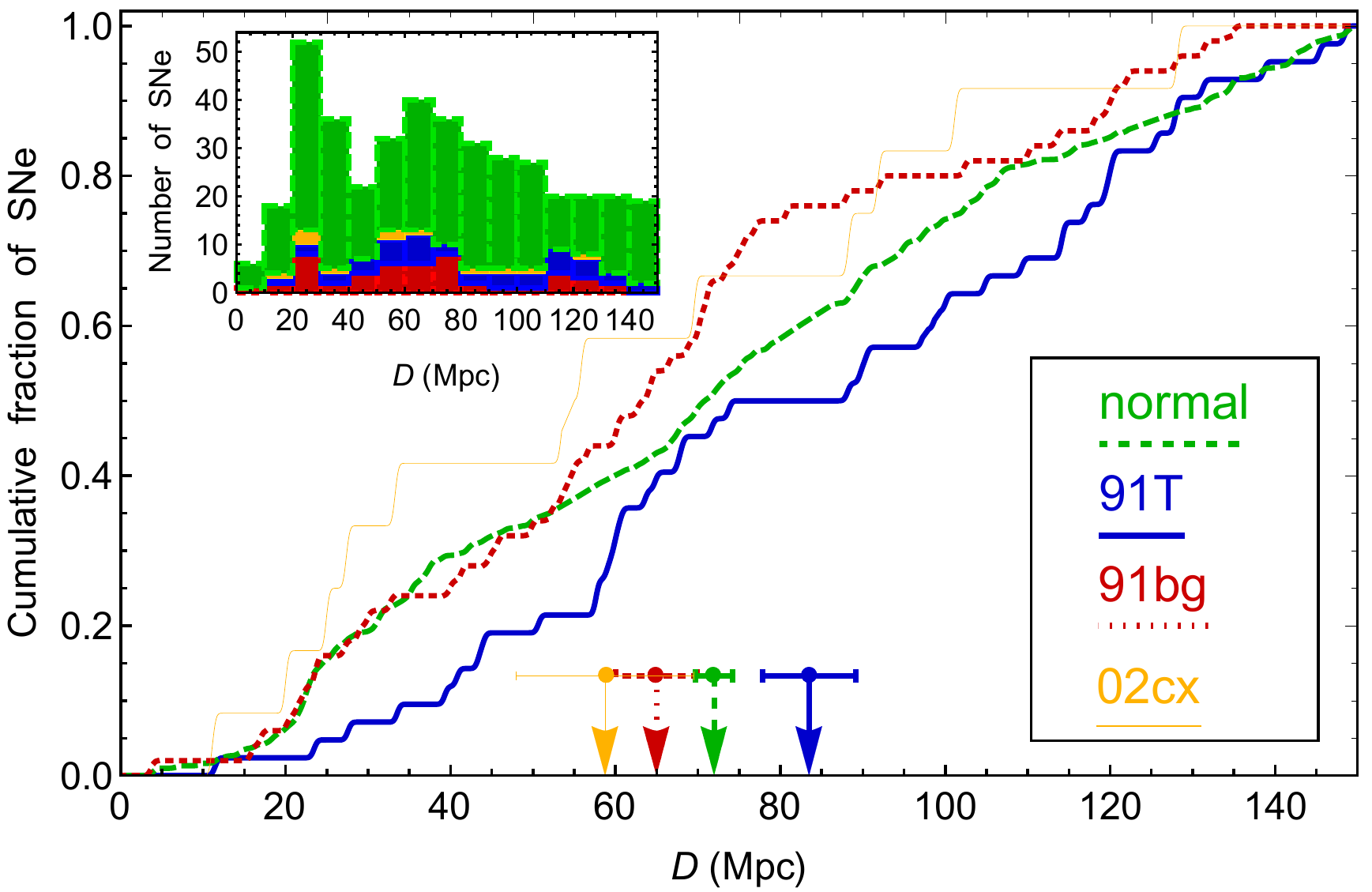}
\end{array}$
\end{center}
\caption{Cumulative distributions and stacked histogram (upper left inset) of
         distances of the subclasses of Type Ia SNe.
         The mean values (with standard errors) of the distributions are shown
         by arrows (with error bars).}
\label{subclassDis}
\end{figure}

To further check the redshift-dependent biases in our SN sample,
in Fig.~\ref{subclassDis} we compared the cumulative distance distributions of the SN subclasses.
In Table~\ref{SNsubKSADDis}, the two-sample KS and AD tests
showed that the distance distributions of normal, 91T-, and 91bg-like events
are not significantly different from one another and could thus be drawn from
the same parent distribution. However, only in the AD statistic,
the distance distributions of 91T- and 02cx-like
SNe are not consistent between each other.
At the same time, the corresponding $P$-value for a MC simulation is slightly higher than
the significance threshold (see Table~\ref{SNsubKSADDis}).
Therefore, in comparison with more luminous SNe,
the detection of 02cx-like events might be biased/complicated at
greater distances (see Fig.~\ref{Delm15vsDis} and
the mean values in Fig.~\ref{subclassDis}),
most probably due to their low intrinsic luminosity at maximum light
\citep[e.g.][]{2017hsn..book..317T}.
With this in mind, we will be cautious in our comparison of the properties of
02cx-like SNe and their host galaxies with those of other SN subclasses.
None of our SN~Ia subsamples (except perhaps 02cx-like) should be
affected by redshift-dependent biases.

\begin{table}
  \centering
  \begin{minipage}{84mm}
  \caption{Comparison of the distributions of distances $D$ (in Mpc) among different SN subclasses.}
  \tabcolsep 2.85pt
  \label{SNsubKSADDis}
    \begin{tabular}{llcllcccc}
    \hline
  \multicolumn{2}{c}{SN subsample~1} & \multicolumn{1}{c}{vs} & \multicolumn{2}{c}{SN subsample~2} &\multicolumn{1}{c}{$P_{\rm KS}$} & \multicolumn{1}{c}{$P_{\rm AD}$} & \multicolumn{1}{c}{$P_{\rm KS}^{\rm MC}$} & \multicolumn{1}{c}{$P_{\rm AD}^{\rm MC}$} \\
  \multicolumn{1}{l}{Subclass} & \multicolumn{1}{c}{$\langle D \rangle$} && \multicolumn{1}{l}{Subclass} & \multicolumn{1}{c}{$\langle D \rangle$} &&&& \\
  \hline
    normal & $72\pm2$ & vs & 91T & $83\pm6$ & 0.094 & 0.071 & 0.089 & 0.074 \\
    normal & $72\pm2$ & vs & 91bg & $65\pm5$ & 0.137 & 0.206 & 0.131 & 0.216 \\
    normal & $72\pm2$ & vs & 02cx & $59\pm11$ & 0.613 & 0.426 & 0.594 & 0.471 \\
    91T & $83\pm6$ & vs & 91bg & $65\pm5$ & 0.060 & 0.055 & 0.056 & 0.061 \\
    91T & $83\pm6$ & vs & 02cx & $59\pm11$ & 0.126 & \textbf{0.038} & 0.104 & 0.051 \\
    91bg & $65\pm5$ & vs & 02cx & $59\pm11$ & 0.863 & 0.843 & 0.822 & 0.867 \\
  \hline
  \end{tabular}
  \parbox{\hsize}{\emph{Notes.} The $P_{\rm KS}$ is the two-sample KS test
                  probability that the two distributions being compared (with respective means
                  and standard errors) are drawn from the same parent distribution.
                  The $P_{\rm KS}^{\rm MC}$ is the KS test probability
                  using a MC simulation with $10^5$ iterations as explained in the text.
                  Analogically, the $P_{\rm AD}$ and $P_{\rm AD}^{\rm MC}$ probabilities are from the AD test.
                  The statistically significant difference between the distributions
                  is highlighted in bold.}
  \end{minipage}
\end{table}

\subsection{SN~Ia host galaxy sample}
\label{samplered2}

Our sample of SN~Ia host galaxies was obtained by
cross-matching the coordinates of our 407 SNe~Ia with
the footprints of the SDSS sixteenth data release \citep[DR16;][]{2020ApJS..249....3A},
the Pan-STARRS second data release \citep[DR2;][]{2016arXiv161205560C}, and
the SkyMapper second data release \citep[DR2;][]{2019PASA...36...33O},
using the techniques described in \citet{2012A&A...544A..81H}.
Note that these surveys cover together the entire sky,
complementing one another, and with some intersection.
As a result, we identified 394 individual host galaxies of all collected SNe~Ia:
1, 2, and 4 SNe are found in 383, 10, and 1 galaxies, respectively.
It should be noted that some of the identified host galaxies from the footprint of SDSS DR16
are already listed in databases of
\citet{2012A&A...544A..81H,2016MNRAS.456.2848H,2017MNRAS.471.1390H,
2018MNRAS.481..566K,2019MNRAS.490..718B}, which are based on older data releases of SDSS.
We nevertheless re-implemented the entire reduction process for all SDSS host galaxies
using the DR16 properties.

Following the approach described in detail in \citet{2012A&A...544A..81H},
we morphologically classified all 394 host galaxies
by visual inspection of images of hosts from the
SDSS\footnote{\href{http://skyserver.sdss.org/dr16/en/tools/chart/listinfo.aspx}{http://skyserver.sdss.org/dr16/en/tools/chart/listinfo.aspx}},
Pan-STARRS\footnote{\href{https://ps1images.stsci.edu/cgi-bin/ps1cutouts}{https://ps1images.stsci.edu/cgi-bin/ps1cutouts}},
and SkyMapper\footnote{\href{http://skymapper.anu.edu.au/sky-viewer/}{http://skymapper.anu.edu.au/sky-viewer/}}
imaging servers, all of which build RGB colour images from the $g$, $r$, and $i$ data channels.
Note that SDSS, Pan-STARRS, and SkyMapper use different $ugriz$, $grizy$,
and $uvgriz$ filters, respectively.
However, the $gri$ sets of Pan-STARRS and SkyMapper are very similar to
the SDSS filters of the same names \citep[][]{2016arXiv161205560C,2018PASA...35...10W}.
The SDSS RGB colour images of typical examples of SNe host galaxies with
morphological classification according to the modified Hubble sequence
(E-E/S0-S0-S0/a-Sa-Sab-Sb-Sbc-Sc-Scd-Sd-Sdm-Sm) can be found in \citet{2012A&A...544A..81H}.
Table~\ref{tabSNhostmorph} displays the distributions of the subclasses of SNe~Ia
among the various morphological types of host galaxies.

\begin{table}
  \centering
  \begin{minipage}{84mm}
  \caption{Numbers of the subclasses of Type Ia SNe at distances $\leq150$~Mpc as a function
           of morphological types of host galaxies.}
  \tabcolsep 2.15pt
  \label{tabSNhostmorph}
  \begin{tabular}{lrrrrrrrrrrrrrr}
  \hline
   &\multicolumn{1}{c}{E}&\multicolumn{1}{c}{E/S0}&\multicolumn{1}{c}{S0}&\multicolumn{1}{c}{S0/a}&\multicolumn{1}{c}{Sa}
  &\multicolumn{1}{c}{Sab}&\multicolumn{1}{c}{Sb}&\multicolumn{1}{c}{Sbc}&\multicolumn{1}{c}{Sc}&\multicolumn{1}{c}{Scd}
  &\multicolumn{1}{c}{Sd}&\multicolumn{1}{c}{Sdm}&\multicolumn{1}{c}{Sm}&\multicolumn{1}{r}{All}\\
  \hline
  normal & 20 & 7 & 20 & 22 & 20 & 15 & 57 & 63 & 47 & 13 & 8 & 7 & 4 & 303 \\
  91T & 0 & 0 & 1 & 1 & 1 & 5 & 7 & 13 & 4 & 4 & 2 & 2 & 2 & 42 \\
  91bg & 17 & 3 & 3 & 8 & 5 & 5 & 1 & 4 & 4 & 0 & 0 & 0 & 0 & 50 \\
  02cx & 0 & 0 & 0 & 1 & 0 & 0 & 2 & 3 & 2 & 1 & 0 & 2 & 1 & 12 \\
  \\
  All & 37 & 10 & 24 & 32 & 26 & 25 & 67 & 83 & 57 & 18 & 10 & 11 & 7 & 407 \\
  \hline
  \end{tabular}
  \end{minipage}
\end{table}

We first applied corrections to transform the Pan-STARRS and
SkyMapper photometry to the SDSS system
\citep[see][]{2016ApJ...822...66F,2018PASA...35...10W,2019PASA...36...33O}.
For all host galaxies, we fitted $25~{\rm mag~arcsec^{-2}}$ elliptical apertures in
the SDSS $g$-band according to the approach presented in \citet{2012A&A...544A..81H}.
Following \citet{2019MNRAS.490..718B}, for each host galaxy the corresponding SDSS $ugriz$ fluxes (apparent
magnitudes\footnote{The magnitudes are in the AB system (for more details, see
\href{https://www.sdss.org/dr16/algorithms/fluxcal/}{https://www.sdss.org/dr16/algorithms/fluxcal/}).})
are measured using the mentioned elliptical apertures in $g$-band.
The $u$-band flux measurements are performed only for host galaxies
located on the footprints of SDSS and SkyMapper survey, which have the corresponding filter support.
During the flux measurements, we masked out bright projected/saturated stars.
The absolute/apparent magnitudes are corrected for
Galactic extinction based on the \citet{2011ApJ...737..103S} recalibration of the
\citet*{1998ApJ...500..525S} dust map.
In addition, these values are corrected for elongation/inclination effects
and for the host galaxy internal extinction according to \citet{1995A&A...296...64B}.
Finally, the colour-based K-corrections \citep*{2010MNRAS.405.1409C}
are mostly negligible ($<0.15$~mag), since the host redshifts are $\lesssim 0.036$.

The database of 407 individual SNe~Ia (SN designation, spectroscopic subclass,
$\Delta m_{15}$, and corresponding sources of the data)
and their 394 hosts (galaxy designation, distance,
corrected $ugriz$ apparent magnitudes, and morphological type)
is available in the online version (Supporting Information) of this article.

\section{Results and discussion}
\label{RESults}

With the aim of clarifying the progenitor natures of normal and
peculiar (91T-, 91bg-, and 02cx-like) SN~Ia subclasses,
in this section we comparatively study the important relations between the LC decline rates
of these SNe~Ia and the global properties of the stellar population of their host galaxies
with different morphological types.

\subsection{Light curve decline rates of SN~Ia subclasses}
\label{RESults1}
\begin{figure}
\begin{center}$
\begin{array}{@{\hspace{0mm}}c@{\hspace{0mm}}}
\includegraphics[width=1\hsize]{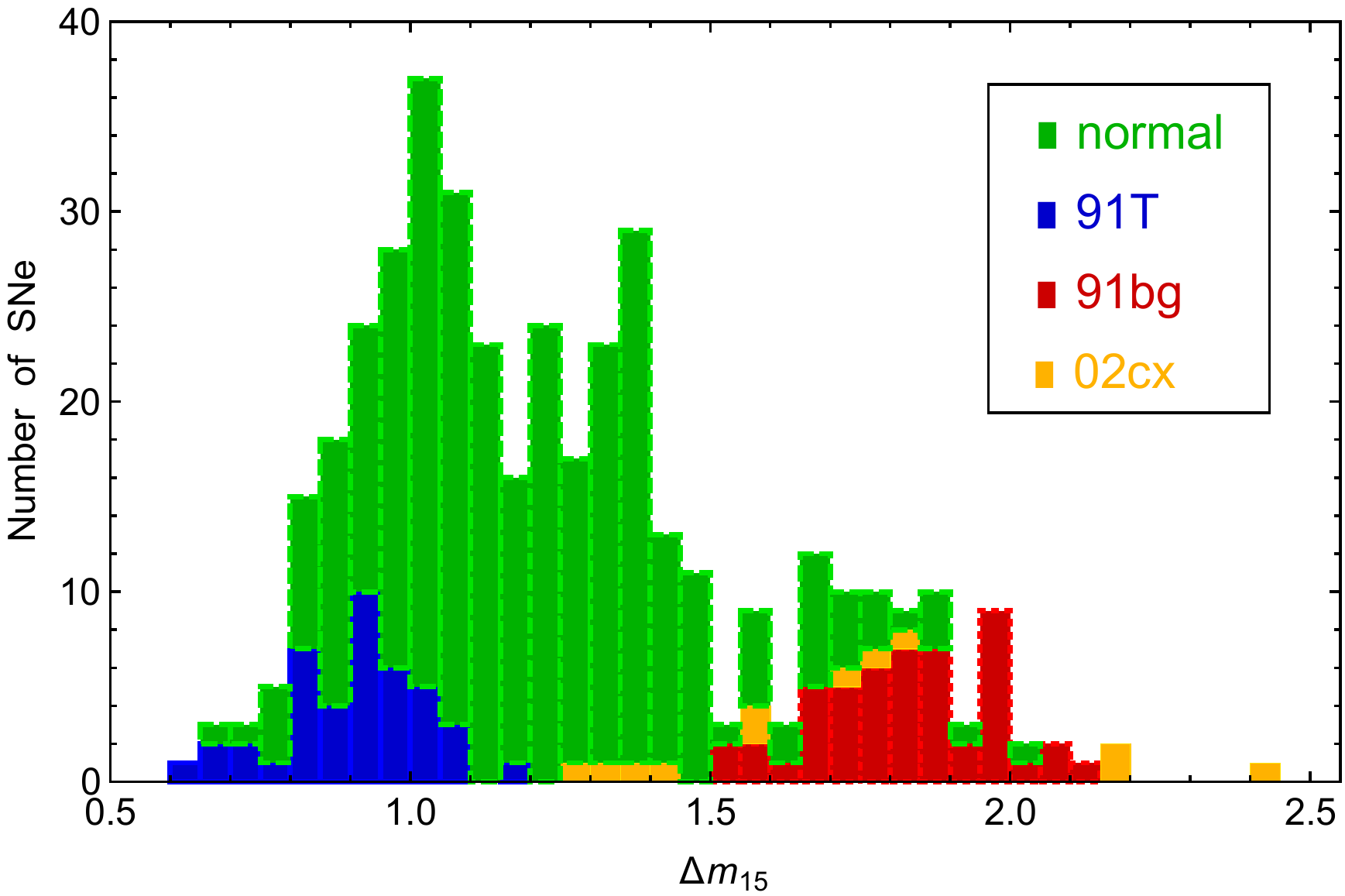}\\
\includegraphics[width=1\hsize]{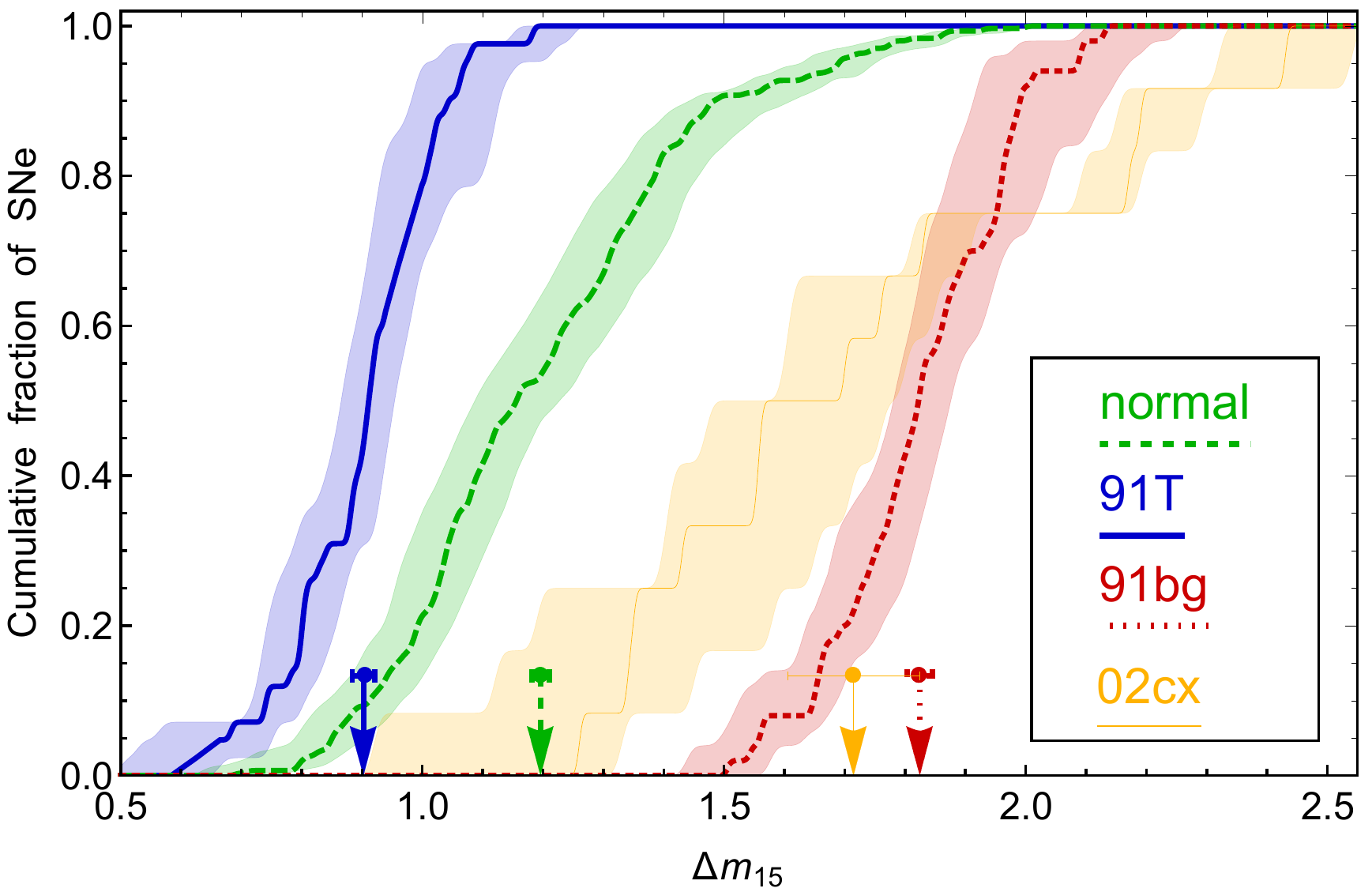}
\end{array}$
\end{center}
\caption{Upper panel: stacked histogram of the $B$-band $\Delta m_{15}$ values
         of different SN~Ia subclasses.
         Bottom panel: cumulative $\Delta m_{15}$ distributions for the SN subclasses.
         The light shaded regions around each cumulative curve show the corresponding
         spreads considering the uncertainties in $\Delta m_{15}$ values.
         The mean values (with standard errors) of the distributions are shown
         by arrows (with error bars).}
\label{Delm15histcum}
\end{figure}
\begin{table*}
  \centering
  \begin{minipage}{123mm}
  \caption{Comparison of the $B$-band $\Delta m_{15}$ distributions among different SN~Ia subclasses.}
  \label{SNsubKSADm15}
    \begin{tabular}{llcllrrrr}
    \hline
  \multicolumn{2}{c}{SN subsample~1} & \multicolumn{1}{c}{vs} & \multicolumn{2}{c}{SN subsample~2} &\multicolumn{1}{c}{$P_{\rm KS}$} & \multicolumn{1}{c}{$P_{\rm AD}$} & \multicolumn{1}{c}{$P_{\rm KS}^{\rm MC}$} & \multicolumn{1}{c}{$P_{\rm AD}^{\rm MC}$}\\
  \multicolumn{1}{l}{Subclass} & \multicolumn{1}{c}{$\langle \Delta m_{15} \rangle$} && \multicolumn{1}{l}{Subclass} & \multicolumn{1}{c}{$\langle \Delta m_{15} \rangle$} &&&& \\
  \hline
    normal & $1.19\pm0.01$ & vs & 91T & $0.90\pm0.02$ & $<$\textbf{0.001} & $<$\textbf{0.001} & $<$\textbf{0.001} & $<$\textbf{0.001}\\
    normal & $1.19\pm0.01$ & vs & 91bg & $1.82\pm0.02$ & $<$\textbf{0.001} & $<$\textbf{0.001} & $<$\textbf{0.001} & $<$\textbf{0.001}\\
    normal & $1.19\pm0.01$ & vs & 02cx & $1.71\pm0.11$ & $<$\textbf{0.001} & $<$\textbf{0.001} & $<$\textbf{0.001} & $<$\textbf{0.001}\\
    91T & $0.90\pm0.02$ & vs & 91bg & $1.82\pm0.02$ & $<$\textbf{0.001} & $<$\textbf{0.001} & $<$\textbf{0.001} & $<$\textbf{0.001}\\
    91T & $0.90\pm0.02$ & vs & 02cx & $1.71\pm0.11$ & $<$\textbf{0.001} & $<$\textbf{0.001} & $<$\textbf{0.001} & $<$\textbf{0.001}\\
    91bg & $1.82\pm0.02$ & vs & 02cx & $1.71\pm0.11$ & \textbf{0.031} & \textbf{0.002} & \textbf{0.025} & \textbf{0.003}\\
  \hline
  \end{tabular}
  \parbox{\hsize}{\emph{Notes.} The explanations for $P$-values
                  are the same as in Table~\ref{SNsubKSADDis}.
                  The statistically significant differences between the distributions
                  are highlighted in bold.}
  \end{minipage}
\end{table*}

Fig.~\ref{Delm15histcum} shows the histogram and cumulative distributions of
$B$-band $\Delta m_{15}$ values for the different SN~Ia subclasses in our sample.
The $\Delta m_{15}$ distribution appears to be bimodal,
with the second (weaker) mode mostly distributed within $\sim 1.5 - 2.1$~mag range.
This faster declining SN range is dominated by 91bg-like (subluminous) events,
while the $\Delta m_{15}$ of 91T-like (overluminous) SNe are distributed only within
the first mode at slower declining range with $\Delta m_{15} \lesssim 1.1$~mag.
This picture of the $\Delta m_{15}$ distributions of faster and slower declining
Type Ia SNe is very similar to the $\Delta m_{15}$ distribution found by \citet{2016MNRAS.460.3529A},
who studied a data set of 165 low redshift ($z < 0.06$) SNe~Ia
\citep[see also][]{2013ApJ...770L...8P,2019MNRAS.486.5785S}.
It should be noted that the separate ranges of $\Delta m_{15}$ distributions
of 91T- and 91bg-like SNe are almost equal to each other ($\sim0.6$~mag),
but about 2.2 times narrower than that of normal SNe~Ia ($\sim1.3$~mag).
At the same time, despite the small number statistics of 02cx-like SNe,
their $B$-band decline rates with some extremes are spread in
the faster side of the $\Delta m_{15}$ distribution of normal SNe~Ia
\citep[see also][]{2015A&A...573A...2S}.
The range of $\Delta m_{15}$ distribution
of these peculiar events is $\sim1.2$~mag.

In Table~\ref{SNsubKSADm15}, two-sample KS and AD tests show that the $\Delta m_{15}$ distributions
are inconsistent significantly between any pairs of SN~Ia subclasses of our sample.
Therefore, the $\Delta m_{15}$ distributions of 91T- and 91bg-like SNe, which cross the
tails of $\Delta m_{15}$ distribution of normal SNe~Ia, without crossing one another
(see Fig.~\ref{Delm15histcum}),
suggest that these SN~Ia subclasses may come from different stellar populations
\citep[e.g.][]{2013ApJ...770L...8P,2016MNRAS.460.3529A}.
A similar idea can be viable also for the progenitor population of 02cx-like SNe
\citep[see][and references therein]{2017hsn..book..375J},
when comparing the $\Delta m_{15}$ properties of normal SNe~Ia with those of
02cx-like events (Fig.~\ref{Delm15histcum} and Table~\ref{SNsubKSADm15}).

In next subsections, we obtain more robust constraints on the SN~Ia progenitor populations,
by studying the host galaxy global properties combined to the distributions of
LC decline rates of the SN subclasses.

\subsection{Morphologies of host galaxies of SN~Ia subclasses}
\label{RESults2}

\begin{figure}
\begin{center}$
\begin{array}{@{\hspace{0mm}}c@{\hspace{0mm}}}
\includegraphics[width=1\hsize]{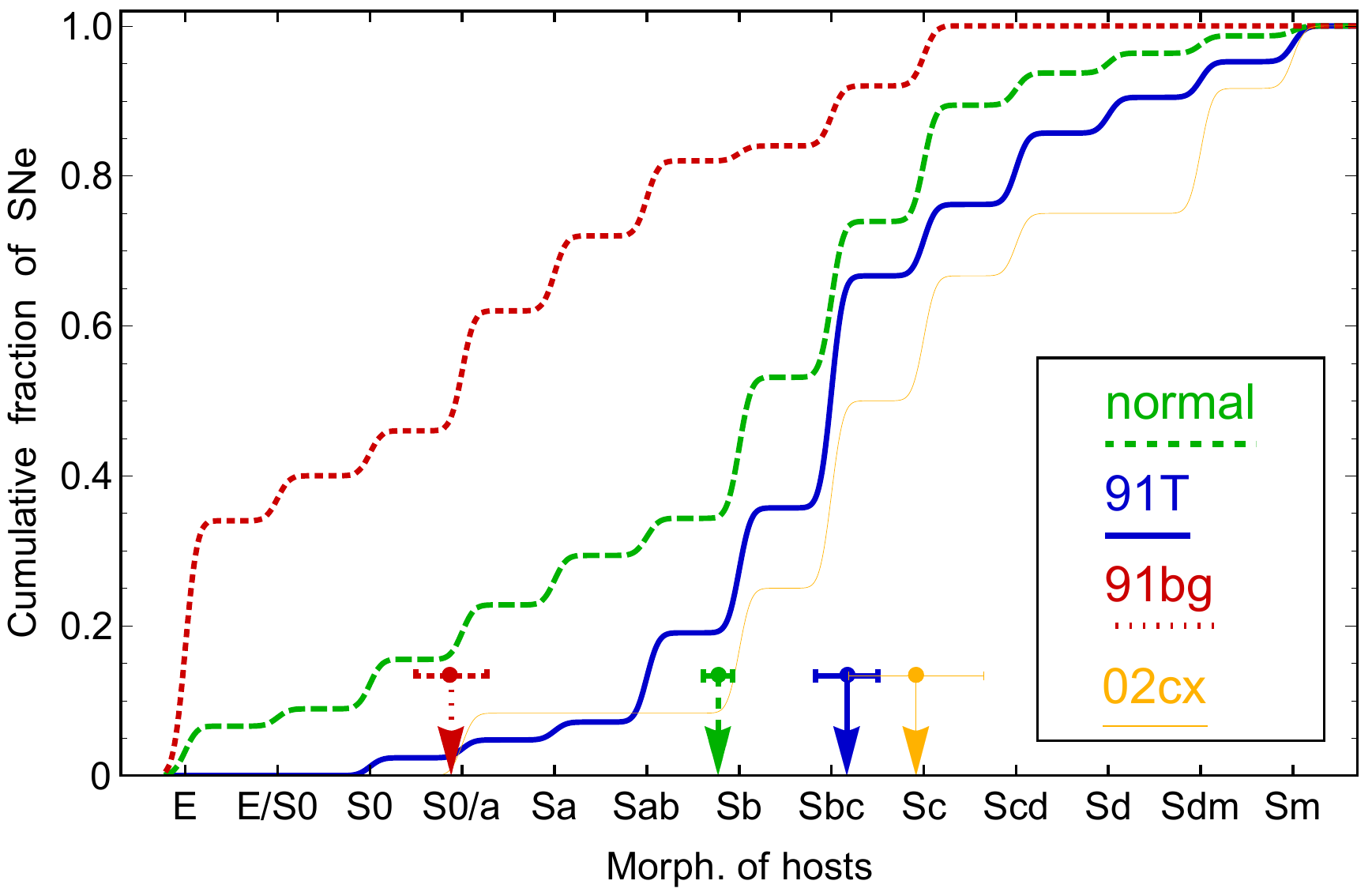}\\
\includegraphics[width=1\hsize]{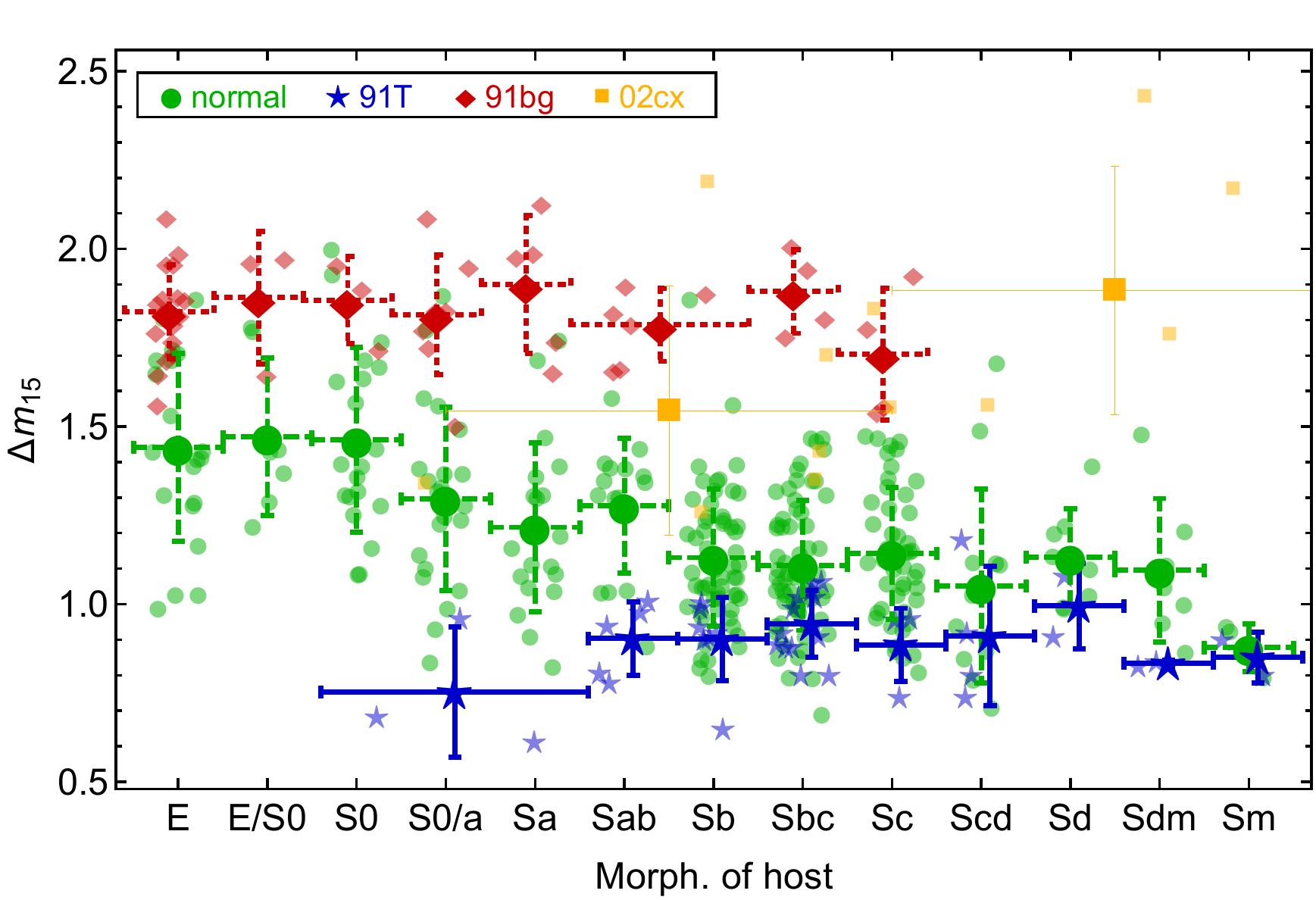}
\end{array}$
\end{center}
\caption{Upper panel: cumulative distributions of host galaxy morphological types of
  the subclasses of Type Ia SNe.
  The mean morphological types (with standard errors) of the host subsamples are
  shown by arrows (with error bars).
  Bottom panel: $B$-band $\Delta m_{15}$ versus host galaxy morphology for
  different SNe~Ia subclasses, displayed as a scatter plot (smaller symbols)
  and averaged in bins of morphological type (bigger symbols).
  The abscissae of the smaller symbols are randomly shifted by $\pm0.3$ in $t$
  for clarity, while they are slightly shifted according to SN~Ia subclass
  for the bigger symbols.
  The S0, S0/a, and Sa morphological bins of 91T-like SNe are merged
  because each of them contains only one 91T-like event.
  For 91bg-like SNe, the Sb bin contains only one SN, therefore it is merged with the Sab bin.
  Similarly, given their small number, the hosts of 02cx-like SNe are merged
  into two broad morphological bins.}
\label{deltm15vsmorph}
\end{figure}
\begin{table*}
  \centering
  \begin{minipage}{129.5mm}
  \caption{Comparison of the distributions of host galaxy morphological types
           among different SN~Ia subclasses.}
  \label{SNsubKSADttype}
    \begin{tabular}{lrclrrrrr}
    \hline
  \multicolumn{2}{c}{Host subsample~1} & \multicolumn{1}{c}{vs} & \multicolumn{2}{c}{Host subsample~2} &\multicolumn{1}{c}{$P_{\rm KS}$} & \multicolumn{1}{c}{$P_{\rm AD}$} & \multicolumn{1}{c}{$P_{\rm KS}^{\rm MC}$} & \multicolumn{1}{c}{$P_{\rm AD}^{\rm MC}$}\\
  \multicolumn{1}{l}{SN subclass} & \multicolumn{1}{c}{$\langle t \rangle$-type} && \multicolumn{1}{l}{SN subclass} & \multicolumn{1}{c}{$\langle t \rangle$-type} &&&& \\
  \hline
    normal & $2.8\pm0.1$ & vs & 91T & $4.2\pm0.3$ & \textbf{0.015} & \textbf{0.004} & \textbf{0.015} & \textbf{0.007}\\
    normal & $2.8\pm0.1$ & vs & 91bg & $-0.1\pm0.4$ & $<$\textbf{0.001} & $<$\textbf{0.001} & $<$\textbf{0.001} & $<$\textbf{0.001}\\
    normal & $2.8\pm0.1$ & vs & 02cx & $4.9\pm0.7$ & 0.134 & \textbf{0.006} & 0.122 & \textbf{0.015}\\
    91T & $4.2\pm0.3$ & vs & 91bg & $-0.1\pm0.4$ & $<$\textbf{0.001} & $<$\textbf{0.001} & $<$\textbf{0.001} & $<$\textbf{0.001}\\
    91T & $4.2\pm0.3$ & vs & 02cx & $4.9\pm0.7$ & 0.673 & 0.470 & 0.668 & 0.475\\
    91bg & $-0.1\pm0.4$ & vs & 02cx & $4.9\pm0.7$ & $<$\textbf{0.001} & $<$\textbf{0.001} & $<$\textbf{0.001} & $<$\textbf{0.001}\\
  \hline
  \end{tabular}
  \parbox{\hsize}{\emph{Notes.} The explanations for $P$-values
                  are the same as in Table~\ref{SNsubKSADDis}.
                  The statistically significant differences between the distributions
                  are highlighted in bold.}
  \end{minipage}
\end{table*}

The upper panel of Fig.~\ref{deltm15vsmorph} presents the cumulative distributions of
host galaxy morphological types of the subclasses of SNe~Ia.
It is clear that the host galaxies of normal, 91T- (overluminous),
and 91bg-like (subluminous) SNe~Ia
have morphological type distributions that are significantly mutually
inconsistent (see Table~\ref{SNsubKSADttype}).
Host galaxies of 91bg-like SNe have, on average, earlier morphological types
$(\langle t \rangle\approx 0)$, with 46 per cent of the events discovered in E--S0 galaxies.
In contrast, host galaxies of 91T-like SNe have, on average, later
morphological types $(\langle t \rangle\approx 4)$,
with a single 91T-like event in E--S0 hosts
(only $\sim 2$ per cent of the subsample, see Table~\ref{tabSNhostmorph}).
The morphological distribution of host galaxies of spectroscopically normal SNe~Ia
occupies an intermediate position between the host morphologies of 91T- and 91bg-like events.
The morphological distribution of 02cx-like SNe hosts is similar to that of
91T-like SNe hosts, though the LC decline rates of the former SN subclass are
significantly larger than those of the latter subclass
(see Fig.~\ref{Delm15histcum} and Table~\ref{SNsubKSADm15}).

\begin{table*}
  \centering
  \begin{minipage}{175mm}
  \caption{Comparison of the $B$-band $\Delta m_{15}$ distributions of SNe~Ia among different subsamples
           of host morphologies.}
  \tabcolsep 5.3pt
  \label{SNhostsubKSADm15}
    \begin{tabular}{llrlcllrlrrrr}
    \hline
  \multicolumn{4}{c}{Subsample~1} & \multicolumn{1}{c}{vs} & \multicolumn{4}{c}{Subsample~2} &\multicolumn{1}{c}{$P_{\rm KS}$} & \multicolumn{1}{c}{$P_{\rm AD}$} & \multicolumn{1}{c}{$P_{\rm KS}^{\rm MC}$} & \multicolumn{1}{c}{$P_{\rm AD}^{\rm MC}$}\\
  \multicolumn{1}{l}{Host} & \multicolumn{1}{l}{SN~subclass} & \multicolumn{1}{c}{$N_{\rm SN}$} & \multicolumn{1}{c}{$\langle \Delta m_{15} \rangle$} && \multicolumn{1}{l}{Host} & \multicolumn{1}{l}{SN~subclass} & \multicolumn{1}{c}{$N_{\rm SN}$} & \multicolumn{1}{c}{$\langle \Delta m_{15} \rangle$} &&&& \\
  \hline
    E--S0 & normal & 47 & $1.45\pm0.04$ & vs & S0/a--Sm & normal & 256 & $1.15\pm0.01$ & $<$\textbf{0.001} & $<$\textbf{0.001} & $<$\textbf{0.001} & $<$\textbf{0.001}\\
    E--S0 & 91bg & 23 & $1.83\pm0.03$ & vs & S0/a--Sm & 91bg & 27 & $1.82\pm0.03$ & 0.876 & 0.840 & 0.806 & 0.842\\
    \\
    E--S0 & normal & 47 & $1.45\pm0.04$ & vs & S0/a--Sbc & normal & 177 & $1.16\pm0.02$ & $<$\textbf{0.001} & $<$\textbf{0.001} & $<$\textbf{0.001} & $<$\textbf{0.001}\\
    E--S0 & normal & 47 & $1.45\pm0.04$ & vs & Sc--Sm & normal & 79 & $1.11\pm0.02$ & $<$\textbf{0.001} & $<$\textbf{0.001} & $<$\textbf{0.001} & $<$\textbf{0.001}\\
    S0/a--Sbc & normal & 177 & $1.16\pm0.02$ & vs & Sc--Sm & normal & 79 & $1.11\pm0.02$ & \textbf{0.050} & 0.067 & \textbf{0.048} & 0.068\\
    S0/a--Sbc & 91T & 27 & $0.91\pm0.02$ & vs & Sc--Sm & 91T & 14 & $0.89\pm0.03$ & 0.608 & 0.322 & 0.557 & 0.366\\
    E--S0 & 91bg & 23 & $1.83\pm0.03$ & vs & S0/a--Sbc & 91bg & 23 & $1.84\pm0.03$ & 0.991 & 0.936 & 0.931 & 0.937\\
  \hline
  \end{tabular}
  \parbox{\hsize}{\emph{Notes.} The explanations for $P$-values
                  are the same as in Table~\ref{SNsubKSADDis}.
                  The statistically significant differences between the distributions
                  are highlighted in bold.
                  Due to small number statistics of 02cx-like SNe, their study in the subsamples of host morphology
                  is statistically useless, and therefore not presented.}
  \end{minipage}
\end{table*}

These results are in good agreement with those of \citet{2014ApJ...795..142G},
who presented a photometric identification technique for
91bg-like SNe, separating them from the normal Type Ia population,
and comparatively studied SNe host galaxy morphologies
\citep[see also][for other SN~Ia subclasses]{2009AJ....138..376F,2009Natur.459..674V,2010Natur.465..322P}.
\citeauthor{2014ApJ...795..142G} showed that the morphological distribution of host galaxies of
91bg-like candidates is significantly earlier ($P_{\rm KS} \simeq 0.002$)
from that of normal SNe~Ia.\footnote{It should be noted that
\citet{2014ApJ...795..142G} also identified other peculiar
SNe~Ia subgroup, which includes 02cx- and 06gz-like events.
However, the results concerning to the hosts morphologies of this peculiar subgroup
are not comparable with those of our case, because their subgroup include nearly equal
numbers of 02cx- and 06gz-like SNe whose hosts morphologies have different distributions
\citep[e.g.][]{2009Natur.459..674V,2011MNRAS.412.2735T,2019MNRAS.490..718B},
thus mixing the morphological types.}
In addition, using a Fisher exact test, \citeauthor{2014ApJ...795..142G} noted that
the number ratio of normal SNe~Ia in passive (E--S0) to star-forming (S0/a--Sm) galaxies
is statistically different (smaller) from the same ratio of 91bg-like SNe
with a probability $P_{\rm F} \simeq 0.002$.
For our SN sample in Table~\ref{tabSNhostmorph}, the behaviour of the ratios is the same
as in \citet{2014ApJ...795..142G}, with $P_{\rm F} < 0.001$.
For the hosts of other SN~Ia subclasses,
\citet{2009AJ....138..376F} and \citet{2010Natur.465..322P} mentioned that
91T- and 02cx-like objects have consistent host morphologies, while hosts of 91T-like SNe
have later morphological types in comparison with normal SNe~Ia.

Taking into account that the mean stellar population age of galaxies is
steadily decreasing along the Hubble sequence from early- to late-type galaxies
\citep[e.g.][]{2015A&A...581A.103G},
the results in the upper panel of Fig.~\ref{deltm15vsmorph} and Table~\ref{SNsubKSADttype}
indicate that the progenitor population age of SN~Ia subclasses in the sequence of
91bg-, normal, and 91T(or 02cx)-like events is decreasing as well.

In the bottom panel of Fig.~\ref{deltm15vsmorph}, we show the distribution
of the $B$-band $\Delta m_{15}$ values as a function of morphological type of host galaxies,
for different SN~Ia subclasses.
When dividing SNe~Ia hosts between E--S0 (galaxies with only old stellar component)
and S0/a--Sm morphological types (galaxies with both old and young stellar components),
Table~\ref{SNhostsubKSADm15} shows that the distributions of $\Delta m_{15}$ values
of 91bg-like SNe are not different between the host subsamples,
being distributed mainly within old ellipticals/lenticulars and early-type spirals
(Fig.~\ref{deltm15vsmorph}).
In contrast, the $\Delta m_{15}$ distribution of normal SNe~Ia in E--S0 hosts
is not consistent with that in S0/a--Sm galaxies.

Using narrow morphological bins, i.e. E--S0, S0/a--Sbc, and Sc--Sm,
shows that the $\Delta m_{15}$ values of normal SNe~Ia are decreasing on average
from early- to late-type galaxies.
Indeed, the Spearman's rank test shows that the significant correlation between
$\Delta m_{15}$ and host $t$-types exists only for normal SNe~Ia
($r_{\rm s}= -0.416$, $P_{\rm s}<0.001$),
although the discrete $t$-type values are not convenient for the test.
The $\Delta m_{15}$ distributions of 91T-like events are not different
in S0/a--Sbc and Sc--Sm galaxies (Table~\ref{SNhostsubKSADm15}),
which lie mostly within spiral hosts (Fig.~\ref{deltm15vsmorph}).
Without a clear separation between the SN~Ia spectroscopic subclasses,
similar results have been obtained in the past with samples of SNe~Ia
and their host galaxies at different redshifts
\citep[e.g.][]{1996AJ....112.2391H,2000AJ....120.1479H,2001ApJ...554L.193H,
2004MNRAS.349.1344A,2005ApJ...634..210G,2016MNRAS.460.3529A,2020arXiv200609433P}.
Note that, because of few data points for 02cx-like SNe,
their study in the subsamples of host morphology is statistically useless,
and therefore not presented in Table~\ref{SNhostsubKSADm15}.

Recall that the intrinsic ranges of $\Delta m_{15}$ distributions of 91T- and
91bg-like SNe are narrower than that of normal SNe~Ia (see Subsection~\ref{RESults1}).
Could these narrow ranges of $\Delta m_{15}$ prevent us from seeing any trends with
host galaxy properties such as morphology?
We probe this issue by searching for trends of $\Delta m_{15}$ and morphology for normal SNe~Ia,
in the corresponding narrow ranges of $\Delta m_{15}$ as for 91T-like events first and then as for 91bg-like SNe.
For 189 normal SNe~Ia with $\Delta m_{15} \lesssim 1.26$~mag, the $r_{\rm s}= -0.177$ and $P_{\rm s}=0.015$,
and for 64 normal events with $\Delta m_{15} \gtrsim 1.38$~mag, the $r_{\rm s}= -0.305$ and $P_{\rm s}=0.014$.
Thus, normal SNe~Ia keep the $\Delta m_{15}$ -- morphology trend direction and significance
even in each of the narrower $\Delta m_{15}$ ranges, showing that the
$\Delta m_{15}$ ranges of 91T- and 91bg-like SNe should
not play a significant role in the absence of the trends for peculiar events.

Along with the mean stellar population ages of galaxies,
the ratio of bulge luminosity (old halo/bulge component) over
disc luminosity (old and young star-forming disc components) is steadily decreasing
along the Hubble sequence \citep[e.g.][]{2009ApJ...705..245O,2011A&A...532A..75D}.
There is no disc component in elliptical galaxies, while the stellar bulge component
is negligible in late-type spiral galaxies where the star-forming disc is prominent.
Therefore, most probably the results above indicate that 91bg-like SNe (subluminous SNe~Ia)
come only from the old stellar component (halo/bulge, old disc) of hosts,
while 91T-like events (overluminous SNe~Ia) originate only from
the young component (star-forming disc) of galaxies.
Faster and slower declining normal SNe~Ia likely come from older and younger
stellar populations of host galaxies, respectively.
Despite their small numbers, 02cx-like SNe typically lie in star-forming galaxies
with late-type morphology (Fig.~\ref{deltm15vsmorph}),
hinting that these events likely originate from young stellar component.

It is instructive to consider counter-examples. The discovery of
SN~2004br \citep{2004IAUC.8340....1G}, a 91T-like event \citep{2012MNRAS.425.1789S},
in NGC~4493 of S0 morphology (see Table~\ref{tabSNhostmorph} and Fig.~\ref{deltm15vsmorph}),
which consists of only old stellar population from a naive point of view.
In NGC~4493, another SN~Ia was also discovered \citep[SN~1994M;][]{1994IAUC.5982....2W}
with a normal spectroscopic classification \citep{2012AJ....143..126B}.
Interestingly, this host galaxy has a distorted stellar disc and shows obvious evidence
of an interaction with neighbor/companion galaxy \citep{2012A&A...544A..81H}.
The age of the younger stellar component in this galaxy is estimated to be down to
about a few hundred Myr \citep{2009MNRAS.397..717S}.
There are many indications that residual star formation episodes
(birth of relatively young stellar component) could also take place in
elliptical or lenticular galaxies, due to galaxy-galaxy interaction with close neighbors
\citep[e.g.][]{2009MNRAS.394.1713K,2014MNRAS.440..889S,2016A&A...588A..68G,2020ApJ...889..132G}.
In such early-type galaxies, in very rare cases, even core-collapse SNe were discovered
\citep[e.g.][]{2008A&A...488..523H,2012A&A...544A..81H,2011ApJ...730..110S,2019ApJ...871...33L}
whose progenitor ages are thought to be up to about hundred Myr
\citep[e.g.][]{2017A&A...601A..29Z}.
Therefore, in our sample, the presence of unique 91T-like SN in the mentioned
interacting lenticular galaxy can be interpreted as a result of such a residual star formation
that could deliver SN~Ia from relatively young progenitors.

\subsection{Colour--mass diagram of SNe~Ia host galaxies}
\label{RESults3}

\begin{figure}
\begin{center}$
\begin{array}{@{\hspace{0mm}}c@{\hspace{0mm}}}
\includegraphics[width=1\hsize]{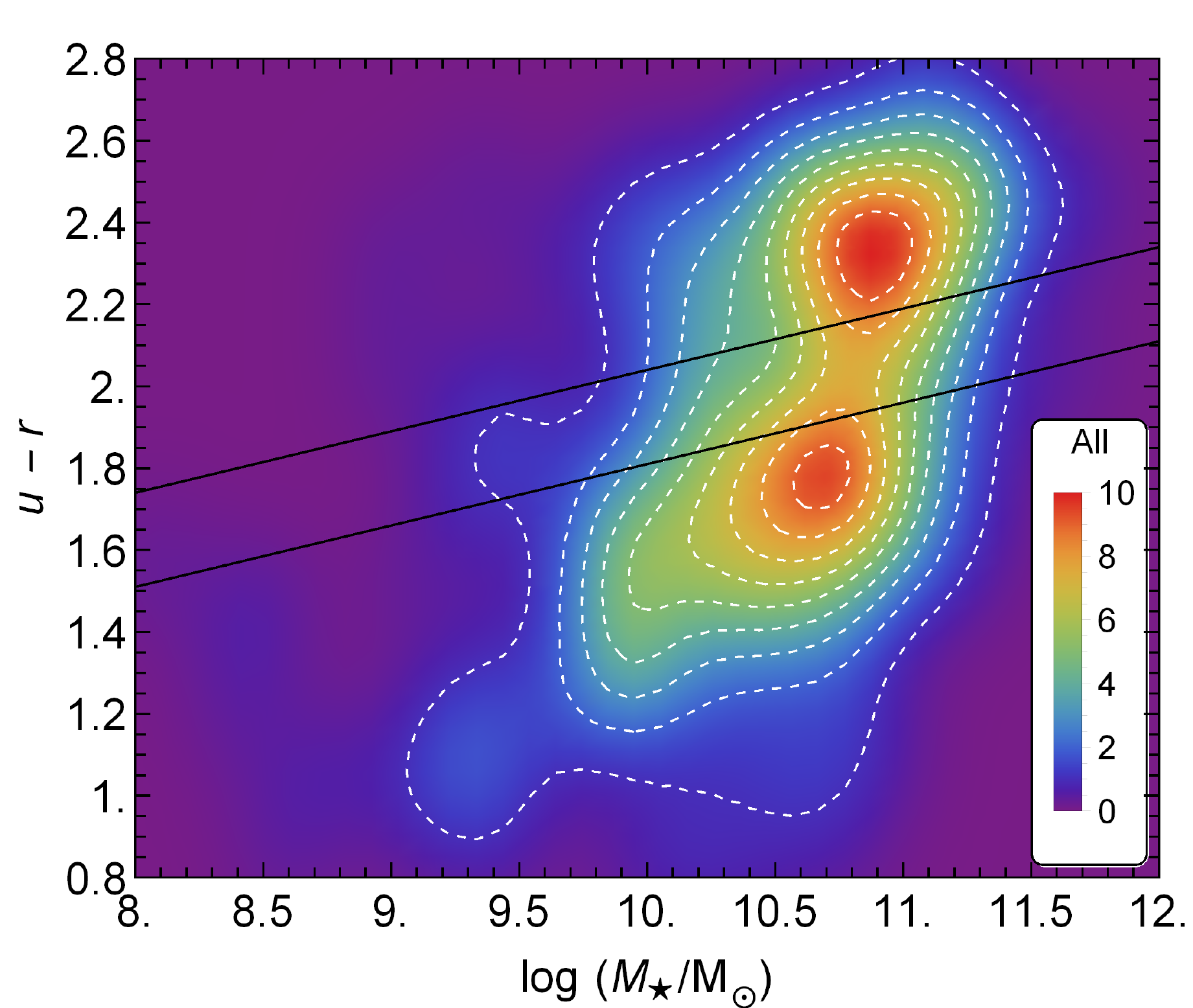}
\end{array}$
\end{center}
\begin{center}$
\begin{array}{@{\hspace{0mm}}c@{\hspace{0.5mm}}c@{\hspace{0mm}}}
\includegraphics[width=0.528\hsize]{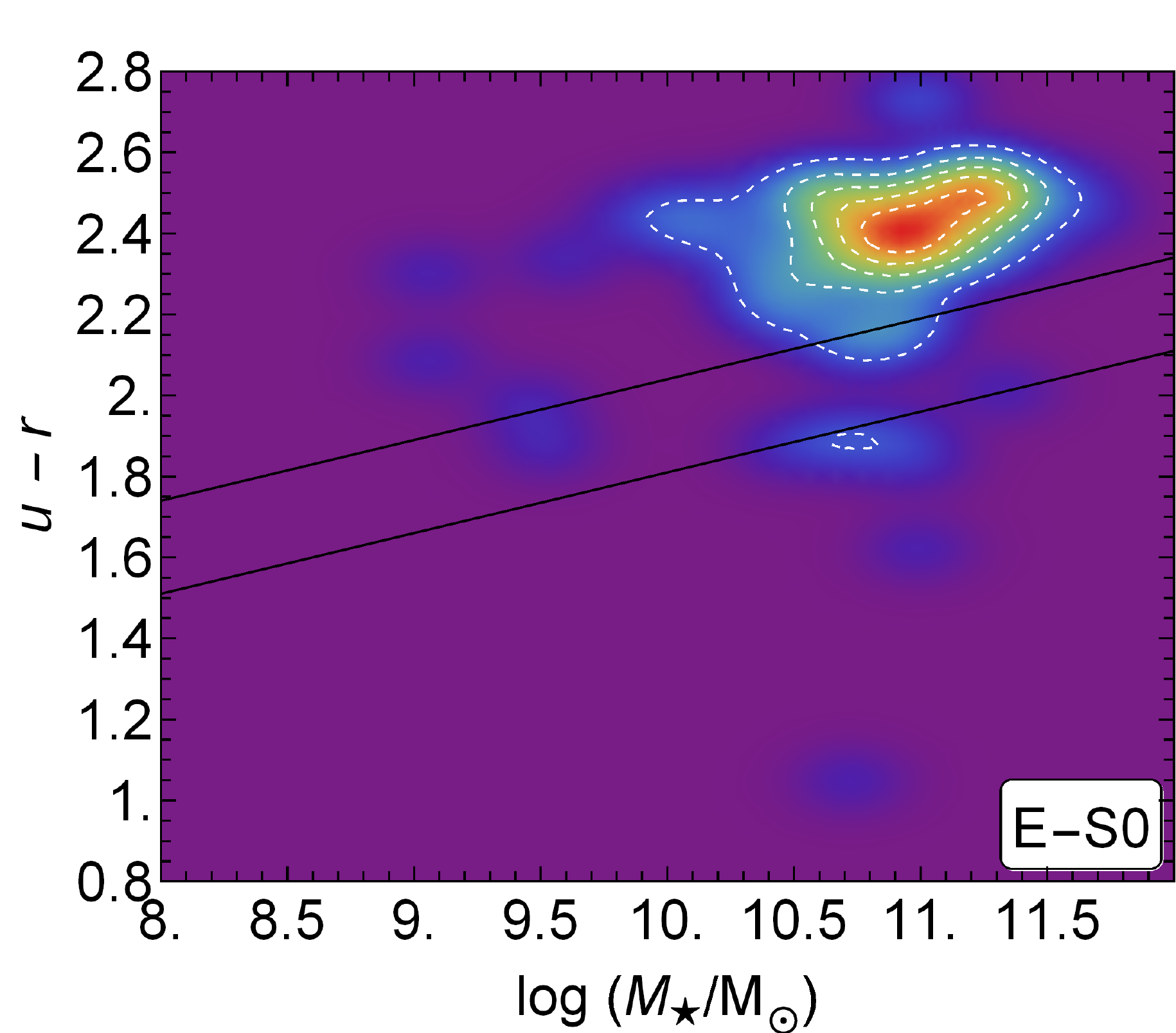} &
\includegraphics[width=0.472\hsize]{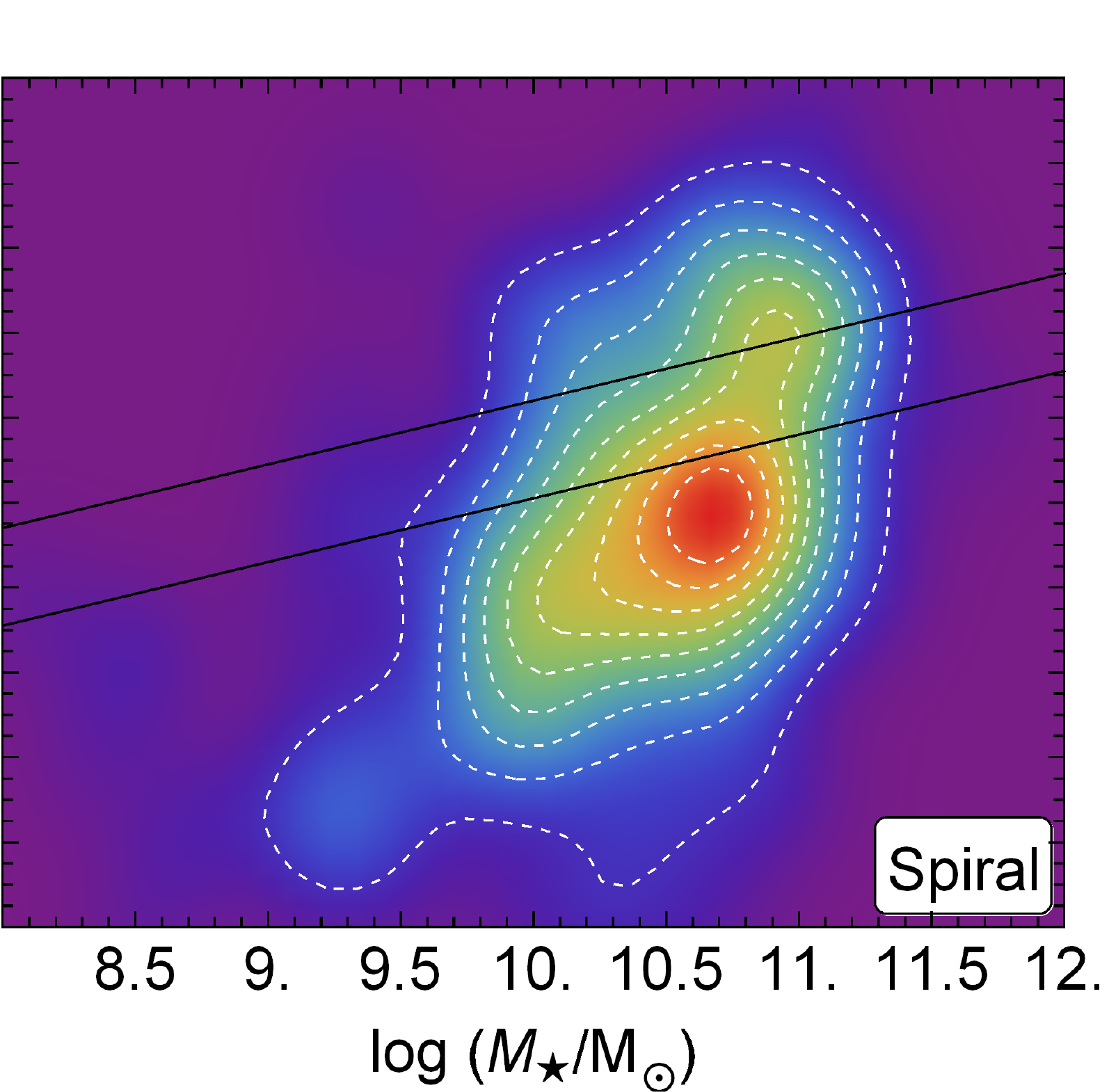}
\end{array}$
\end{center}
\caption{Colour--mass diagrams of SNe~Ia host galaxies viewed as density and contours.
         The region between two solid lines marks the Green Valley (see text for details).
         Upper panel: all 326 SNe~Ia host galaxies with
         measured $u$ and $r$ magnitudes. The colour bar shows the linear
         (arbitrary) units of density.
         Bottom panels: same for the E--S0 (\emph{left}) and
         spiral (\emph{right}) host morphologies.}
\label{CMDDENSit}
\end{figure}

Galaxy colours represent a more quantitative measure of galaxy
classes, albeit spectral classes.
Therefore, in this subsection we only use the photometric data of SNe~Ia hosts,
accompanied with the values of SN LC decline rates.
To analyse the distribution of SN host galaxies on
the colour--mass diagram, we estimate the stellar masses ($M_{\ast}$)
of our sample galaxies, using a simple empirical relation of \citet[][]{2011MNRAS.418.1587T}
between $\log(M_{\ast}/{\rm M_{\odot}})$, $g-i$ colour and $i$-band absolute magnitude, $M_i$
\citep[see][for details]{2019MNRAS.490..718B}.
While this mass estimate is rudimentary and may be biased relative to
more refined mass measurements, it suffices for our aim to.
In the colour--mass diagram (Fig.~\ref{CMDDENSit}),
we prefer to use $u-r$ colours, which present the largest contrast among
optical colours between the inputs of young and old stellar populations
\citep[e.g.][]{2014MNRAS.440..889S,2019MNRAS.490..718B}.
Recall that $u$-band measurements are available only for 326 SNe~Ia
host galaxies (80 per cent of the sample)
located in the footprints of the SDSS and SkyMapper survey,
which provide the corresponding fits images (see Subsection~\ref{samplered2}).

In the upper panel of Fig.~\ref{CMDDENSit},
the $u - r$ colour--mass diagram clearly displays a bimodal distribution
of colours of SNe~Ia host galaxies \citep[see also][for the $g - i$ colours]{2020arXiv200901242P},
as seen in general galaxy samples \citep[e.g.][]{2006MNRAS.373..469B}. More precisely,
galaxies with dominant old stellar populations and low specific SFRs (mostly massive galaxies)
lie in the so-called Red Sequence of the diagram, which is located at $u-r \gtrsim 2$~mag,
with a tail of about 10 per cent of the population reaching down to $u-r \approx 1.5$~mag
\citep[e.g.][]{2006MNRAS.373..469B,2014MNRAS.440..889S}.
Star-forming galaxies with prominent young stellar population are located in the so-called
Blue Cloud of the diagram, mostly at $u-r \lesssim 2$~mag, with a tail to redder colours.

The bottom left panel of Fig.~\ref{CMDDENSit} shows the colour--mass relation
for our E--S0 hosts with 83 SNe~Ia.
Similarly, spiral galaxies of the sample that host 243 SNe~Ia are presented
in the bottom right panel of Fig.~\ref{CMDDENSit}.
The bimodality in the diagram is due to the superposition of colours of
these two distinct populations of galaxies \citep[e.g.][]{2014MNRAS.440..889S}.
A dip between bimodal (red and blue) colours in the colour--mass diagram is called Green Valley
\citep[the region between two solid lines in Fig.~\ref{CMDDENSit}, see also][]{2019MNRAS.490..718B}.
This region is thought to include galaxies that are in transitional stage of
the evolution between star-forming galaxies in the Blue Cloud and passively evolving
quenched galaxies in the Red Sequence
\citep[e.g.][]{2006MNRAS.373..469B,2011ApJ...736..110M,2014MNRAS.440..889S,2014MNRAS.442..533M}.

\begin{figure*}
\begin{center}$
\begin{array}{@{\hspace{0mm}}c@{\hspace{0mm}}}
\includegraphics[width=0.75\hsize]{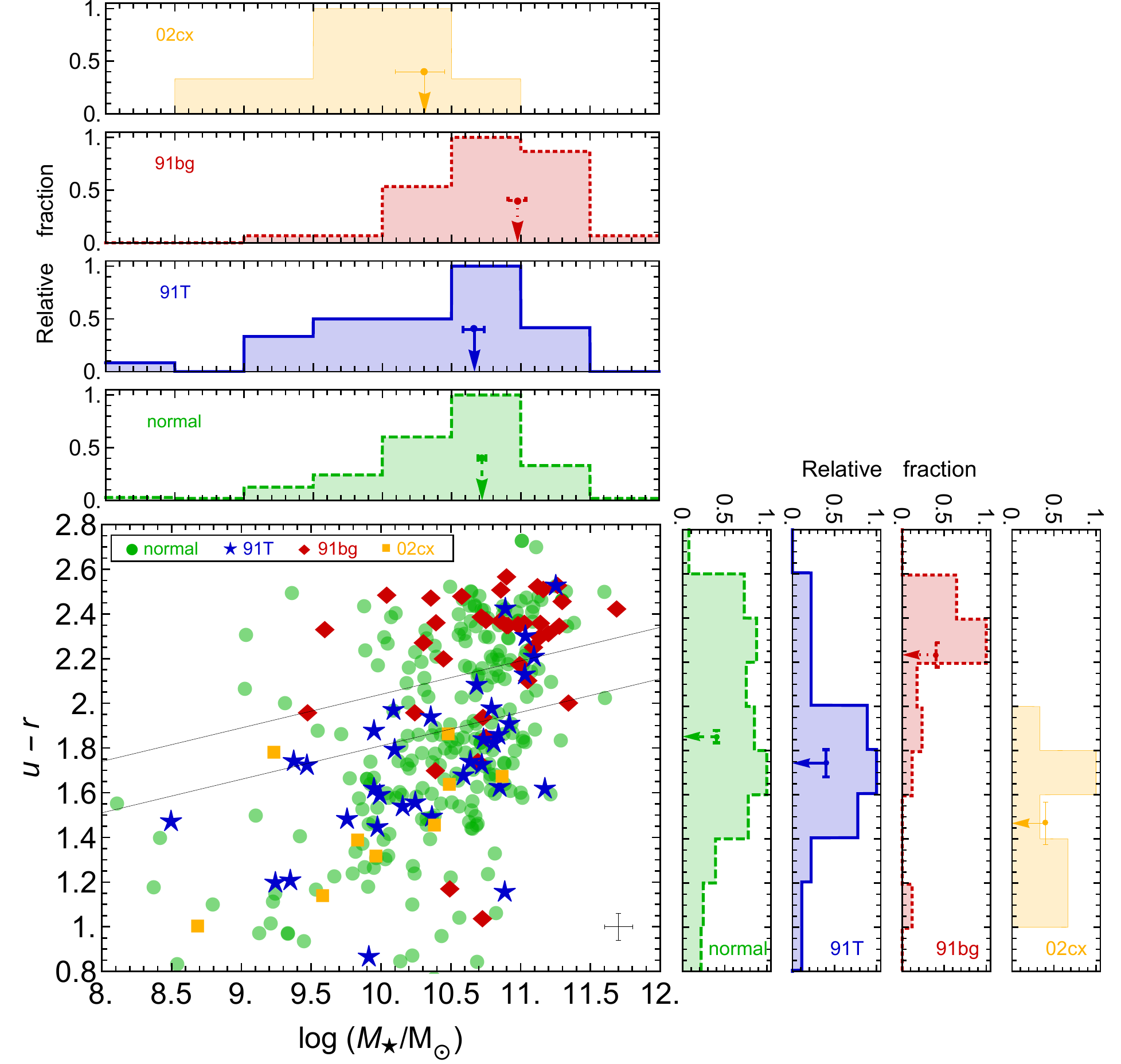}
\end{array}$
\end{center}
\caption{Colour--mass diagram for 326 SNe~Ia host galaxies with
         measured $u$ and $r$ magnitudes, displayed as scatter plots and distributions.
         In the bottom right corner of the lower left panel,
         the error bar represents the characteristic errors
         in our estimations of colours and masses of galaxies.
         The region between two solid lines marks the Green Valley (see text for details).
         For host galaxies of different SN subclasses, the right and upper panels show
         separately the histograms (distributions of relative fractions) of the colours and masses,
         respectively. The mean values (with standard errors) of the distributions are shown by arrows
         (with error bars).}
\label{colMagALL}
\end{figure*}
\begin{table*}
  \centering
  \begin{minipage}{161mm}
  \caption{Comparison of the distributions of host galaxy $u-r$ colours and stellar masses
           among different SN~Ia subclasses.}
  \label{SNsubKSADurLogM}
    \begin{tabular}{lrcclrcrrrr}
    \hline
  \multicolumn{3}{c}{Host subsample~1} & \multicolumn{1}{c}{vs} & \multicolumn{3}{c}{Host subsample~2} &\multicolumn{1}{c}{$P_{\rm KS}$} & \multicolumn{1}{c}{$P_{\rm AD}$} & \multicolumn{1}{c}{$P_{\rm KS}^{\rm MC}$} & \multicolumn{1}{c}{$P_{\rm AD}^{\rm MC}$}\\
  \multicolumn{1}{l}{SN subclass} & \multicolumn{1}{l}{$N_{\rm SN}$} & \multicolumn{1}{c}{$\langle u-r \rangle$} && \multicolumn{1}{l}{SN subclass} & \multicolumn{1}{l}{$N_{\rm SN}$} & \multicolumn{1}{c}{$\langle u-r \rangle$} &&&& \\
  \hline
    normal & 244 & $1.86\pm0.03$ & vs & 91T & 34 & $1.74\pm0.06$ & 0.064 & 0.092 & 0.060 & 0.101\\
    normal & 244 & $1.86\pm0.03$ & vs & 91bg & 39 & $2.23\pm0.05$ & $<$\textbf{0.001} & $<$\textbf{0.001} & $<$\textbf{0.001} & $<$\textbf{0.001}\\
    normal & 244 & $1.86\pm0.03$ & vs & 02cx & 9 & $1.47\pm0.10$ & \textbf{0.020} & \textbf{0.007} & \textbf{0.016} & \textbf{0.013}\\
    91T & 34 & $1.74\pm0.06$ & vs & 91bg & 39 & $2.23\pm0.05$ & $<$\textbf{0.001} & $<$\textbf{0.001} & $<$\textbf{0.001} & $<$\textbf{0.001}\\
    91T & 34 & $1.74\pm0.06$ & vs & 02cx & 9 & $1.47\pm0.10$ & 0.140 & \textbf{0.041} & 0.117 & 0.059\\
    91bg & 39 & $2.23\pm0.05$ & vs & 02cx & 9 & $1.47\pm0.10$ & $<$\textbf{0.001} & $<$\textbf{0.001} & $<$\textbf{0.001} & $<$\textbf{0.001}\\
    \\
   && \multicolumn{1}{c}{$\langle\log(M_{\ast}/{\rm M_{\odot}})\rangle$} &&&& \multicolumn{1}{c}{$\langle\log(M_{\ast}/{\rm M_{\odot}})\rangle$} &&&& \\
    normal & 244 & $10.72^{+0.03}_{-0.03}$ & vs & 91T & 34 & $10.66^{+0.07}_{-0.08}$ & 0.342 & 0.432 & 0.325 & 0.444\\
    normal & 244 & $10.72^{+0.03}_{-0.03}$ & vs & 91bg & 39 & $10.98^{+0.06}_{-0.07}$ & $<$\textbf{0.001} & $<$\textbf{0.001} & $<$\textbf{0.001} & $<$\textbf{0.001}\\
    normal & 244 & $10.72^{+0.03}_{-0.03}$ & vs & 02cx & 9 & $10.30^{+0.14}_{-0.21}$ & \textbf{0.030} & \textbf{0.014} & \textbf{0.027} & \textbf{0.024}\\
    91T & 34 & $10.66^{+0.07}_{-0.08}$ & vs & 91bg & 39 & $10.98^{+0.06}_{-0.07}$ & \textbf{0.019} & \textbf{0.001} & \textbf{0.016} & \textbf{0.001}\\
    91T & 34 & $10.66^{+0.07}_{-0.08}$ & vs & 02cx & 9 & $10.30^{+0.14}_{-0.21}$ & 0.175 & 0.130 & 0.143 & 0.169\\
    91bg & 39 & $10.98^{+0.06}_{-0.07}$ & vs & 02cx & 9 & $10.30^{+0.14}_{-0.21}$ & \textbf{0.001} & $<$\textbf{0.001} & $<$\textbf{0.001} & \textbf{0.001}\\
  \hline
  \end{tabular}
  \parbox{\hsize}{\emph{Notes.} The explanations for $P$-values
                  are the same as in Table~\ref{SNsubKSADDis}.
                  The statistically significant differences between the distributions
                  are highlighted in bold.}
  \end{minipage}
\end{table*}
\begin{figure*}
\begin{center}$
\begin{array}{@{\hspace{0mm}}c@{\hspace{0mm}}c@{\hspace{0mm}}}
\includegraphics[width=0.493\hsize]{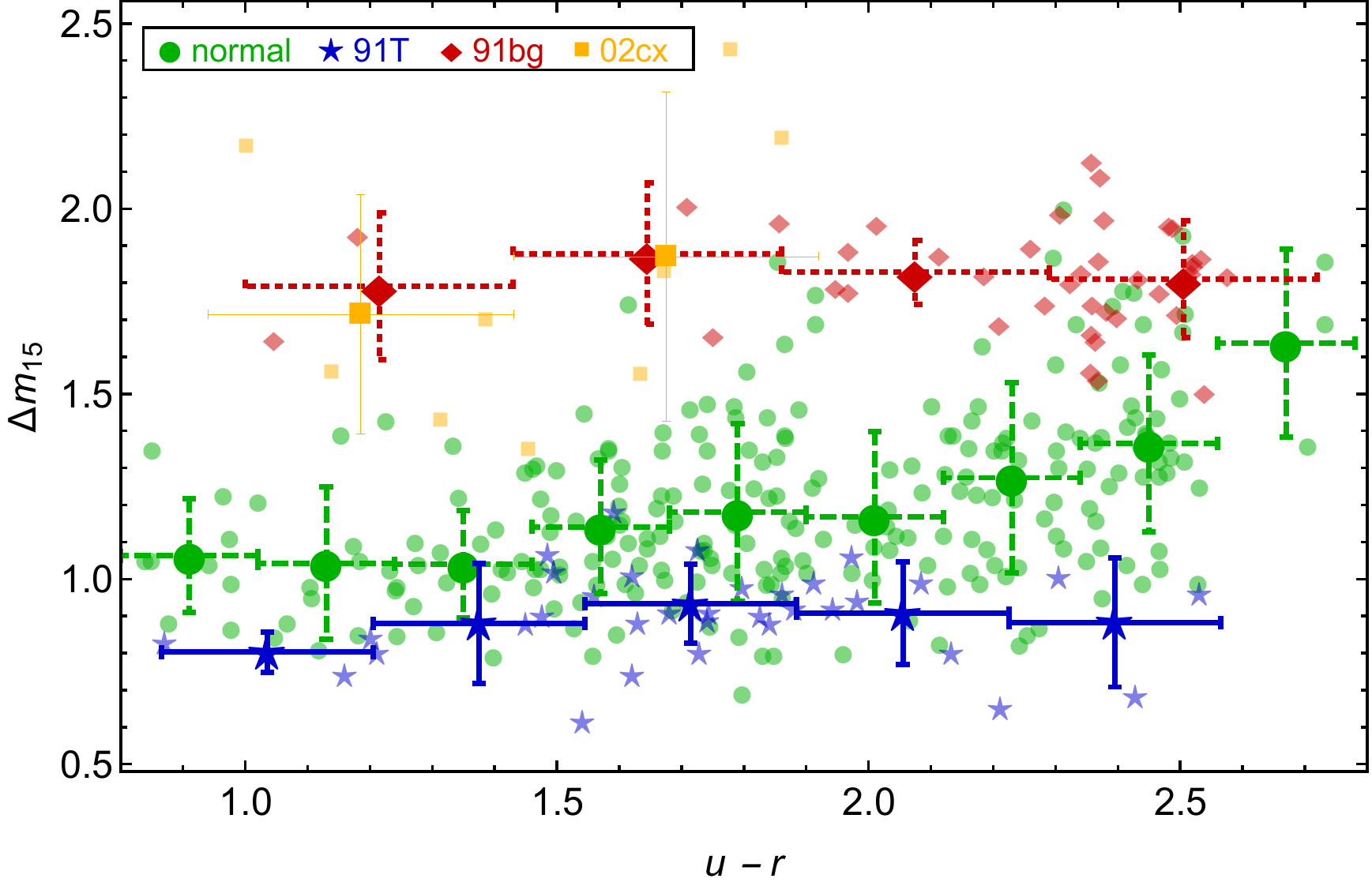} &
\includegraphics[width=0.5\hsize]{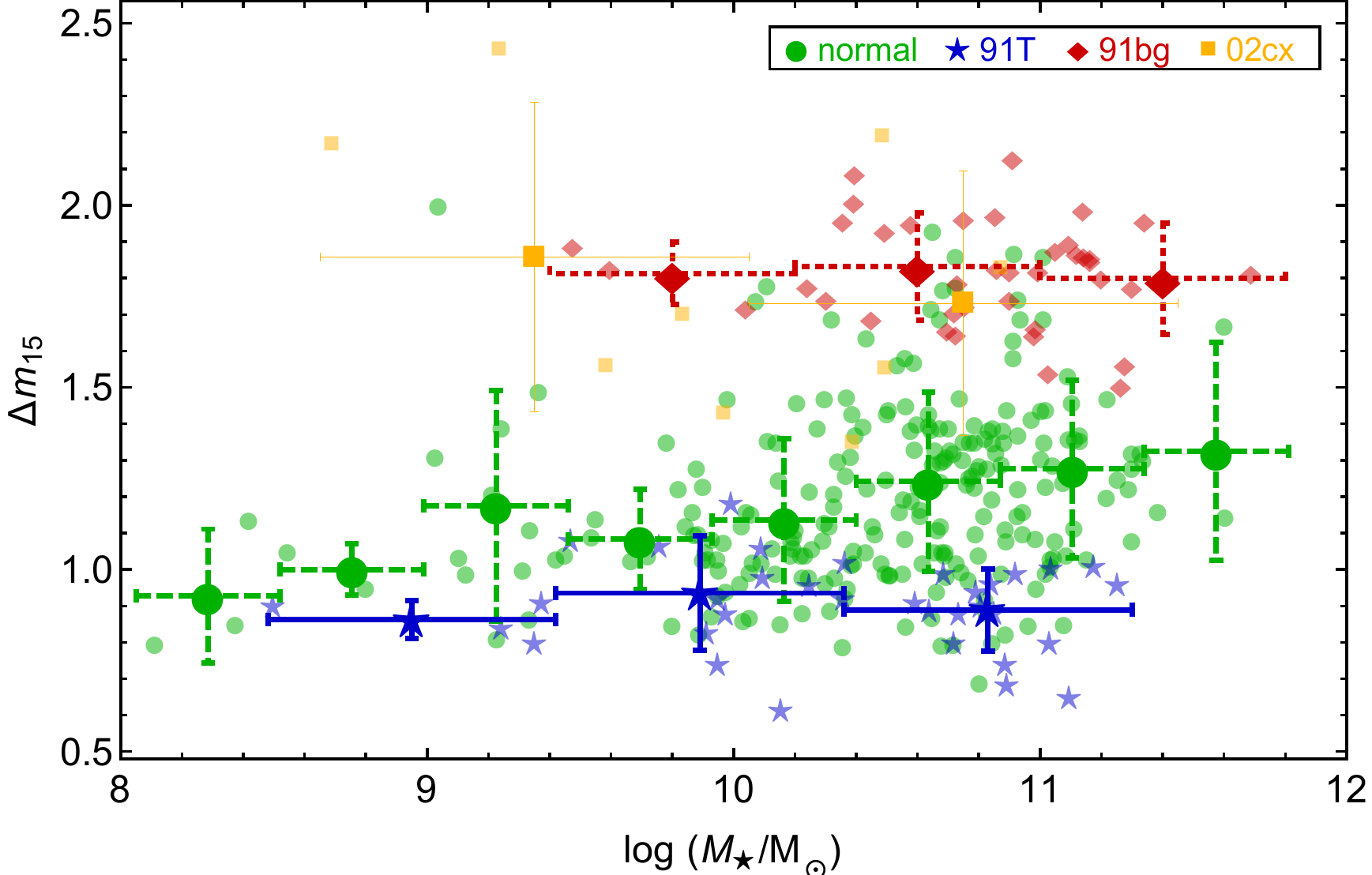}
\end{array}$
\end{center}
\caption{$B$-band $\Delta m_{15}$ values of the SN~Ia subclasses versus host galaxy colour
         (\emph{left}) and host galaxy stellar mass (\emph{right}).
         Binned and averaged values of $\Delta m_{15}$ (bigger symbols) are superposed on
         the original distributions (smaller symbols). Horizontal bars show the bin sizes.
         Depending on the numbers of SNe~Ia for certain subclasses, different binning sizes
         are selected to include sufficient numbers of the objects.}
\label{dm15colLogM}
\end{figure*}

Fig.~\ref{colMagALL} shows how the host galaxies of different SN~Ia
subclasses are distributed in the colour--mass diagram.
The figure also displays the distributions of
$u - r$ colours and masses for host galaxies for the different SN subclasses.
Host galaxies of 91bg-like SNe are clearly located in the Red Sequence,
and most of them have $u - r$ colours $\gtrsim 2$~mag (i.e. above the Green Valley).
In comparison with hosts of normal, 91T-, and 02cx-like SNe, the colour distribution
of host galaxies of 91bg-like SNe are significantly redder (see Table~\ref{SNsubKSADurLogM}).
Also, the bulk of hosts of 91bg-like SNe are significantly massive
$(\log(M_{\ast}/{\rm M_{\odot}}) > 10.5)$.
The distribution of host masses is significantly inconsistent with those of the other
SN~Ia subclasses (Table~\ref{SNsubKSADurLogM}).
At the same time, the colour (resp. mass) distributions are not statistically different
between hosts of normal and 91T-like SNe, spanning almost the entire ranges
of host colour (resp. mass).
Finally, although the small number statistics,
all the host galaxies of 02cx-like SNe are positioned in the Blue Cloud,
mostly below the Green Valley in Fig.~\ref{colMagALL}.
Their colour (mass) distribution is significantly bluer (lower)
in comparison with that of normal SNe~Ia host galaxies,
but closer to that of 91T-like SNe hosts (Table~\ref{SNsubKSADurLogM}).

In Fig.~\ref{dm15colLogM}, we show the distribution of the $B$-band $\Delta m_{15}$ values of
the SN~Ia subclasses as a function of $u - r$ colours and stellar masses of host galaxies.
Interestingly, only normal SNe~Ia show average systematic increase in $\Delta m_{15}$ values
with an increase in colour (mass) values of hosts.
The results of Spearman's rank test, shown in Table~\ref{Dm15vsURLogMSpRo},
confirm that only for normal SNe~Ia we do see a significant correlation between LC
decline rates and host galaxy colours (masses).
We note that similar dependence of SNe~Ia LC properties, such as decline rates,
on the stellar masses of host galaxies have been shown in the literature
using various SN~Ia/host samples \citep[e.g.][]{2009ApJ...691..661H,2009ApJ...707.1449N,
2010MNRAS.406..782S,2011ApJ...727..107G,2014MNRAS.438.1391P,
2016MNRAS.457.3470C,2017ApJ...848...56U,2018A&A...615A..68R,2020arXiv200812101K}.
However, instead of $\Delta m_{15}$, these studies used the SN~Ia LC
shape parameter $\Delta$ (adding and subtracting template LC shapes)
or stretch $x_1$ (stretching or compressing the time axis of the LC by a single ``stretch factor'')
obtained from two commonly used LC fitters in cosmology: MLCS2k2 \citep*{2007ApJ...659..122J}
and SALT2 \citep{2007A&A...466...11G}, respectively.
These parameters ($\Delta$ and $x_1$)
show different correlations with the observed $\Delta m_{15}$
\citep[e.g.][]{2019MNRAS.486.5785S}.
In addition, these studies did not perform a clear separation between
normal and peculiar (91T-, 91bg-, and 02cx-like) SN~Ia spectroscopic subclasses.
\begin{table}
  \centering
  \begin{minipage}{84mm}
  \caption{Results of Spearman's rank correlation test for the $B$-band $\Delta m_{15}$ values
           of the SN~Ia subclasses versus $u-r$ colours and stellar masses of host galaxies.}
  \tabcolsep 3.3pt
  \label{Dm15vsURLogMSpRo}
  \begin{tabular}{lrcccrr}
  \hline
    \multicolumn{1}{c}{SN~subclass}&\multicolumn{1}{c}{$N_{\rm SN}$}&
    \multicolumn{1}{c}{Variable~1}&\multicolumn{1}{c}{vs}&
    \multicolumn{1}{c}{Variable~2}&\multicolumn{1}{c}{$r_{\rm s}$}&
    \multicolumn{1}{c}{$P_{\rm s}$}\\
  \hline
    normal &244&$\Delta m_{15}$& vs &$u-r$&0.429&$<$\textbf{0.001}\\
    normal$^a$ &149&$\Delta m_{15}$& vs &$u-r$&0.210&\textbf{0.010}\\
    normal$^b$ &53&$\Delta m_{15}$& vs &$u-r$&0.285&\textbf{0.038}\\
    91T &34&$\Delta m_{15}$& vs &$u-r$&0.143&0.421\\
    91bg &39&$\Delta m_{15}$& vs &$u-r$&$-$0.092&0.576\\
    02cx &9&$\Delta m_{15}$& vs &$u-r$&0.417&0.264\\
    \\
    normal &244&$\Delta m_{15}$& vs &$\log(M_{\ast}/{\rm M_{\odot}})$&0.286&$<$\textbf{0.001}\\
    normal$^a$ &149&$\Delta m_{15}$& vs &$\log(M_{\ast}/{\rm M_{\odot}})$&0.159&0.052\\
    normal$^b$ &53&$\Delta m_{15}$& vs &$\log(M_{\ast}/{\rm M_{\odot}})$&0.062&0.657\\
    91T &34&$\Delta m_{15}$& vs &$\log(M_{\ast}/{\rm M_{\odot}})$&$-$0.052&0.770\\
    91bg &39&$\Delta m_{15}$& vs &$\log(M_{\ast}/{\rm M_{\odot}})$&$-$0.091&0.582\\
    02cx &9&$\Delta m_{15}$& vs &$\log(M_{\ast}/{\rm M_{\odot}})$&$-$0.300&0.433\\
  \hline
  \end{tabular}
  \parbox{\hsize}{\emph{Notes.} The explanations for $r_{\rm s}$ and $P_{\rm s}$-values
                  are the same as in Table~\ref{Delm15vsDisSpRo}.
                  $^a \Delta m_{15} \lesssim 1.26$~mag, similar to 91T-like SNe.
                  $^b \Delta m_{15} \gtrsim 1.38$~mag, similar to 91bg-like events.
                  Statistically significant correlations between the variables are highlighted in bold.}
\end{minipage}
\end{table}
The dependence of SNe~Ia LC properties on the $u - r$ colours of host galaxies was not directly
studied in the literature,\footnote{But see \citet[][]{2020arXiv200812101K},
for indirectly obtained $U - R$ colours of SNe~Ia host galaxies.}
although it is expected, because similar
dependencies on host morphology and other colours were known \citep[e.g.][]{1996AJ....112.2391H,
2000AJ....120.1479H,2001ApJ...554L.193H,2005ApJ...634..210G,2007ApJ...659..122J,
2016MNRAS.460.3529A,2018A&A...615A..68R,2020arXiv200609433P},
again without clear separation between the spectroscopic subclasses of SNe~Ia.

\begin{table*}
  \centering
  \begin{minipage}{175mm}
  \caption{Comparison of the $B$-band $\Delta m_{15}$ distributions of only normal SNe~Ia among different subsamples
           of host parameters.}
  \tabcolsep 4.8pt
  \label{SNurLogMhstKSADm15}
    \begin{tabular}{llrlcllrlrrrr}
    \hline
  \multicolumn{4}{c}{Subsample~1} & \multicolumn{1}{c}{vs} & \multicolumn{4}{c}{Subsample~2} &\multicolumn{1}{c}{$P_{\rm KS}$} & \multicolumn{1}{c}{$P_{\rm AD}$} & \multicolumn{1}{c}{$P_{\rm KS}^{\rm MC}$} & \multicolumn{1}{c}{$P_{\rm AD}^{\rm MC}$}\\
  \multicolumn{2}{c}{Host parameter} & \multicolumn{1}{c}{$N_{\rm SN}$} & \multicolumn{1}{c}{$\langle \Delta m_{15} \rangle$} && \multicolumn{2}{c}{Host parameter} & \multicolumn{1}{c}{$N_{\rm SN}$} & \multicolumn{1}{c}{$\langle \Delta m_{15} \rangle$} &&&& \\
  \hline
    $u-r$ & $>2$~mag & 100 & $1.30\pm0.02$ & vs & $u-r$ & $\leq2$~mag & 144 & $1.13\pm0.02$ & $<$\textbf{0.001} & $<$\textbf{0.001} & $<$\textbf{0.001} & $<$\textbf{0.001}\\
    $\log(M_{\ast}/{\rm M_{\odot}})$ & $>10.5$ & 139 & $1.26\pm0.02$ & vs & $\log(M_{\ast}/{\rm M_{\odot}})$ & $\leq10.5$ & 105 & $1.13\pm0.02$ & $<$\textbf{0.001} & $<$\textbf{0.001} & $<$\textbf{0.001} & $<$\textbf{0.001}\\
    \multicolumn{2}{l}{above Green Valley} & 81 & $1.33\pm0.03$ & vs & \multicolumn{2}{l}{below/in Green Valley} & 163 & $1.14\pm0.02$ & $<$\textbf{0.001} & $<$\textbf{0.001} & $<$\textbf{0.001} & $<$\textbf{0.001}\\
    \multicolumn{2}{l}{above Green Valley} & 81 & $1.33\pm0.03$ & vs & \multicolumn{2}{l}{below Green Valley} & 133 & $1.13\pm0.02$ & $<$\textbf{0.001} & $<$\textbf{0.001} & $<$\textbf{0.001} & $<$\textbf{0.001}\\
  \hline
  \end{tabular}
  \parbox{\hsize}{\emph{Notes.} The explanations for $P$-values
                  are the same as in Table~\ref{SNsubKSADDis}.
                  A 0.15~mag and 0.25~dex variations in colour and mass demarcations, respectively,
                  do not change the statistical behaviour of the tests.
                  The statistically significant differences between the distributions
                  are highlighted in bold.}
  \end{minipage}
\end{table*}

As in Subsection~\ref{RESults2},
for normal SNe~Ia we check for the correlations between their $\Delta m_{15}$
and host galaxy properties in two separate $\Delta m_{15}$ ranges corresponding to the sizes of
the $\Delta m_{15}$ ranges of 91T- and 91bg-like SNe.
Table~\ref{Dm15vsURLogMSpRo} shows that normal SNe~Ia keep
the $\Delta m_{15}$ -- colour trend direction and significance
in each of the narrower $\Delta m_{15}$ ranges, while
the $\Delta m_{15}$ -- $\log(M_{\ast}/{\rm M_{\odot}})$ trend becomes insignificant,
hinting that the $\Delta m_{15}$ ranges
might play a role in the absence of the $\Delta m_{15}$ -- $\log(M_{\ast}/{\rm M_{\odot}})$
trend for peculiar SNe~Ia.

Without mixing the SN~Ia subclasses, we compare in Table~\ref{SNurLogMhstKSADm15}
the distributions of LC decline rates of normal SNe~Ia between two distinct populations
of host galaxies in the colour--mass diagram.
The separations between two populations of galaxies are done
according to $u - r$ colour \citep[e.g.][]{2014MNRAS.440..889S},
stellar mass \citep[e.g.][]{2017ApJ...848...56U},
or according to locations of galaxies above and below the Green Valley.
Indeed, the Green Valley separates best the two populations of
passively evolving quenched (older) galaxies and of star-forming (younger) hosts.
Table~\ref{SNurLogMhstKSADm15} shows that the distributions of $\Delta m_{15}$ values
of normal SNe~Ia are significantly different between all the mentioned host subsamples.
On average, those normal SNe~Ia that are in galaxies above the Green Valley
have faster declining LCs compared to those in galaxies below the Green Valley.
This result suggests that the correlation seen in Fig.~\ref{dm15colLogM} and
Table~\ref{Dm15vsURLogMSpRo} is due to a superposition of faster and slower declining
normal SNe~Ia from old and young stellar populations of host galaxies,
respectively dominating in the Red Sequence and Blue Cloud of the colour--mass
diagram (see Fig.~\ref{colMagALL}).
In addition to the findings in Subsection~\ref{RESults2},
the colour--mass study of SN~Ia hosts also suggest that there could be at least
two distinct populations of normal SNe~Ia: the halo/bulge and old disc components
of galaxies most likely host faster declining normal SNe~Ia,
while star-forming component of galaxies hosts their slower declining counterparts.

Taken separately, the LC decline rates of 91bg-like SNe (subluminous SNe~Ia)
and 91T-like events (overluminous SNe~Ia) do not show dependencies on
host galaxy colour (Fig.~\ref{dm15colLogM} and Table~\ref{Dm15vsURLogMSpRo}).
At the same time, the distribution of hosts on the colour--mass diagram confirms the known tendency for
91bg-like SNe to occur in globally red/old galaxies while 91T-like events prefer blue/younger hosts.
Therefore, the results in Figs.~\ref{CMDDENSit}, \ref{colMagALL}, and Table~\ref{SNsubKSADurLogM}
reinforce those obtained in Subsection~\ref{RESults2} that among the considered SN~Ia subclasses,
91bg-like SNe come only from old stellar population of hosts
(halo/bulge and old disc components), while 91T-like events
originate only from young population of galaxies (star-forming component).
Probably, the decline rates of 02cx-like SNe
also do not show dependencies on host galaxy properties
(Table~\ref{Dm15vsURLogMSpRo}), although we note that statistics of these events
is based on small numbers.
Complementing the results of Subsection~\ref{RESults2}, the positions of 02cx-like SNe hosts
in the colour--mass diagram suggest that these events may all originate from
the young stellar component, although they diverge from 91T-like events in
their significantly higher LC decline rate.

Since the $g$- and $i$-bands photometry is available for
all 407 SNe~Ia host galaxies, we also check the stellar mass-based
results in Tables~\ref{SNsubKSADurLogM}-\ref{SNurLogMhstKSADm15} using the entire SN~Ia subsamples.
As a result, all behaviours of the dependencies
$(\Delta m_{15}$ vs $\log(M_{\ast}/{\rm M_{\odot}}))$
and statistical significances of the tests remain almost unchanged.

Interestingly, some studies suggested that the correlations
between the optically-based LC properties and host galaxy mass
might be due to differences in dust extinction in galaxies
with different masses \citep[e.g.][]{2018ApJ...869...56B,2020arXiv200410206B}.
However, most recently \citet{2020arXiv200615164U} observed near-constant correlations
between SN~Ia LC properties and host galaxy mass across near-infrared,
which are less sensitive to dust, and optical bands.
They suggested that dust extinction might not play a significant role in
the observed correlation \citep[see also][]{2020arXiv200613803P}.
In this respect, the combination of the lack/non-significance
of $\Delta m_{15}$ -- host stellar mass
(morphology, colour) correlations
for peculiar SNe~Ia and the observed significant correlation for normal SNe~Ia
(Figs.~\ref{deltm15vsmorph}, \ref{dm15colLogM} and Table~\ref{Dm15vsURLogMSpRo})
also suggest that dust may not be the dominant process in the correlations of
LCs and host properties for normal SNe~Ia.
If the impact of dust were strong, then one would expect that, contrary to what we
found in Fig.~\ref{deltm15vsmorph}, the $B$-band $\Delta m_{15}$
of 91T-like SNe would be correlated
with host galaxy properties, because the bulk of
these events are discovered in star-forming galaxies
with different masses and dust properties.

\begin{figure}
\begin{center}$
\begin{array}{@{\hspace{0mm}}c@{\hspace{0mm}}}
\includegraphics[width=1\hsize]{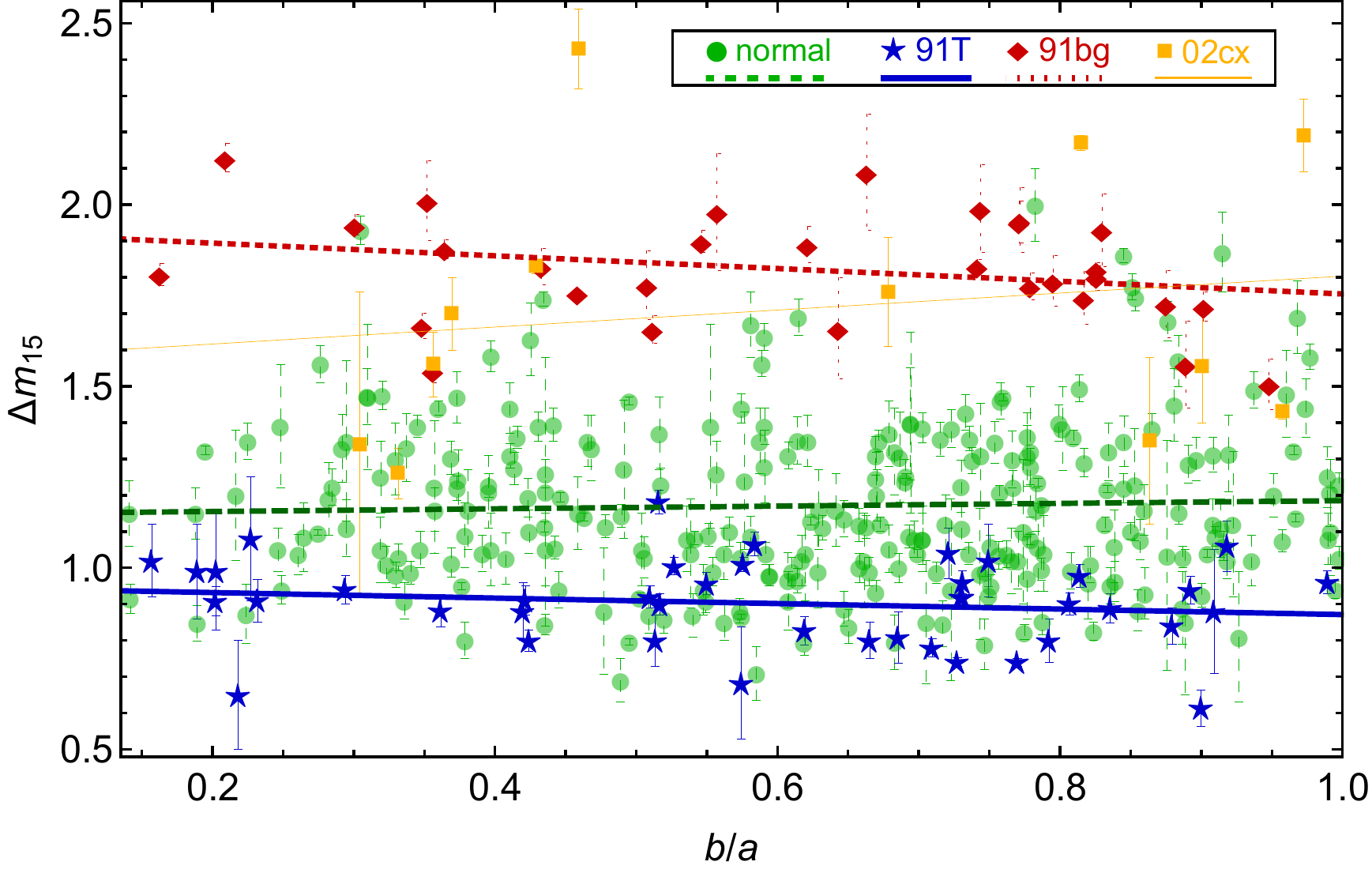}
\end{array}$
\end{center}
\caption{Distributions of the $B$-band $\Delta m_{15}$ values of different subclasses of
         360 SNe~Ia as a function of $b/a$ of S0--Sm host galaxies
         (a measure of galaxy inclination), and their best-fitting lines.}
\label{Delm15vsba}
\end{figure}

Indeed, one could also expect that SNe~Ia in strongly inclined
hosts are more extinguished and will show smaller
$\Delta m_{15}$ (see Subsection~\ref{samplered1}).
Fig.~\ref{Delm15vsba} shows the distribution of $\Delta m_{15}$ versus $b/a$ of S0--Sm host
galaxies,\footnote{The elongations ($b/a$) of host galaxies are obtained from the
fitted $25~{\rm mag~arcsec^{-2}}$ elliptical apertures in the $g$-band (see Subsection~\ref{samplered2}).}
which is a measure of galaxy inclination.
However, the Spearman's rank test shows that these variables are not correlated:
the $P_{\rm s}$-values are 0.745, 0.419, 0.123, and 0.275 for normal,
91T-, 91bg-, and 02cx-like SNe, respectively
\citep[see also][]{2017ApJ...848...56U}.
Therefore, we believe that the observed normal SN~Ia-host relations
are dominated by the diversity of SNe~Ia progenitors.

\subsection{Luminosity-weighted ages of SNe~Ia host galaxies}
\label{RESults4}

In order to move from the qualitative description of the progenitor population ages
of the SN~Ia subclasses to the quantitative ones, we determine the
luminosity-weighted ages of their host galaxies.\footnote{The luminosity-weighted ages
of host galaxies are available in the online version of this article (Supporting Information).}
To conform to values used in the previous paper of our series \citep[][]{2019MNRAS.490..718B},
we use the fixed redshifts of the galaxies to fit the \texttt{P\'{E}GASE}.2 templates library
\citep[][]{1997A&A...326..950F,1999astro.ph.12179F},
comprised of various morphological/spectral types, to the measured $ugriz$ photometry of SN hosts.
Five photometric values of a host galaxy (in some cases only four $griz$ band values are available,
see Subsection~\ref{samplered2}) and its fixed redshift are used
to select the best locations of the values on the spectral energy distribution (SED) templates.
Accordingly, the best matched SED model and corresponding luminosity-weighted age
can be used from the collection of all possible synthetic templates for different galaxy ages (up to 19~Gyr).
The luminosity-weighted stellar population ages are very sensitive to recent star formation,
i.e. bright young stellar populations get more weight in the estimated ages.
An example of SN~Ia host galaxy photometry with its best template,
and more information on the SED fitting method can be found in \citet{2019MNRAS.490..718B}.

\begin{figure}
\begin{center}$
\begin{array}{@{\hspace{0mm}}c@{\hspace{0mm}}}
\includegraphics[width=1\hsize]{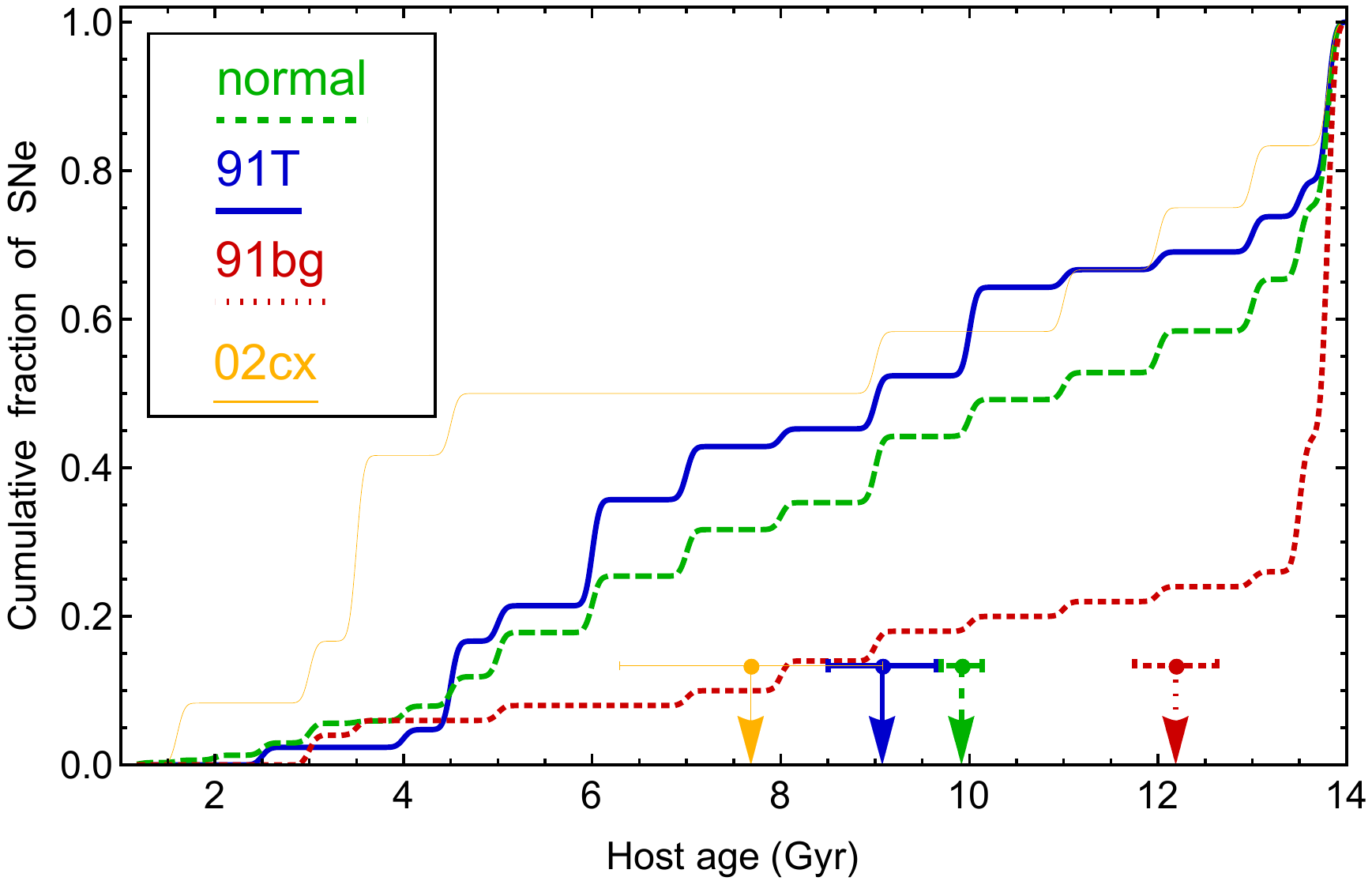}\\
\includegraphics[width=1\hsize]{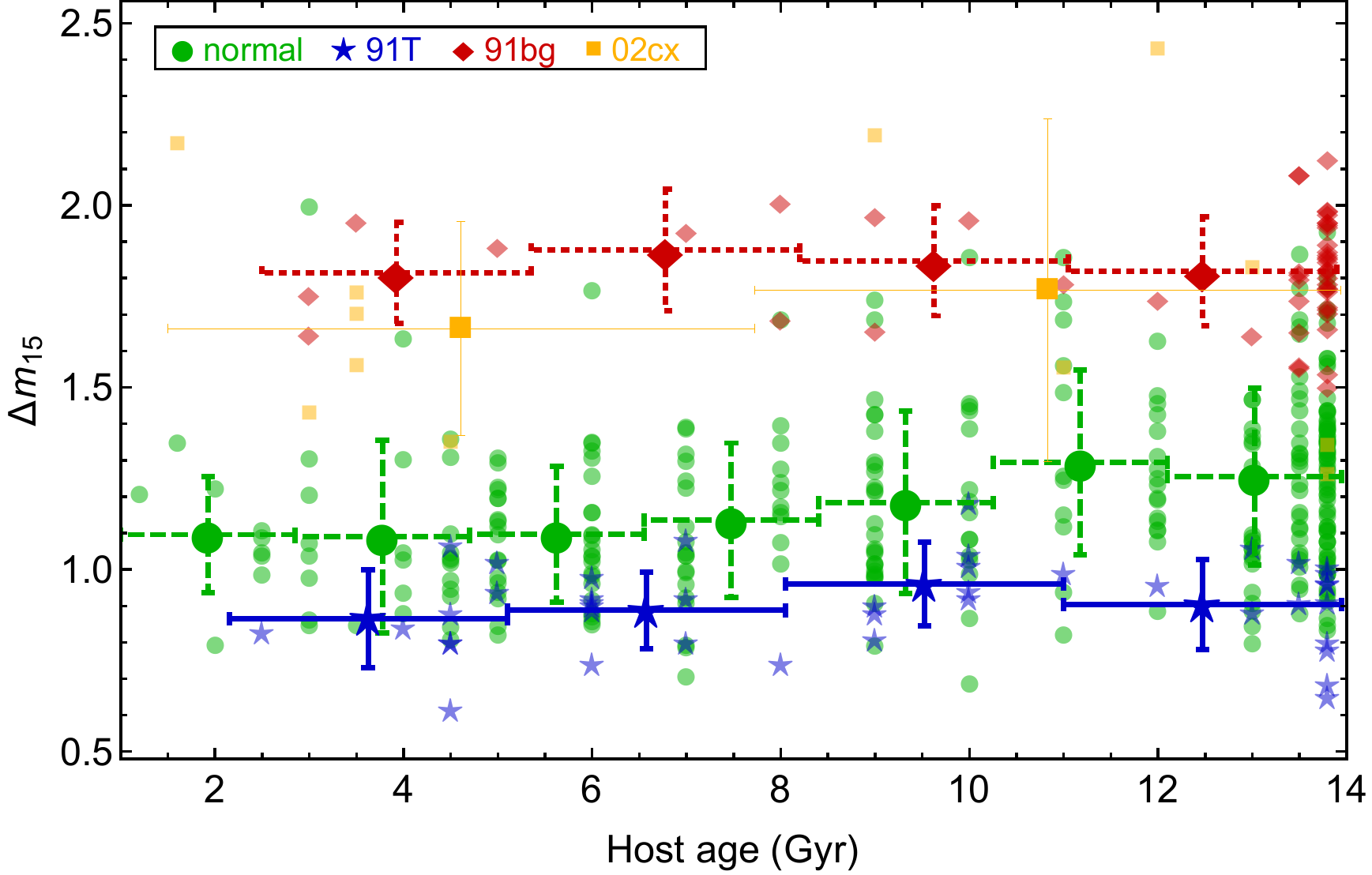}
\end{array}$
\end{center}
\caption{Upper panel: cumulative distributions of SNe~Ia host galaxy luminosity-weighted ages.
         The mean ages (with standard errors) of the host subsamples are shown by arrows (with error bars).
         Bottom panel: $B$-band $\Delta m_{15}$ values of the SN~Ia subclasses
         versus luminosity-weighted ages of host galaxies.
         Binned and averaged values of $\Delta m_{15}$ (bigger symbols) are superposed on
         the original distributions (smaller symbols). Horizontal bars show the bin sizes.
         Depending on the numbers of SNe~Ia for certain subclasses, different binning sizes
         are selected to include sufficient numbers of the objects.}
\label{Delm15agecumetc}
\end{figure}
\begin{table*}
  \centering
  \begin{minipage}{161mm}
  \caption{Comparison of the distributions of host galaxy luminosity-weighted ages
           among different SN~Ia subclasses.}
  \tabcolsep 7.7pt
  \label{SNsubKSADhostage}
    \begin{tabular}{lrrclrrrrrr}
    \hline
  \multicolumn{3}{c}{Host subsample~1} & \multicolumn{1}{c}{vs} & \multicolumn{3}{c}{Host subsample~2} &\multicolumn{1}{c}{$P_{\rm KS}$} & \multicolumn{1}{c}{$P_{\rm AD}$} & \multicolumn{1}{c}{$P_{\rm KS}^{\rm MC}$} & \multicolumn{1}{c}{$P_{\rm AD}^{\rm MC}$}\\
  \multicolumn{1}{l}{SN subclass} & \multicolumn{1}{l}{$N_{\rm SN}$} & \multicolumn{1}{c}{$\langle$host age$\rangle$} && \multicolumn{1}{l}{SN subclass} & \multicolumn{1}{l}{$N_{\rm SN}$} & \multicolumn{1}{c}{$\langle$host age$\rangle$} &&&& \\
  \hline
    normal & 303 & $9.9\pm0.2$ & vs & 91T & 42 & $9.1\pm0.6$ & 0.211 & 0.310 & 0.211 & 0.325\\
    normal & 303 & $9.9\pm0.2$ & vs & 91bg & 50 & $12.2\pm0.4$ & $<$\textbf{0.001} & $<$\textbf{0.001} & $<$\textbf{0.001} & $<$\textbf{0.001}\\
    normal & 303 & $9.9\pm0.2$ & vs & 02cx & 12 & $7.7\pm1.4$ & \textbf{0.029} & \textbf{0.016} & \textbf{0.030} & \textbf{0.030}\\
    91T & 42 & $9.1\pm0.6$ & vs & 91bg & 50 & $12.2\pm0.4$ & $<$\textbf{0.001} & $<$\textbf{0.001} & $<$\textbf{0.001} & $<$\textbf{0.001}\\
    91T & 42 & $9.1\pm0.6$ & vs & 02cx & 12 & $7.7\pm1.4$ & 0.052 & \textbf{0.038} & 0.051 & \textbf{0.049}\\
    91bg & 50 & $12.2\pm0.4$ & vs & 02cx & 12 & $7.7\pm1.4$ & \textbf{0.001} & $<$\textbf{0.001} & \textbf{0.001} & \textbf{0.002}\\
  \hline
  \end{tabular}
  \parbox{\hsize}{\emph{Notes.} The explanations for $P$-values
                  are the same as in Table~\ref{SNsubKSADDis}.
                  The statistically significant differences between the distributions
                  are highlighted in bold.}
  \end{minipage}
\end{table*}

In the upper panel of Fig.~\ref{Delm15agecumetc},
we present the cumulative distributions of luminosity-weighted ages
of normal and peculiar (91T-, 91bg-, and 02cx-like) SNe~Ia host galaxies.
Note that only about 10 per cent of the host galaxies
in our sample have very late-type morphologies (Scd--Sm, see Table~\ref{tabSNhostmorph})
that generally exhibit stellar populations younger than a few Gyr \citep[e.g.][]{2015A&A...581A.103G}.
Therefore, we have a deficit of hosts with ages less than a few Gyr in our age calculations.
The two-sample KS and AD tests in Table~\ref{SNsubKSADhostage} show that the
age distribution of normal SNe~Ia hosts is consistent with that of 91T-like SNe hosts,
but significantly inconsistent with those of 91bg- and 02cx-like SNe hosts.
In comparison with hosts of normal and 91T-like SNe,
the ages of 91bg-like SNe hosts are older on average,
while the ages of hosts of 02cx-like events are younger.
These results are expected in light of those in Fig.~\ref{colMagALL} and Table~\ref{SNsubKSADurLogM},
when considering the $u - r$ colour as a proxy for
recent star formation or stellar population ages of galaxies.

\begin{table}
  \centering
  \begin{minipage}{84mm}
  \caption{Results of Spearman's rank correlation test for the $B$-band $\Delta m_{15}$ values
           of the SN~Ia subclasses versus luminosity-weighted ages of host galaxies.}
  \tabcolsep 4.4pt
  \label{Dm15vsAgeSpRo}
  \begin{tabular}{lrcccrr}
  \hline
    \multicolumn{1}{c}{SN~subclass}&\multicolumn{1}{c}{$N_{\rm SN}$}&
    \multicolumn{1}{c}{Variable~1}&\multicolumn{1}{c}{vs}&
    \multicolumn{1}{c}{Variable~2}&\multicolumn{1}{c}{$r_{\rm s}$}&
    \multicolumn{1}{c}{$P_{\rm s}$}\\
  \hline
    normal &303&$\Delta m_{15}$& vs & host age &0.314&$<$\textbf{0.001}\\
    normal$^a$ &189&$\Delta m_{15}$& vs & host age &0.161&\textbf{0.027}\\
    normal$^b$ &64&$\Delta m_{15}$& vs & host age &$-$0.059&0.644\\
    91T &42&$\Delta m_{15}$& vs & host age &0.154&0.330\\
    91bg &50&$\Delta m_{15}$& vs & host age &0.082&0.569\\
    02cx &12&$\Delta m_{15}$& vs & host age &$-$0.264&0.406\\
  \hline
  \end{tabular}
  \parbox{\hsize}{\emph{Notes.} The explanations for $r_{\rm s}$ and $P_{\rm s}$-values
                  are the same as in Table~\ref{Delm15vsDisSpRo}.
                  $^a \Delta m_{15} \lesssim 1.26$~mag, similar to 91T-like SNe.
                  $^b \Delta m_{15} \gtrsim 1.38$~mag, similar to 91bg-like events.
                  Statistically significant correlations between the variables are highlighted in bold.}
\end{minipage}
\end{table}

In the bottom panel of Fig.~\ref{Delm15agecumetc},
we show the distribution of the $B$-band $\Delta m_{15}$ values of the SN~Ia subclasses
as a function of luminosity-weighted ages of host galaxies.
In analogy with the results in Table~\ref{Dm15vsURLogMSpRo}, the Spearman's rank test,
presented in Table~\ref{Dm15vsAgeSpRo}, shows that the significant correlation exists
only between normal SNe~Ia LC decline rates and stellar population ages of their host galaxies.
These SNe~Ia show an average increase in $\Delta m_{15}$ values with an increase in host's age.
Using the SN~Ia LC stretch instead of $\Delta m_{15}$,
similar correlations between the SN LC and host parameters have been shown many times in the literature
\citep[e.g.][]{2009ApJ...707.1449N,2011ApJ...740...92G,2013MNRAS.435.1680J,2014MNRAS.438.1391P,
2016MNRAS.457.3470C,2020ApJ...889....8K}.
The samples of these studies mostly include normal SNe~Ia events, thus not
studying separately the correlations for peculiar (91T-, 91bg-, and 02cx-like) SN~Ia subclasses.
While, importantly we show that these peculiar SNe~Ia subclasses originate from host galaxies
diverging in age (Table~\ref{SNsubKSADhostage}), however taken separately their LC decline rates
do not show dependencies on host galaxy age, contrary to normal SNe~Ia events (Table~\ref{Dm15vsAgeSpRo}).
These results are in good support of those obtained in Subsections~\ref{RESults2}-\ref{RESults3} that
91bg-like SNe come only from old stellar population of hosts, while 91T-like events originate
only from young population of galaxies, thus likely belonging to two unique classes of progenitors.

In Table~\ref{Dm15vsAgeSpRo},
for normal SNe~Ia we check for the $\Delta m_{15}$ – age correlation
in the two separate and narrower $\Delta m_{15}$ ranges.
In the slower declining range, normal SNe~Ia keep
the $\Delta m_{15}$ -- age trend direction and significance, while
in the faster declining range, the trend becomes insignificant,
hinting that the $\Delta m_{15}$ ranges and/or small number statistics might play some role in
the absence of the $\Delta m_{15}$ – age trend for 91T- and 91bg-like SNe.

Interestingly, to study the connection between SNe~Ia and their host characteristics,
\citet*{2019ApJ...874...32R} estimated the hosts' global stellar population ages
as well as the local environment ages around the sites of
photometrically and spectroscopically classified SDSS~II SNe~Ia (103 events, $z < 0.2$).
The authors noted that the significant correlation between SN~Ia LC properties/luminosity
and host stellar population age might be an effect of age step at $\sim8$~Gyr
(corresponding to the jump in the average LC properties).
\citeauthor{2019ApJ...874...32R} also noted that both the local environment age and global one
show the equivalent correlations with the SNe properties, however, as expected,
the age of the local stellar population is younger than that of the global
one for SN~Ia in star-forming environments.
With that in mind, we split the sample of normal SNe~Ia into two subsamples according to the
host age step at 8~Gyr, and compare the $\Delta m_{15}$ distributions between the subsamples
($N_{\rm SN}=107$, $\langle\Delta m_{15}\rangle = 1.10\pm0.02$
for host age $\leq 8$~Gyr versus $N_{\rm SN}=196$,
$\langle\Delta m_{15}\rangle = 1.24\pm0.02$ for host age $> 8$~Gyr).
The two-sample KS and AD tests show that the $\Delta m_{15}$ distributions are significantly
inconsistent between each other $(P_{\rm KS}^{\rm MC} < 0.001$, $P_{\rm AD}^{\rm MC} < 0.001)$.
As expected from Fig.~\ref{Delm15agecumetc} and Table~\ref{Dm15vsAgeSpRo},
galaxies with ages $\leq 8$~Gyr host on average slower declining/brighter
SNe~Ia (spectroscopically normal) in comparison with older hosts \citep*[see also][]{2017ApJ...851L..50S}.
Note that a 3~Gyr variation in the host age step does not change
the statistical behaviour of the tests.
Thus, the significant correlation seen in Table~\ref{Dm15vsAgeSpRo}
is due to a superposition of faster and slower declining normal SNe~Ia from
older and younger stellar components of hosts, respectively
\citep[see also][as measured by the environmental H$\alpha$ emission at the positions of SNe~Ia]{2013A&A...560A..66R}.

This result can be explained considering the correlation between $^{56}{\rm Ni}$ mass
synthesized in SN~Ia and host galaxy age \citep[e.g.][]{2009ApJ...691..661H}.
In this respect, \citet{2009ApJ...707.1449N} suggested that
the details of the $^{56}{\rm Ni}$ mass -- age correlations
in their study and in \citet{2009ApJ...691..661H} imply an age step/threshold of about several Gyr
for SN~Ia hosts, above which galaxies are less likely to produce SNe~Ia with $^{56}{\rm Ni}$ masses
$\gtrsim 0.5 M_{\odot}$, i.e. slower declining/brighter events.

It is important also to mention that the $\Delta m_{15}$ -- morphology, $\Delta m_{15}$ -- colour,
and $\Delta m_{15}$ -- $\log(M_{\ast}/{\rm M_{\odot}})$ correlations,
we showed in previous subsections for spectroscopically normal SN~Ia,
and the $\Delta m_{15}$ – age correlation agree with one another
in the sense that host galaxy morphology, colour, mass, and age are correlated,
with older galaxies mainly being earlier type, redder, and more massive
\citep[e.g.][]{2015A&A...581A.103G}.
Therefore, the previously reported correlations between LC properties of SNe~Ia
and physical parameters of their host galaxies are most likely originated from the differences
in host stellar population ages \citep[see also earlier discussions in][]{2011ApJ...740...92G,
2013A&A...560A..66R,2014MNRAS.445.1898C,2019ApJ...874...32R,2020ApJ...889....8K}.

\subsection{Constraints on delay time distributions of SNe~Ia}
\label{RESults5}

Current theoretical models of spectroscopically normal SNe~Ia progenitors
predict that most likely their DTD peaks below/close to 1~Gyr,
corresponding to the young/prompt component,
then declines fast by over two orders of magnitude  at several Gyr
\citep[e.g.][]{2000ApJ...528..108Y,2005A&A...441.1055G,2014MNRAS.445.1898C}.
At redshifts near zero (as in our sample), the SN~Ia rate in evolved galaxies,
from a large volume, results in a bimodal shape of
the distribution of SN~Ia progenitor ages,
with assumption that galaxies have nearly constant SFR \citep[e.g.][]{2014MNRAS.445.1898C}.
In such a bimodal age distribution,
the second peak includes old/delayed events at current epoch,
which are naturally absent at high redshifts.
Young/prompt SNe~Ia originate in actively star-forming galaxies, predominantly from progenitors
whose ages correspond to the first peak of the SN~Ia progenitor age distribution.
While old/delayed SNe~Ia originate in galaxies that already have an old stellar component or
they are entirely old whose star formation ceased in the past.
In such old stellar systems, the DTD is lacking young/prompt events, meaning that
delayed SNe~Ia occur from different age group of progenitors
\citep[e.g.][]{2014MNRAS.445.1898C,2014ARA&A..52..107M}.
In this respect, the significant correlation between the LC decline rates and the global
host galaxy ages of normal SNe~Ia (see Subsection~\ref{RESults4})
is likely caused by the bimodal behaviour of
the age distribution of normal SN~Ia progenitors,
which includes both the prompt events from young progenitors (slower declining SNe~Ia)
and delayed SNe~Ia from old systems (faster declining events).
On the other hand, the absence (non-significance)
of the mentioned correlation for peculiar (91T-, 91bg-, and 02cx-like) SNe~Ia
and the observed differences in the properties of their host galaxies
(see Subsections~\ref{RESults2} and \ref{RESults3}) can be considered as evidences
for distinct (single-mode) behaviours of their progenitor age distribution,
thus distinct progenitor channels.

Our results on peculiar 91bg-like events
agree with those of \citet{2011ApJ...727..107G},
who found that subluminous/91bg-like SNe (18 events with $z \leq 0.6$) in their
photometrically identified SNe sample are mainly found in early-type hosts with
almost no star formation, and argued that these SNe come from a delayed progenitor component
with a Gaussian DTD centered around 6~Gyr, but with large uncertainty up to $\sim 11$~Gyr.
Recently, \citet[][]{2019PASA...36...31P} studied integral field observations of the hosts'
explosion sites of 11 spectroscopically identified 91bg-like events $(z \leq 0.04)$
and found that the majority of the stellar populations in the vicinity of these SNe locations
are dominated by old stars with a lack of recent star formation.
The authors concluded that the DTD of 91bg-like SNe is likely weighted toward
long delay times, larger than $\sim6$~Gyr \citep[see also][]{2017NatAs...1E.135C}.
Most recently, in \citet{2019MNRAS.490..718B} we found that in elliptical host galaxies of SNe~Ia,
the age distribution of 91bg-like SNe hosts ($\gtrsim 8$~Gyr) does not extend down to the stellar ages
that produce significant $u$-band fluxes of their early-type hosts, contrary
to the hosts of normal SNe~Ia.
Therefore, younger stars in elliptical galaxies do not produce 91bg-like SNe.

Our study includes the largest sample of 50 subluminous 91bg-like SNe
in a much wider morphological range of host galaxies than previous studies,
and shows even more clearly that the progenitor population age of these events
is strongly weighted toward very old ages
(oldest among the other SN~Ia subclasses, see Table~\ref{SNsubKSADhostage}
and Fig.~\ref{Delm15agecumetc}).
More than 85 per cent of 91bg-like SNe hosts are older than 8~Gyr.
As already mentioned, such delay times are much longer than the delays of normal Type Ia SNe from
star-forming environments, whose DTD peaks between several hundred Myr and $\sim 1$~Gyr
(a sharp initial peak) with a low tail at higher delay times \citep[e.g.][]{2014MNRAS.445.1898C}.
Therefore, most likely 91bg-like events originate exclusively from old stellar population
of host galaxies, and thus these peculiar SNe~Ia have no prompt component.

As in \citet{2019MNRAS.490..718B}, we favour the DD channel
for the progenitors of these peculiar events that
belong to the old stellar components of galaxies
\citep[see also][]{2009ApJ...699.2026R,2010Natur.463..924G,2011NewA...16..250L}:
in particular the progenitor models such as He-ignited violent mergers
(CO WD primary with He WD companion) might be appropriate for 91bg-like events
\citep[e.g.][]{2013ApJ...770L...8P,2017NatAs...1E.135C}.
These models predict very long delay times for
subluminous 91bg-like SNe ($\gtrsim$ several Gyr).
On the other hand, the DD channel with CO WD primary and CO WD companion
\citep[e.g.][]{2013ApJ...770L...8P}
can provide an appropriate DTD for normal SNe~Ia with an initial peak below/close to 1~Gyr
and a tail up to about 10~Gyr, while most SD models predict few or no SNe~Ia at long delays
\citep[e.g.][]{2014MNRAS.445.1898C,2014ARA&A..52..107M}.
In addition, \citet{2019arXiv191007532P} showed that mergers of CO WD with
hybrid He-CO WD could also give rise to normal Type Ia SNe whose synthetic LCs
and spectra are consistent with those of observed SNe~Ia
\citep*[see also][for the properties of hybrid He-CO WDs]{2019MNRAS.482.1135Z}.
These authors noted that together with the contribution from mergers of massive
double CO WDs that give rise to more luminous SNe Ia, their models can potentially
reproduce the full range of normal SNe~Ia, their DTD and rate.

Relatively younger global host ages of overluminous 91T-like SNe
(Fig.~\ref{Delm15agecumetc})
can be explained in the context of shorter DTD in comparison with that of 91bg-like events.
In addition, 91T-like SNe strongly prefer blue (Fig.~\ref{colMagALL}) and late-type hosts
(Fig.~\ref{deltm15vsmorph}), and no such an event is observed in old elliptical galaxies
(Table~\ref{tabSNhostmorph}), thus supporting shorter, i.e. relatively prompt DTD for these SNe.
In this sense, the DTD of the SD channel is different from that of DD one
and can be appropriate for 91T-like SNe \citep[e.g.][]{2015ApJ...805..150F}.
The DTD of SD channel has a sharp cutoff at 2-3~Gyr, because in its original formulation
a donor/companion star in an SD system is a main sequence or red giant star with only a narrow range of masses
that could provide mass accretion at the sufficient rate resulting in a stable burning and an explosion of WD
\citep[e.g.][]{2004MNRAS.350.1301H}.
Also, note that massive CO WD primaries in DD models are naturally explode in more luminous SNe~Ia
(e.g. overluminous/91T-like events).
Such massive WDs, which originated from more massive intermediate-mass stars, in DD systems
have delay times $\lesssim$ several hundred Myr
\citep[e.g. the violent WDs merger scenario;][]{2013MNRAS.429.1425R},
and is therefore expected in late-type star-forming galaxies.

These DTDs may seem too young in comparison with the global age distribution of hosts
(see Fig.~\ref{Delm15agecumetc} and Table~\ref{SNsubKSADhostage}),
however, as mentioned in Subsection~\ref{RESults4} for normal SN~Ia in star-forming environments,
the age of the local stellar population at the SN position
is younger (on average) than the global age of its host galaxies \citep[e.g.][]{2019ApJ...874...32R}.
Therefore, in an ideal case, the stellar population age obtained from the location of the SN
in the host galaxy would be preferable to the global host age that we estimate in the current study
\citep[see also discussions in][]{2011ApJ...740...92G,2015ApJ...805..150F}.
Consequently, additional local age-estimations are needed for 91T-like SNe to better test
the DTDs of the SD and DD progenitors models.

Despite the small number statistics of 02cx-like SNe, on average their host galaxies
have the youngest global ages (Fig.~\ref{Delm15agecumetc}), latest morphology (Fig.~\ref{deltm15vsmorph}),
bluest colour and lowest masses (Fig.~\ref{colMagALL})
among the other SN~Ia subclasses, again strongly pointing to a shorter DTD for the events.
In our sample the host galaxy properties of 02cx-like SNe are close to those of star-forming hosts
of 91T-like SNe \citep[see also][]{2009AJ....138..376F,2010Natur.465..322P},
however, their LC decline rates are significantly different
from those of 91T-like events (Fig.~\ref{Delm15histcum} and Table~\ref{SNsubKSADm15})
and fall off the SN~Ia LC width-luminosity relation \citep[see e.g.][]{2017hsn..book..317T}.
The properties of 02cx-like SNe and evidence of their short DTD suggest
a binary system where 02cx-like SN arises from massive CO WD that quickly accretes helium from
an He-star donor \citep[e.g.][]{2014LRR....17....3P}.
Then accretion-triggered explosion of a Chandrasekhar mass WD does not necessarily fully disrupt the star.
Among a variety of proposed scenarios, this is now the leading model
\citep[see review by][and references therein]{2017hsn..book..375J}.
This channel might be the dominant one for delays of up to a few hundred Myr, above which hydrogen-accreting SD
and DD systems dominate \citep[e.g.][]{2014A&A...563A..83C}.
Again, the stellar population ages obtained at the locations of 02cx-like SNe in the host galaxies
would be preferable to the global host ages to better test the DTD of the leading model.

\section{Conclusions}
\label{DISconc}

In the seventh paper of a series, using a well-defined sample of Type Ia SNe and
their host galaxies, we comparatively analyse the relations between
the LC decline rates and the global properties of hosts of various SN~Ia subclasses
to better understand the diverse nature of SNe~Ia progenitors.
The spectroscopic subclasses of SNe~Ia (normal, 91T-, 91bg-, and 02cx-like)
and their $B$-band LC decline rates ($\Delta m_{15}$)
are carefully compiled from the available literature,
while the data of the SNe~Ia host galaxies is a homogeneous set of consistent measurements
performed by the authors of this study.
Our sample consists of 394 relatively nearby ($\leq 150$~Mpc,
the mean distance is 72~Mpc) E--Sm galaxies, which host 407 SNe~Ia in total.

There is no strong redshift bias within our sample,
which could drive the observed relations between global properties of hosts and
extinction-independent LC decline rates of the SN~Ia subclasses.
In addition, the representation of SN subclasses in our compilation and
in nearly complete volume-limited sample of the LOSS are not different statistically.
However, due to the small number statistics of 02cx-like SNe, the results on these peculiar
events and their hosts should be considered with caution.

All the obtained results and their interpretations are summarized below.

\begin{enumerate}
\item In general, the $B$-band $\Delta m_{15}$ distribution of SNe~Ia seems to be bimodal
      (Fig.~\ref{Delm15histcum}),
      with the second (weaker) mode mostly distributed within $\sim 1.5$ to 2.1~mag.
      This faster declining range is generally occupied by 91bg-like (subluminous) events,
      while the $\Delta m_{15}$ of 91T-like (overluminous) SNe are distributed only within
      the first mode at slower declining range ($\Delta m_{15} \lesssim 1.1$~mag).
      The decline rates of 02cx-like SNe are spread on the faster side of
      the $\Delta m_{15}$ distribution of normal SNe~Ia.
      Statistically, all these distributions are significantly different from one another
      (Table~\ref{SNsubKSADm15}).
\item The host galaxies of normal, 91T-, and 91bg-like SNe~Ia
      have morphological type distributions that are significantly inconsistent
      between one another (Fig.~\ref{deltm15vsmorph} and Table~\ref{SNsubKSADttype}).
      Hosts of 91bg-like SNe have, on average, earlier morphological types
      $(\langle t \rangle\approx 0)$
      with a large number of the events discovered in E--S0 galaxies.
      While hosts of 91T-like SNe have on average later morphological types
      $(\langle t \rangle\approx 4)$
      with a single 91T-like event in the E--S0 bin (Table~\ref{tabSNhostmorph}).
      The same distribution of hosts of normal SNe~Ia
      occupies an intermediate position between the host morphologies of 91T- and 91bg-like events.
      The morphological distribution of 02cx-like SNe hosts is similar to that of
      91T-like SNe hosts.
\item As for galaxies in general, the distribution of SNe~Ia hosts in
      the $u - r$ colour--mass diagram is bimodal (Fig.~\ref{CMDDENSit}).
      The hosts of 91bg-like SNe are located in the Red Sequence of the diagram,
      most of them have $u - r$ colours $\gtrsim 2$~mag (i.e. above the Green Valley).
      In comparison with hosts of normal, 91T-, and 02cx-like SNe, the colour distribution
      of hosts of 91bg-like SNe are significantly redder (Table~\ref{SNsubKSADurLogM}).
      Importantly, the bulk of hosts of 91bg-like SNe are significantly massive
      $(\log(M_{\ast}/{\rm M_{\odot}}) > 10.5)$ and old (more than 10~Gyr).
      The hosts' mass (age) distribution is significantly inconsistent with those of the other
      SN~Ia subclasses (Tables~\ref{SNsubKSADurLogM} and \ref{SNsubKSADhostage}).
      At the same time, the colour (mass, age) distributions are not statistically different
      between hosts of normal and 91T-like SNe.
      Finally, all the host galaxies of 02cx-like SNe are positioned in the Blue Cloud
      of the colour--mass diagram, mostly below the Green Valley (Fig.~\ref{colMagALL}).
      Their colour (mass, age) distribution is significantly bluer (lower, younger)
      in comparison with that of normal SNe~Ia hosts,
      but closer to that of 91T-like SNe hosts.
\item As previously shown with smaller nearby SN~Ia samples, there is a significant correlation
      between normal SNe~Ia LC decline rates and global ages (morphologies, colours, and masses)
      of their host galaxies (Tables~\ref{Dm15vsURLogMSpRo} and \ref{Dm15vsAgeSpRo}).
      On average, those normal SNe~Ia that are in galaxies above the Green Valley,
      i.e. in early-type, red, massive, and old hosts,
      have faster declining LCs in comparison with those in galaxies below the Green Valley,
      i.e. in late-type, blue, less massive, and younger hosts
      (Tables~\ref{SNhostsubKSADm15} and \ref{SNurLogMhstKSADm15}).
      The results suggest that the observed correlations between
      normal SNe~Ia LC decline rates and global properties of host galaxies are due to
      a superposition of at least two distinct populations of faster and slower declining
      normal SNe~Ia from old (halo/bulge and old disc) and young (star-forming disc) components
      of hosts, respectively dominating in the Red Sequence and Blue Cloud of the colour--mass diagram.
\item For the first time we show that the LC decline rates of
      subluminous/91bg-like SNe and overluminous/91T-like events do not show
      dependencies on the host galaxy morphology and colour
      (Figs.~\ref{deltm15vsmorph}, \ref{dm15colLogM}, and Table~\ref{Dm15vsURLogMSpRo}).
      The distribution of hosts on the colour--mass diagram (Fig.~\ref{colMagALL})
      confirms the known tendency for 91bg-like SNe to occur in globally red/old galaxies
      (from halo/bulge and old disc components)
      while 91T-like events prefer blue/younger hosts (related to star-forming component).
      Probably, the decline rates of 02cx-like SNe also do not show dependencies on hosts' properties.
      On average, the youngest global ages of 02cx-like SNe hosts and their
      positions in the colour--mass diagram
      hint that these events likely originate from the young stellar component,
      but they differ from 91T-like events in the LC decline rate.
\item As in our previous study \citep{2019MNRAS.490..718B}, for the progenitors of 91bg-like events
      we favour the DD channel, in particular
      the progenitor models such as He-ignited violent mergers
      \citep[CO WD primary with He WD companion, e.g.][]{2013ApJ...770L...8P,2017NatAs...1E.135C}.
      In agreement with our findings, these models predict very long delay times for
      subluminous 91bg-like SNe ($\gtrsim$ several Gyr).
      In addition, the DD channel with CO WD primary and CO (or hybrid He-CO) WD companion
      \citep[e.g.][]{2013ApJ...770L...8P,2019arXiv191007532P}
      can provide an appropriate DTD for normal SNe~Ia with an initial peak below/close to 1~Gyr
      and a tail up to about 10~Gyr \citep[e.g.][]{2014MNRAS.445.1898C,2014ARA&A..52..107M}.
      On the other hand, the DTD of the SD channel has a sharp cutoff at 2-3~Gyr
      \citep[e.g.][]{2004MNRAS.350.1301H} and can be appropriate for 91T-like SNe.
      Note that massive CO WD primaries in DD models are naturally explode in more luminous SNe~Ia
      and also can be appropriate for these overluminous events
      \citep[e.g.][]{2013MNRAS.429.1425R}. Also,
      we show evidence of short DTD for 02cx-like SNe, which can be interpreted within
      the leading model for the events
      \citep[a binary system where SN arises from
      massive CO WD that quickly accretes helium from
      an He-star donor, and then explosion of WD does not necessarily
      fully disrupt the star, e.g.][]{2014LRR....17....3P}.
\end{enumerate}

We stress again that, in an ideal case, the stellar population properties
obtained from the SN location in the host galaxy would be preferable to
the global host properties that we estimated in our study
\citep[e.g.][]{2013A&A...560A..66R,2019ApJ...874...32R,2019PASA...36...31P}.
Therefore, additional local age-estimations are needed for
normal and peculiar (91T-, 91bg-, and 02cx-like) SN~Ia subclasses to more precisely test
the DTDs of the SD and DD progenitor models.
Hopefully, the forthcoming Vera C.~Rubin Observatory (formerly the Large
Synoptic Survey Telescope) will provide much bigger
spectroscopic and photometric sample of relatively nearby SNe~Ia that will give
an opportunity to better constrain the nature of their progenitors.

\section*{Acknowledgements}

We are grateful to our referee for his/her constructive comments.
This work was supported by the RA MES State Committee of Science,
in the frame of the research project number 15T--1C129.
This work was made possible in part by a research grant from the
Armenian National Science and Education Fund (ANSEF)
based in New York, USA.
MT is partially supported by the PRIN-INAF~2016 with the project Towards
the SKA and CTA era: discovery, localisation, and physics of transient
sources (P.I. M.~Giroletti).
Funding for the Sloan Digital Sky Survey IV has been provided by
the Alfred P.~Sloan Foundation, the US Department of Energy Office of Science,
and the Participating Institutions. SDSS-IV acknowledges
support and resources from the Center for High-Performance Computing at
the University of Utah. The SDSS web site is \href{http://www.sdss.org/}{www.sdss.org}.
SDSS-IV is managed by the Astrophysical Research Consortium for the
Participating Institutions of the SDSS Collaboration including the
Brazilian Participation Group, the Carnegie Institution for Science,
Carnegie Mellon University, the Chilean Participation Group, the French Participation Group,
Harvard-Smithsonian Center for Astrophysics,
Instituto de Astrof\'isica de Canarias, The Johns Hopkins University,
Kavli Institute for the Physics and Mathematics of the Universe (IPMU) /
University of Tokyo, the Korean Participation Group, Lawrence Berkeley National Laboratory,
Leibniz Institut f\"ur Astrophysik Potsdam (AIP),
Max-Planck-Institut f\"ur Astronomie (MPIA Heidelberg),
Max-Planck-Institut f\"ur Astrophysik (MPA Garching),
Max-Planck-Institut f\"ur Extraterrestrische Physik (MPE),
National Astronomical Observatories of China, New Mexico State University,
New York University, University of Notre Dame,
Observat\'ario Nacional / MCTI, The Ohio State University,
Pennsylvania State University, Shanghai Astronomical Observatory,
United Kingdom Participation Group,
Universidad Nacional Aut\'onoma de M\'exico, University of Arizona,
University of Colorado Boulder, University of Oxford, University of Portsmouth,
University of Utah, University of Virginia, University of Washington, University of Wisconsin,
Vanderbilt University, and Yale University.


\section*{Data Availability}

The database underlying this study is available in the online version
(Supporting Information) of the article.
The first 10 entries of our database of 407 individual SNe~Ia
(SN designation, spectroscopic subclass,
$\Delta m_{15}$, and corresponding sources of the data)
and their 394 hosts (galaxy designation, distance,
corrected $ugriz$ apparent magnitudes, morphological type, and luminosity-weighted age)
are shown in Table~\ref{databasetab}.
The full table is available electronically in an CSV file format.
A portion is shown in Table~\ref{databasetab} for guidance regarding its content and format.

\begin{sidewaystable}
  \caption{The database of 407 individual SNe~Ia and their 394 host galaxies.
           Only the first 10 entries are shown. The full table is available in
           the online version of this article.}
  \label{databasetab}
  \begin{tabular}{lllcllrrrrrrlr}
    \hline
    \multicolumn{1}{c}{SN} & \multicolumn{1}{c}{Subclass} & \multicolumn{1}{c}{Source} & \multicolumn{1}{c}{$\Delta m_{15}$} & \multicolumn{1}{c}{Source} & \multicolumn{1}{c}{Host} & \multicolumn{1}{c}{Dist.} & \multicolumn{1}{c}{$u$} & \multicolumn{1}{c}{$g$} & \multicolumn{1}{c}{$r$} & \multicolumn{1}{c}{$i$} & \multicolumn{1}{c}{$z$} & \multicolumn{1}{c}{Morph.} & \multicolumn{1}{c}{Age}\\
     & & \multicolumn{1}{c}{Bibcode} & \multicolumn{1}{c}{mag} & \multicolumn{1}{c}{Bibcode} & & \multicolumn{1}{c}{Mpc} & \multicolumn{1}{c}{mag} & \multicolumn{1}{c}{mag} & \multicolumn{1}{c}{mag} & \multicolumn{1}{c}{mag} & \multicolumn{1}{c}{mag} & & \multicolumn{1}{c}{Gyr} \\
    \hline
    1980N & norm & 1991AJ....102..208H & $1.28\pm0.05$ & 2009ApJ...700..331 & NGC1316 & 22.228 & 10.93 & 9.03 & 8.46 & 8.06 & 7.91 & S0/a: & 13.8\\
    1981B & norm & 2009ApJ...699L.139W & $1.10\pm0.06$ & 2009ApJ...700..331 & NGC4536 & 25.514 & 11.26 & 10.17 & 9.64 & 9.36 & 9.12 & Sbc & 6.0\\
    1981D & norm & 1991AJ....102..208H & $1.32\pm0.18$ & 2009ApJ...700..331 & NGC1316 & 22.228 & 10.93 & 9.03 & 8.46 & 8.06 & 7.91 & S0/a: & 13.8\\
    1983G & norm & 1996ApJ...465...73B & $1.37\pm0.10$ & 2005ApJ...623.1011B & NGC4753 & 18.602 & 11.65 & 10.01 & 9.29 & 8.88 & 8.65 & S0/a & 13.8\\
    1984A & norm & 2009ApJ...699L.139W & $1.21\pm0.10$ & 2005ApJ...623.1011B & NGC4419 & 16.095 & 12.82 & 11.25 & 10.52 & 10.10 & 9.82 & Sa & 13.8\\
    1986G & 91bg & 2017hsn..book..317T & $1.65\pm0.09$ & 2009ApJ...700..331 & NGC5128 & 3.714 & 5.98 & 5.29 & 4.93 & 4.79 & 4.41 & E & 3.0\\
    1989A & norm & 2012MNRAS.425.1789S & $1.06\pm0.10$ & 2005ApJ...623.1011B & NGC3687 & 36.559 & 13.85 & 12.64 & 12.10 & 11.82 & 11.58 & SBc & 13.0\\
    1989B & norm & 2012MNRAS.425.1789S & $1.02\pm0.13$ & 2009ApJ...700..331 & NGC3627 & 9.592 & 10.10 & 8.80 & 8.20 & 7.87 & 7.67 & SBb & 9.0\\
    1990N & norm & 2012MNRAS.425.1789S & $1.04\pm0.13$ & 2009ApJ...700..331 & NGC4639 & 22.471 & 12.70 & 11.51 & 10.95 & 10.63 & 10.46 & SBbc & 7.0\\
    1990O & norm & 2012MNRAS.425.1789S & $0.96\pm0.10$ & 1996AJ....112.2408H & MCG+03-44-003 & 131.369 & -- & 14.31 & 13.60 & 13.31 & 13.02 & SBbc & 13.5\\
    \hline
  \end{tabular}
\end{sidewaystable}


\bibliography{diversSNIabib}

\newcommand{\SortNoop}[1]{}
\begin{thebibliography}{}
\makeatletter
\relax
\def\mn@urlcharsother{\let\do\@makeother \do\$\do\&\do\#\do\^\do\_\do\%\do\~}
\def\mn@doi{\begingroup\mn@urlcharsother \@ifnextchar [ {\mn@doi@}
  {\mn@doi@[]}}
\def\mn@doi@[#1]#2{\def\@tempa{#1}\ifx\@tempa\@empty \href
  {http://dx.doi.org/#2} {doi:#2}\else \href {http://dx.doi.org/#2} {#1}\fi
  \endgroup}
\def\mn@eprint#1#2{\mn@eprint@#1:#2::\@nil}
\def\mn@eprint@arXiv#1{\href {http://arxiv.org/abs/#1} {{\tt arXiv:#1}}}
\def\mn@eprint@dblp#1{\href {http://dblp.uni-trier.de/rec/bibtex/#1.xml}
  {dblp:#1}}
\def\mn@eprint@#1:#2:#3:#4\@nil{\def\@tempa {#1}\def\@tempb {#2}\def\@tempc
  {#3}\ifx \@tempc \@empty \let \@tempc \@tempb \let \@tempb \@tempa \fi \ifx
  \@tempb \@empty \def\@tempb {arXiv}\fi \@ifundefined
  {mn@eprint@\@tempb}{\@tempb:\@tempc}{\expandafter \expandafter \csname
  mn@eprint@\@tempb\endcsname \expandafter{\@tempc}}}

\bibitem[\protect\citeauthoryear{{Ahumada} et~al.,}{{Ahumada}
  et~al.}{2020}]{2020ApJS..249....3A}
{Ahumada} R.,  et~al., 2020, \mn@doi [\apjs] {10.3847/1538-4365/ab929e}, \href
  {https://ui.adsabs.harvard.edu/abs/2020ApJS..249....3A} {249, 3}

\bibitem[\protect\citeauthoryear{{Altavilla} et~al.,}{{Altavilla}
  et~al.}{2004}]{2004MNRAS.349.1344A}
{Altavilla} G.,  et~al., 2004, \mn@doi [\mnras]
  {10.1111/j.1365-2966.2004.07616.x}, \href
  {https://ui.adsabs.harvard.edu/abs/2004MNRAS.349.1344A} {349, 1344}

\bibitem[\protect\citeauthoryear{{Arnett}}{{Arnett}}{1982}]{1982ApJ...253..785A}
{Arnett} W.~D.,  1982, \mn@doi [\apj] {10.1086/159681}, \href
  {https://ui.adsabs.harvard.edu/abs/1982ApJ...253..785A} {253, 785}

\bibitem[\protect\citeauthoryear{{Ashall}, {Mazzali}, {Sasdelli}  \&
  {Prentice}}{{Ashall} et~al.}{2016}]{2016MNRAS.460.3529A}
{Ashall} C.,  {Mazzali} P.,  {Sasdelli} M.,   {Prentice} S.~J.,  2016, \mn@doi
  [\mnras] {10.1093/mnras/stw1214}, \href
  {https://ui.adsabs.harvard.edu/abs/2016MNRAS.460.3529A} {460, 3529}

\bibitem[\protect\citeauthoryear{{Baldry}, {Balogh}, {Bower}, {Glazebrook},
  {Nichol}, {Bamford}  \& {Budavari}}{{Baldry}
  et~al.}{2006}]{2006MNRAS.373..469B}
{Baldry} I.~K.,  {Balogh} M.~L.,  {Bower} R.~G.,  {Glazebrook} K.,  {Nichol}
  R.~C.,  {Bamford} S.~P.,   {Budavari} T.,  2006, \mn@doi [\mnras]
  {10.1111/j.1365-2966.2006.11081.x}, \href
  {https://ui.adsabs.harvard.edu/abs/2006MNRAS.373..469B} {373, 469}

\bibitem[\protect\citeauthoryear{{Barkhudaryan}, {Hakobyan}, {Karapetyan},
  {Mamon}, {Kunth}, {Adibekyan}  \& {Turatto}}{{Barkhudaryan}
  et~al.}{2019}]{2019MNRAS.490..718B}
{Barkhudaryan} L.~V.,  {Hakobyan} A.~A.,  {Karapetyan} A.~G.,  {Mamon} G.~A.,
  {Kunth} D.,  {Adibekyan} V.,   {Turatto} M.,  2019, \mn@doi [\mnras]
  {10.1093/mnras/stz2585}, \href
  {https://ui.adsabs.harvard.edu/abs/2019MNRAS.490..718B} {490, 718}

\bibitem[\protect\citeauthoryear{{Blondin} et~al.,}{{Blondin}
  et~al.}{2012}]{2012AJ....143..126B}
{Blondin} S.,  et~al., 2012, \mn@doi [\aj] {10.1088/0004-6256/143/5/126}, \href
  {https://ui.adsabs.harvard.edu/abs/2012AJ....143..126B} {143, 126}

\bibitem[\protect\citeauthoryear{{Bottinelli}, {Gouguenheim}, {Paturel}  \&
  {Teerikorpi}}{{Bottinelli} et~al.}{1995}]{1995A&A...296...64B}
{Bottinelli} L.,  {Gouguenheim} L.,  {Paturel} G.,   {Teerikorpi} P.,  1995,
  \aap, \href {https://ui.adsabs.harvard.edu/abs/1995A&A...296...64B} {296, 64}

\bibitem[\protect\citeauthoryear{{Branch}, {Fisher}  \& {Nugent}}{{Branch}
  et~al.}{1993}]{1993AJ....106.2383B}
{Branch} D.,  {Fisher} A.,   {Nugent} P.,  1993, \mn@doi [\aj]
  {10.1086/116810}, \href
  {https://ui.adsabs.harvard.edu/abs/1993AJ....106.2383B} {106, 2383}

\bibitem[\protect\citeauthoryear{{Brout} \& {Scolnic}}{{Brout} \&
  {Scolnic}}{2020}]{2020arXiv200410206B}
{Brout} D.,  {Scolnic} D.,  2020, preprint, \href
  {https://ui.adsabs.harvard.edu/abs/2020arXiv200410206B} {} (\mn@eprint
  {arXiv} {2004.10206})

\bibitem[\protect\citeauthoryear{{Burns} et~al.,}{{Burns}
  et~al.}{2018}]{2018ApJ...869...56B}
{Burns} C.~R.,  et~al., 2018, \mn@doi [\apj] {10.3847/1538-4357/aae51c}, \href
  {https://ui.adsabs.harvard.edu/abs/2018ApJ...869...56B} {869, 56}

\bibitem[\protect\citeauthoryear{{Cameron}}{{Cameron}}{2011}]{2011PASA...28..128C}
{Cameron} E.,  2011, \mn@doi [\pasa] {10.1071/AS10046}, \href
  {https://ui.adsabs.harvard.edu/abs/2011PASA...28..128C} {28, 128}

\bibitem[\protect\citeauthoryear{{Campbell}, {Fraser}  \& {Gilmore}}{{Campbell}
  et~al.}{2016}]{2016MNRAS.457.3470C}
{Campbell} H.,  {Fraser} M.,   {Gilmore} G.,  2016, \mn@doi [\mnras]
  {10.1093/mnras/stw115}, \href
  {https://ui.adsabs.harvard.edu/abs/2016MNRAS.457.3470C} {457, 3470}

\bibitem[\protect\citeauthoryear{{Chambers} et~al.,}{{Chambers}
  et~al.}{2016}]{2016arXiv161205560C}
{Chambers} K.~C.,  et~al., 2016, preprint, \href
  {https://ui.adsabs.harvard.edu/abs/2016arXiv161205560C} {} (\mn@eprint
  {arXiv} {1612.05560})

\bibitem[\protect\citeauthoryear{{Childress}, {Wolf}  \& {Zahid}}{{Childress}
  et~al.}{2014}]{2014MNRAS.445.1898C}
{Childress} M.~J.,  {Wolf} C.,   {Zahid} H.~J.,  2014, \mn@doi [\mnras]
  {10.1093/mnras/stu1892}, \href
  {https://ui.adsabs.harvard.edu/abs/2014MNRAS.445.1898C} {445, 1898}

\bibitem[\protect\citeauthoryear{{Chilingarian}, {Melchior}  \&
  {Zolotukhin}}{{Chilingarian} et~al.}{2010}]{2010MNRAS.405.1409C}
{Chilingarian} I.~V.,  {Melchior} A.-L.,   {Zolotukhin} I.~Y.,  2010, \mn@doi
  [\mnras] {10.1111/j.1365-2966.2010.16506.x}, \href
  {https://ui.adsabs.harvard.edu/abs/2010MNRAS.405.1409C} {405, 1409}

\bibitem[\protect\citeauthoryear{{Claeys}, {Pols}, {Izzard}, {Vink}  \&
  {Verbunt}}{{Claeys} et~al.}{2014}]{2014A&A...563A..83C}
{Claeys} J.~S.~W.,  {Pols} O.~R.,  {Izzard} R.~G.,  {Vink} J.,   {Verbunt}
  F.~W.~M.,  2014, \mn@doi [\aap] {10.1051/0004-6361/201322714}, \href
  {https://ui.adsabs.harvard.edu/abs/2014A&A...563A..83C} {563, A83}

\bibitem[\protect\citeauthoryear{{Coelho}, {Calv{\~a}o}, {Reis}  \&
  {Siffert}}{{Coelho} et~al.}{2015}]{2015EJPh...36a5007C}
{Coelho} R. C.~V.,  {Calv{\~a}o} M.~O.,  {Reis} R. R.~R.,   {Siffert} B.~B.,
  2015, \mn@doi [Eur. J. Phys.] {10.1088/0143-0807/36/1/015007}, \href
  {https://ui.adsabs.harvard.edu/abs/2015EJPh...36a5007C} {36, 015007}

\bibitem[\protect\citeauthoryear{{Crocker} et~al.,}{{Crocker}
  et~al.}{2017}]{2017NatAs...1E.135C}
{Crocker} R.~M.,  et~al., 2017, \mn@doi [Nat. Astron.]
  {10.1038/s41550-017-0135}, \href
  {https://ui.adsabs.harvard.edu/abs/2017NatAs...1E.135C} {1, 0135}

\bibitem[\protect\citeauthoryear{{\SortNoop{De Lapparent}}de~Lapparent,
  {Baillard}  \& {Bertin}}{{\SortNoop{De Lapparent}}de~Lapparent
  et~al.}{2011}]{2011A&A...532A..75D}
{\SortNoop{De Lapparent}}de~Lapparent V.,  {Baillard} A.,   {Bertin} E.,  2011,
  \mn@doi [\aap] {10.1051/0004-6361/201016424}, \href
  {https://ui.adsabs.harvard.edu/abs/2011A&A...532A..75D} {532, A75}

\bibitem[\protect\citeauthoryear{{Feigelson} \& {Babu}}{{Feigelson} \&
  {Babu}}{2012}]{FeigelsonBabu2012}
{Feigelson} E.~D.,  {Babu} G.~J.,  2012, {Modern Statistical Methods for
  Astronomy. Cambridge Univ. Press, Cambridge, UK}

\bibitem[\protect\citeauthoryear{{Filippenko} et~al.,}{{Filippenko}
  et~al.}{1992a}]{1992AJ....104.1543F}
{Filippenko} A.~V.,  et~al., 1992a, \mn@doi [\aj] {10.1086/116339}, \href
  {https://ui.adsabs.harvard.edu/abs/1992AJ....104.1543F} {104, 1543}

\bibitem[\protect\citeauthoryear{{Filippenko} et~al.,}{{Filippenko}
  et~al.}{1992b}]{1992ApJ...384L..15F}
{Filippenko} A.~V.,  et~al., 1992b, \mn@doi [\apjl] {10.1086/186252}, \href
  {https://ui.adsabs.harvard.edu/abs/1992ApJ...384L..15F} {384, L15}

\bibitem[\protect\citeauthoryear{{Finkbeiner} et~al.,}{{Finkbeiner}
  et~al.}{2016}]{2016ApJ...822...66F}
{Finkbeiner} D.~P.,  et~al., 2016, \mn@doi [\apj] {10.3847/0004-637X/822/2/66},
  \href {https://ui.adsabs.harvard.edu/abs/2016ApJ...822...66F} {822, 66}

\bibitem[\protect\citeauthoryear{{Fioc} \& {Rocca-Volmerange}}{{Fioc} \&
  {Rocca-Volmerange}}{1997}]{1997A&A...326..950F}
{Fioc} M.,  {Rocca-Volmerange} B.,  1997, \aap, \href
  {https://ui.adsabs.harvard.edu/abs/1997A&A...326..950F} {326, 950}

\bibitem[\protect\citeauthoryear{{Fioc} \& {Rocca-Volmerange}}{{Fioc} \&
  {Rocca-Volmerange}}{1999}]{1999astro.ph.12179F}
{Fioc} M.,  {Rocca-Volmerange} B.,  1999, preprint, \href
  {https://ui.adsabs.harvard.edu/abs/1999astro.ph.12179F} {} (\mn@eprint
  {arXiv} {astro-ph/9912179})

\bibitem[\protect\citeauthoryear{{Fisher} \& {Jumper}}{{Fisher} \&
  {Jumper}}{2015}]{2015ApJ...805..150F}
{Fisher} R.,  {Jumper} K.,  2015, \mn@doi [\apj] {10.1088/0004-637X/805/2/150},
  \href {https://ui.adsabs.harvard.edu/abs/2015ApJ...805..150F} {805, 150}

\bibitem[\protect\citeauthoryear{{Folatelli} et~al.,}{{Folatelli}
  et~al.}{2013}]{2013ApJ...773...53F}
{Folatelli} G.,  et~al., 2013, \mn@doi [\apj] {10.1088/0004-637X/773/1/53},
  \href {https://ui.adsabs.harvard.edu/abs/2013ApJ...773...53F} {773, 53}

\bibitem[\protect\citeauthoryear{{Foley} et~al.,}{{Foley}
  et~al.}{2009}]{2009AJ....138..376F}
{Foley} R.~J.,  et~al., 2009, \mn@doi [\aj] {10.1088/0004-6256/138/2/376},
  \href {https://ui.adsabs.harvard.edu/abs/2009AJ....138..376F} {138, 376}

\bibitem[\protect\citeauthoryear{{Foley} et~al.,}{{Foley}
  et~al.}{2013}]{2013ApJ...767...57F}
{Foley} R.~J.,  et~al., 2013, \mn@doi [\apj] {10.1088/0004-637X/767/1/57},
  \href {https://ui.adsabs.harvard.edu/abs/2013ApJ...767...57F} {767, 57}

\bibitem[\protect\citeauthoryear{{Gallagher}, {Garnavich}, {Berlind},
  {Challis}, {Jha}  \& {Kirshner}}{{Gallagher}
  et~al.}{2005}]{2005ApJ...634..210G}
{Gallagher} J.~S.,  {Garnavich} P.~M.,  {Berlind} P.,  {Challis} P.,  {Jha} S.,
    {Kirshner} R.~P.,  2005, \mn@doi [\apj] {10.1086/491664}, \href
  {https://ui.adsabs.harvard.edu/abs/2005ApJ...634..210G} {634, 210}

\bibitem[\protect\citeauthoryear{{Gallagher}, {Garnavich}, {Caldwell},
  {Kirshner}, {Jha}, {Li}, {Ganeshalingam}  \& {Filippenko}}{{Gallagher}
  et~al.}{2008}]{2008ApJ...685..752G}
{Gallagher} J.~S.,  {Garnavich} P.~M.,  {Caldwell} N.,  {Kirshner} R.~P.,
  {Jha} S.~W.,  {Li} W.,  {Ganeshalingam} M.,   {Filippenko} A.~V.,  2008,
  \mn@doi [\apj] {10.1086/590659}, \href
  {https://ui.adsabs.harvard.edu/abs/2008ApJ...685..752G} {685, 752}

\bibitem[\protect\citeauthoryear{{Ganeshalingam} et~al.,}{{Ganeshalingam}
  et~al.}{2010}]{2010ApJS..190..418G}
{Ganeshalingam} M.,  et~al., 2010, \mn@doi [\apjs]
  {10.1088/0067-0049/190/2/418}, \href
  {https://ui.adsabs.harvard.edu/abs/2010ApJS..190..418G} {190, 418}

\bibitem[\protect\citeauthoryear{{Ge}, {Gu}, {Garc{\'\i}a-Benito}, {Xiao}  \&
  {Li}}{{Ge} et~al.}{2020}]{2020ApJ...889..132G}
{Ge} X.,  {Gu} Q.-S.,  {Garc{\'\i}a-Benito} R.,  {Xiao} M.-Y.,   {Li} Z.-N.,
  2020, \mn@doi [\apj] {10.3847/1538-4357/ab65f6}, \href
  {https://ui.adsabs.harvard.edu/abs/2020ApJ...889..132G} {889, 132}

\bibitem[\protect\citeauthoryear{{Gilfanov} \& {Bogd{\'a}n}}{{Gilfanov} \&
  {Bogd{\'a}n}}{2010}]{2010Natur.463..924G}
{Gilfanov} M.,  {Bogd{\'a}n} {\'A}.,  2010, \mn@doi [\nat]
  {10.1038/nature08685}, \href
  {https://ui.adsabs.harvard.edu/abs/2010Natur.463..924G} {463, 924}

\bibitem[\protect\citeauthoryear{{Gomes} et~al.,}{{Gomes}
  et~al.}{2016}]{2016A&A...588A..68G}
{Gomes} J.~M.,  et~al., 2016, \mn@doi [\aap] {10.1051/0004-6361/201525976},
  \href {https://ui.adsabs.harvard.edu/abs/2016A&A...588A..68G} {588, A68}

\bibitem[\protect\citeauthoryear{{Gonz{\'a}lez Delgado} et~al.,}{{Gonz{\'a}lez
  Delgado} et~al.}{2015}]{2015A&A...581A.103G}
{Gonz{\'a}lez Delgado} R.~M.,  et~al., 2015, \mn@doi [\aap]
  {10.1051/0004-6361/201525938}, \href
  {https://ui.adsabs.harvard.edu/abs/2015A&A...581A.103G} {581, A103}

\bibitem[\protect\citeauthoryear{{Gonz{\'a}lez-Gait{\'a}n}
  et~al.,}{{Gonz{\'a}lez-Gait{\'a}n} et~al.}{2011}]{2011ApJ...727..107G}
{Gonz{\'a}lez-Gait{\'a}n} S.,  et~al., 2011, \mn@doi [\apj]
  {10.1088/0004-637X/727/2/107}, \href
  {https://ui.adsabs.harvard.edu/abs/2011ApJ...727..107G} {727, 107}

\bibitem[\protect\citeauthoryear{{Gonz{\'a}lez-Gait{\'a}n}
  et~al.,}{{Gonz{\'a}lez-Gait{\'a}n} et~al.}{2014}]{2014ApJ...795..142G}
{Gonz{\'a}lez-Gait{\'a}n} S.,  et~al., 2014, \mn@doi [\apj]
  {10.1088/0004-637X/795/2/142}, \href
  {https://ui.adsabs.harvard.edu/abs/2014ApJ...795..142G} {795, 142}

\bibitem[\protect\citeauthoryear{{Graham} \& {Foley}}{{Graham} \&
  {Foley}}{2004}]{2004IAUC.8340....1G}
{Graham} J.,  {Foley} R.~J.,  2004, \iaucirc, \href
  {https://ui.adsabs.harvard.edu/abs/2004IAUC.8340....1G} {8340, 1}

\bibitem[\protect\citeauthoryear{{Graur}, {Bianco}, {Modjaz}, {Shivvers},
  {Filippenko}, {Li}  \& {Smith}}{{Graur} et~al.}{2017}]{2017ApJ...837..121G}
{Graur} O.,  {Bianco} F.~B.,  {Modjaz} M.,  {Shivvers} I.,  {Filippenko} A.~V.,
   {Li} W.,   {Smith} N.,  2017, \mn@doi [\apj] {10.3847/1538-4357/aa5eb7},
  \href {https://ui.adsabs.harvard.edu/abs/2017ApJ...837..121G} {837, 121}

\bibitem[\protect\citeauthoryear{{Greggio}}{{Greggio}}{2005}]{2005A&A...441.1055G}
{Greggio} L.,  2005, \mn@doi [\aap] {10.1051/0004-6361:20052926}, \href
  {https://ui.adsabs.harvard.edu/abs/2005A&A...441.1055G} {441, 1055}

\bibitem[\protect\citeauthoryear{{Guillochon}, {Parrent}, {Kelley}  \&
  {Margutti}}{{Guillochon} et~al.}{2017}]{2017ApJ...835...64G}
{Guillochon} J.,  {Parrent} J.,  {Kelley} L.~Z.,   {Margutti} R.,  2017,
  \mn@doi [\apj] {10.3847/1538-4357/835/1/64}, \href
  {https://ui.adsabs.harvard.edu/abs/2017ApJ...835...64G} {835, 64}

\bibitem[\protect\citeauthoryear{{Gupta} et~al.,}{{Gupta}
  et~al.}{2011}]{2011ApJ...740...92G}
{Gupta} R.~R.,  et~al., 2011, \mn@doi [\apj] {10.1088/0004-637X/740/2/92},
  \href {https://ui.adsabs.harvard.edu/abs/2011ApJ...740...92G} {740, 92}

\bibitem[\protect\citeauthoryear{{Guy} et~al.,}{{Guy}
  et~al.}{2007}]{2007A&A...466...11G}
{Guy} J.,  et~al., 2007, \mn@doi [\aap] {10.1051/0004-6361:20066930}, \href
  {https://ui.adsabs.harvard.edu/abs/2007A&A...466...11G} {466, 11}

\bibitem[\protect\citeauthoryear{{Hakobyan}, {Petrosian}, {McLean}, {Kunth},
  {Allen}, {Turatto}  \& {Barbon}}{{Hakobyan}
  et~al.}{2008}]{2008A&A...488..523H}
{Hakobyan} A.~A.,  {Petrosian} A.~R.,  {McLean} B.,  {Kunth} D.,  {Allen}
  R.~J.,  {Turatto} M.,   {Barbon} R.,  2008, \mn@doi [\aap]
  {10.1051/0004-6361:200809817}, \href
  {https://ui.adsabs.harvard.edu/abs/2008A&A...488..523H} {488, 523}

\bibitem[\protect\citeauthoryear{{Hakobyan}, {Adibekyan}, {Aramyan},
  {Petrosian}, {Gomes}, {Mamon}, {Kunth}  \& {Turatto}}{{Hakobyan}
  et~al.}{2012}]{2012A&A...544A..81H}
{Hakobyan} A.~A.,  {Adibekyan} V.~Z.,  {Aramyan} L.~S.,  {Petrosian} A.~R.,
  {Gomes} J.~M.,  {Mamon} G.~A.,  {Kunth} D.,   {Turatto} M.,  2012, \mn@doi
  [\aap] {10.1051/0004-6361/201219541}, \href
  {https://ui.adsabs.harvard.edu/abs/2012A&A...544A..81H} {544, A81}

\bibitem[\protect\citeauthoryear{{Hakobyan} et~al.,}{{Hakobyan}
  et~al.}{2016}]{2016MNRAS.456.2848H}
{Hakobyan} A.~A.,  et~al., 2016, \mn@doi [\mnras] {10.1093/mnras/stv2853},
  \href {https://ui.adsabs.harvard.edu/abs/2016MNRAS.456.2848H} {456, 2848}

\bibitem[\protect\citeauthoryear{{Hakobyan} et~al.,}{{Hakobyan}
  et~al.}{2017}]{2017MNRAS.471.1390H}
{Hakobyan} A.~A.,  et~al., 2017, \mn@doi [\mnras] {10.1093/mnras/stx1608},
  \href {https://ui.adsabs.harvard.edu/abs/2017MNRAS.471.1390H} {471, 1390}

\bibitem[\protect\citeauthoryear{{Hamuy}, {Phillips}, {Suntzeff}, {Schommer},
  {Maza}  \& {Aviles}}{{Hamuy} et~al.}{1996}]{1996AJ....112.2391H}
{Hamuy} M.,  {Phillips} M.~M.,  {Suntzeff} N.~B.,  {Schommer} R.~A.,  {Maza}
  J.,   {Aviles} R.,  1996, \mn@doi [\aj] {10.1086/118190}, \href
  {https://ui.adsabs.harvard.edu/abs/1996AJ....112.2391H} {112, 2391}

\bibitem[\protect\citeauthoryear{{Hamuy}, {Trager}, {Pinto}, {Phillips},
  {Schommer}, {Ivanov}  \& {Suntzeff}}{{Hamuy}
  et~al.}{2000}]{2000AJ....120.1479H}
{Hamuy} M.,  {Trager} S.~C.,  {Pinto} P.~A.,  {Phillips} M.~M.,  {Schommer}
  R.~A.,  {Ivanov} V.,   {Suntzeff} N.~B.,  2000, \mn@doi [\aj]
  {10.1086/301527}, \href
  {https://ui.adsabs.harvard.edu/abs/2000AJ....120.1479H} {120, 1479}

\bibitem[\protect\citeauthoryear{{Han} \& {Podsiadlowski}}{{Han} \&
  {Podsiadlowski}}{2004}]{2004MNRAS.350.1301H}
{Han} Z.,  {Podsiadlowski} P.,  2004, \mn@doi [\mnras]
  {10.1111/j.1365-2966.2004.07713.x}, \href
  {https://ui.adsabs.harvard.edu/abs/2004MNRAS.350.1301H} {350, 1301}

\bibitem[\protect\citeauthoryear{{Hicken} et~al.,}{{Hicken}
  et~al.}{2009}]{2009ApJ...700..331H}
{Hicken} M.,  et~al., 2009, \mn@doi [\apj] {10.1088/0004-637X/700/1/331}, \href
  {https://ui.adsabs.harvard.edu/abs/2009ApJ...700..331H} {700, 331}

\bibitem[\protect\citeauthoryear{{Hillebrandt}, {Kromer}, {R{\"o}pke}  \&
  {Ruiter}}{{Hillebrandt} et~al.}{2013}]{2013FrPhy...8..116H}
{Hillebrandt} W.,  {Kromer} M.,  {R{\"o}pke} F.~K.,   {Ruiter} A.~J.,  2013,
  \mn@doi [Front. Phys.] {10.1007/s11467-013-0303-2}, \href
  {https://ui.adsabs.harvard.edu/abs/2013FrPhy...8..116H} {8, 116}

\bibitem[\protect\citeauthoryear{{Howell}}{{Howell}}{2001}]{2001ApJ...554L.193H}
{Howell} D.~A.,  2001, \mn@doi [\apjl] {10.1086/321702}, \href
  {https://ui.adsabs.harvard.edu/abs/2001ApJ...554L.193H} {554, L193}

\bibitem[\protect\citeauthoryear{{Howell} et~al.,}{{Howell}
  et~al.}{2009}]{2009ApJ...691..661H}
{Howell} D.~A.,  et~al., 2009, \mn@doi [\apj] {10.1088/0004-637X/691/1/661},
  \href {https://ui.adsabs.harvard.edu/abs/2009ApJ...691..661H} {691, 661}

\bibitem[\protect\citeauthoryear{{Iben} \& {Tutukov}}{{Iben} \&
  {Tutukov}}{1984}]{1984ApJS...54..335I}
{Iben} I. J.,  {Tutukov} A.~V.,  1984, \mn@doi [\apjs] {10.1086/190932}, \href
  {https://ui.adsabs.harvard.edu/abs/1984ApJS...54..335I} {54, 335}

\bibitem[\protect\citeauthoryear{{Iben} \& {Tutukov}}{{Iben} \&
  {Tutukov}}{1985}]{1985ApJS...58..661I}
{Iben} I. J.,  {Tutukov} A.~V.,  1985, \mn@doi [\apjs] {10.1086/191054}, \href
  {https://ui.adsabs.harvard.edu/abs/1985ApJS...58..661I} {58, 661}

\bibitem[\protect\citeauthoryear{{Jha}}{{Jha}}{2017}]{2017hsn..book..375J}
{Jha} S.~W.,  2017, {in Alsabti~A.~W., Murdin~P., eds, Type Iax Supernovae,
  Handbook of Supernovae. Springer, Cham, p.~375}

\bibitem[\protect\citeauthoryear{{Jha}, {Riess}  \& {Kirshner}}{{Jha}
  et~al.}{2007}]{2007ApJ...659..122J}
{Jha} S.,  {Riess} A.~G.,   {Kirshner} R.~P.,  2007, \mn@doi [\apj]
  {10.1086/512054}, \href
  {https://ui.adsabs.harvard.edu/abs/2007ApJ...659..122J} {659, 122}

\bibitem[\protect\citeauthoryear{{Johansson} et~al.,}{{Johansson}
  et~al.}{2013}]{2013MNRAS.435.1680J}
{Johansson} J.,  et~al., 2013, \mn@doi [\mnras] {10.1093/mnras/stt1408}, \href
  {https://ui.adsabs.harvard.edu/abs/2013MNRAS.435.1680J} {435, 1680}

\bibitem[\protect\citeauthoryear{{Kang}, {Lee}, {Kim}, {Chung}  \&
  {Ree}}{{Kang} et~al.}{2020}]{2020ApJ...889....8K}
{Kang} Y.,  {Lee} Y.-W.,  {Kim} Y.-L.,  {Chung} C.,   {Ree} C.~H.,  2020,
  \mn@doi [\apj] {10.3847/1538-4357/ab5afc}, \href
  {https://ui.adsabs.harvard.edu/abs/2020ApJ...889....8K} {889, 8}

\bibitem[\protect\citeauthoryear{{Karapetyan}, {Hakobyan}, {Barkhudaryan},
  {Mamon}, {Kunth}, {Adibekyan}  \& {Turatto}}{{Karapetyan}
  et~al.}{2018}]{2018MNRAS.481..566K}
{Karapetyan} A.~G.,  {Hakobyan} A.~A.,  {Barkhudaryan} L.~V.,  {Mamon} G.~A.,
  {Kunth} D.,  {Adibekyan} V.,   {Turatto} M.,  2018, \mn@doi [\mnras]
  {10.1093/mnras/sty2291}, \href
  {https://ui.adsabs.harvard.edu/abs/2018MNRAS.481..566K} {481, 566}

\bibitem[\protect\citeauthoryear{{Kaviraj}, {Peirani}, {Khochfar}, {Silk}  \&
  {Kay}}{{Kaviraj} et~al.}{2009}]{2009MNRAS.394.1713K}
{Kaviraj} S.,  {Peirani} S.,  {Khochfar} S.,  {Silk} J.,   {Kay} S.,  2009,
  \mn@doi [\mnras] {10.1111/j.1365-2966.2009.14403.x}, \href
  {https://ui.adsabs.harvard.edu/abs/2009MNRAS.394.1713K} {394, 1713}

\bibitem[\protect\citeauthoryear{{Kelsey} et~al.,}{{Kelsey}
  et~al.}{2020}]{2020arXiv200812101K}
{Kelsey} L.,  et~al., 2020, preprint, \href
  {https://ui.adsabs.harvard.edu/abs/2020arXiv200812101K} {} (\mn@eprint
  {arXiv} {2008.12101})

\bibitem[\protect\citeauthoryear{{K{\"o}nyves-T{\'o}th}
  et~al.,}{{K{\"o}nyves-T{\'o}th} et~al.}{2020}]{2020ApJ...892..121K}
{K{\"o}nyves-T{\'o}th} R.,  et~al., 2020, \mn@doi [\apj]
  {10.3847/1538-4357/ab76bb}, \href
  {https://ui.adsabs.harvard.edu/abs/2020ApJ...892..121K} {892, 121}

\bibitem[\protect\citeauthoryear{{Lee}, {Jang}  \& {Kang}}{{Lee}
  et~al.}{2019}]{2019ApJ...871...33L}
{Lee} M.~G.,  {Jang} I.~S.,   {Kang} J.,  2019, \mn@doi [\apj]
  {10.3847/1538-4357/aaf72c}, \href
  {https://ui.adsabs.harvard.edu/abs/2019ApJ...871...33L} {871, 33}

\bibitem[\protect\citeauthoryear{{Leibundgut} et~al.,}{{Leibundgut}
  et~al.}{1993}]{1993AJ....105..301L}
{Leibundgut} B.,  et~al., 1993, \mn@doi [\aj] {10.1086/116427}, \href
  {https://ui.adsabs.harvard.edu/abs/1993AJ....105..301L} {105, 301}

\bibitem[\protect\citeauthoryear{{Li} et~al.,}{{Li}
  et~al.}{2003}]{2003PASP..115..453L}
{Li} W.,  et~al., 2003, \mn@doi [\pasp] {10.1086/374200}, \href
  {https://ui.adsabs.harvard.edu/abs/2003PASP..115..453L} {115, 453}

\bibitem[\protect\citeauthoryear{{Li} et~al.,}{{Li}
  et~al.}{2011}]{2011MNRAS.412.1441L}
{Li} W.,  et~al., 2011, \mn@doi [\mnras] {10.1111/j.1365-2966.2011.18160.x},
  \href {https://ui.adsabs.harvard.edu/abs/2011MNRAS.412.1441L} {412, 1441}

\bibitem[\protect\citeauthoryear{{Lipunov}, {Panchenko}  \&
  {Pruzhinskaya}}{{Lipunov} et~al.}{2011}]{2011NewA...16..250L}
{Lipunov} V.~M.,  {Panchenko} I.~E.,   {Pruzhinskaya} M.~V.,  2011, \mn@doi
  [\na] {10.1016/j.newast.2010.09.001}, \href
  {https://ui.adsabs.harvard.edu/abs/2011NewA...16..250L} {16, 250}

\bibitem[\protect\citeauthoryear{{Livio} \& {Mazzali}}{{Livio} \&
  {Mazzali}}{2018}]{2018PhR...736....1L}
{Livio} M.,  {Mazzali} P.,  2018, \mn@doi [\physrep]
  {10.1016/j.physrep.2018.02.002}, \href
  {https://ui.adsabs.harvard.edu/abs/2018PhR...736....1L} {736, 1}

\bibitem[\protect\citeauthoryear{{Maeda} \& {Terada}}{{Maeda} \&
  {Terada}}{2016}]{2016IJMPD..2530024M}
{Maeda} K.,  {Terada} Y.,  2016, \mn@doi [Int. J. Mod. Phys. D]
  {10.1142/S021827181630024X}, \href
  {https://ui.adsabs.harvard.edu/abs/2016IJMPD..2530024M} {25, 1630024}

\bibitem[\protect\citeauthoryear{{Maoz}, {Mannucci}  \& {Nelemans}}{{Maoz}
  et~al.}{2014}]{2014ARA&A..52..107M}
{Maoz} D.,  {Mannucci} F.,   {Nelemans} G.,  2014, \mn@doi [\araa]
  {10.1146/annurev-astro-082812-141031}, \href
  {https://ui.adsabs.harvard.edu/abs/2014ARA&A..52..107M} {52, 107}

\bibitem[\protect\citeauthoryear{{Matteucci} \& {Greggio}}{{Matteucci} \&
  {Greggio}}{1986}]{1986A&A...154..279M}
{Matteucci} F.,  {Greggio} L.,  1986, \aap, \href
  {https://ui.adsabs.harvard.edu/abs/1986A&A...154..279M} {154, 279}

\bibitem[\protect\citeauthoryear{{Mazzali}, {R{\"o}pke}, {Benetti}  \&
  {Hillebrandt}}{{Mazzali} et~al.}{2007}]{2007Sci...315..825M}
{Mazzali} P.~A.,  {R{\"o}pke} F.~K.,  {Benetti} S.,   {Hillebrandt} W.,  2007,
  \mn@doi [Science] {10.1126/science.1136259}, \href
  {https://ui.adsabs.harvard.edu/abs/2007Sci...315..825M} {315, 825}

\bibitem[\protect\citeauthoryear{{McIntosh} et~al.,}{{McIntosh}
  et~al.}{2014}]{2014MNRAS.442..533M}
{McIntosh} D.~H.,  et~al., 2014, \mn@doi [\mnras] {10.1093/mnras/stu808}, \href
  {https://ui.adsabs.harvard.edu/abs/2014MNRAS.442..533M} {442, 533}

\bibitem[\protect\citeauthoryear{{Mendez}, {Coil}, {Lotz}, {Salim}, {Moustakas}
   \& {Simard}}{{Mendez} et~al.}{2011}]{2011ApJ...736..110M}
{Mendez} A.~J.,  {Coil} A.~L.,  {Lotz} J.,  {Salim} S.,  {Moustakas} J.,
  {Simard} L.,  2011, \mn@doi [\apj] {10.1088/0004-637X/736/2/110}, \href
  {https://ui.adsabs.harvard.edu/abs/2011ApJ...736..110M} {736, 110}

\bibitem[\protect\citeauthoryear{{Meng}}{{Meng}}{2019}]{2019ApJ...886...58M}
{Meng} X.-C.,  2019, \mn@doi [\apj] {10.3847/1538-4357/ab4e10}, \href
  {https://ui.adsabs.harvard.edu/abs/2019ApJ...886...58M} {886, 58}

\bibitem[\protect\citeauthoryear{{Neill} et~al.,}{{Neill}
  et~al.}{2009}]{2009ApJ...707.1449N}
{Neill} J.~D.,  et~al., 2009, \mn@doi [\apj] {10.1088/0004-637X/707/2/1449},
  \href {https://ui.adsabs.harvard.edu/abs/2009ApJ...707.1449N} {707, 1449}

\bibitem[\protect\citeauthoryear{{Nomoto}, {Iwamoto}  \& {Kishimoto}}{{Nomoto}
  et~al.}{1997}]{1997Sci...276.1378N}
{Nomoto} K.,  {Iwamoto} K.,   {Kishimoto} N.,  1997, \mn@doi [Science]
  {10.1126/science.276.5317.1378}, \href
  {https://ui.adsabs.harvard.edu/abs/1997Sci...276.1378N} {276, 1378}

\bibitem[\protect\citeauthoryear{{Onken} et~al.,}{{Onken}
  et~al.}{2019}]{2019PASA...36...33O}
{Onken} C.~A.,  et~al., 2019, \mn@doi [\pasa] {10.1017/pasa.2019.27}, \href
  {https://ui.adsabs.harvard.edu/abs/2019PASA...36...33O} {36, e033}

\bibitem[\protect\citeauthoryear{{Oohama}, {Okamura}, {Fukugita}, {Yasuda}  \&
  {Nakamura}}{{Oohama} et~al.}{2009}]{2009ApJ...705..245O}
{Oohama} N.,  {Okamura} S.,  {Fukugita} M.,  {Yasuda} N.,   {Nakamura} O.,
  2009, \mn@doi [\apj] {10.1088/0004-637X/705/1/245}, \href
  {https://ui.adsabs.harvard.edu/abs/2009ApJ...705..245O} {705, 245}

\bibitem[\protect\citeauthoryear{{Pakmor}, {Kromer}, {Taubenberger}  \&
  {Springel}}{{Pakmor} et~al.}{2013}]{2013ApJ...770L...8P}
{Pakmor} R.,  {Kromer} M.,  {Taubenberger} S.,   {Springel} V.,  2013, \mn@doi
  [\apjl] {10.1088/2041-8205/770/1/L8}, \href
  {https://ui.adsabs.harvard.edu/abs/2013ApJ...770L...8P} {770, L8}

\bibitem[\protect\citeauthoryear{{Pan} et~al.,}{{Pan}
  et~al.}{2014}]{2014MNRAS.438.1391P}
{Pan} Y.~C.,  et~al., 2014, \mn@doi [\mnras] {10.1093/mnras/stt2287}, \href
  {https://ui.adsabs.harvard.edu/abs/2014MNRAS.438.1391P} {438, 1391}

\bibitem[\protect\citeauthoryear{{Panther}, {Seitenzahl}, {Ruiter}, {Crocker},
  {Lidman}, {Wang}, {Tucker}  \& {Groves}}{{Panther}
  et~al.}{2019}]{2019PASA...36...31P}
{Panther} F.~H.,  {Seitenzahl} I.~R.,  {Ruiter} A.~J.,  {Crocker} R.~M.,
  {Lidman} C.,  {Wang} E.~X.,  {Tucker} B.~E.,   {Groves} B.,  2019, \mn@doi
  [\pasa] {10.1017/pasa.2019.24}, \href
  {https://ui.adsabs.harvard.edu/abs/2019PASA...36...31P} {36, e031}

\bibitem[\protect\citeauthoryear{{Perets} et~al.,}{{Perets}
  et~al.}{2010}]{2010Natur.465..322P}
{Perets} H.~B.,  et~al., 2010, \mn@doi [\nat] {10.1038/nature09056}, \href
  {https://ui.adsabs.harvard.edu/abs/2010Natur.465..322P} {465, 322}

\bibitem[\protect\citeauthoryear{{Perets}, {Zenati}, {Toonen}  \&
  {Bobrick}}{{Perets} et~al.}{2019}]{2019arXiv191007532P}
{Perets} H.~B.,  {Zenati} Y.,  {Toonen} S.,   {Bobrick} A.,  2019, preprint,
  \href {https://ui.adsabs.harvard.edu/abs/2019arXiv191007532P} {} (\mn@eprint
  {arXiv} {1910.07532})

\bibitem[\protect\citeauthoryear{{Perley} et~al.,}{{Perley}
  et~al.}{2020}]{2020arXiv200901242P}
{Perley} D.~A.,  et~al., 2020, preprint, \href
  {https://ui.adsabs.harvard.edu/abs/2020arXiv200901242P} {} (\mn@eprint
  {arXiv} {2009.01242})

\bibitem[\protect\citeauthoryear{{Perlmutter} et~al.,}{{Perlmutter}
  et~al.}{1999}]{1999ApJ...517..565P}
{Perlmutter} S.,  et~al., 1999, \mn@doi [\apj] {10.1086/307221}, \href
  {https://ui.adsabs.harvard.edu/abs/1999ApJ...517..565P} {517, 565}

\bibitem[\protect\citeauthoryear{{Phillips}}{{Phillips}}{1993}]{1993ApJ...413L.105P}
{Phillips} M.~M.,  1993, \mn@doi [\apjl] {10.1086/186970}, \href
  {https://ui.adsabs.harvard.edu/abs/1993ApJ...413L.105P} {413, L105}

\bibitem[\protect\citeauthoryear{{Phillips}, {Wells}, {Suntzeff}, {Hamuy},
  {Leibundgut}, {Kirshner}  \& {Foltz}}{{Phillips}
  et~al.}{1992}]{1992AJ....103.1632P}
{Phillips} M.~M.,  {Wells} L.~A.,  {Suntzeff} N.~B.,  {Hamuy} M.,  {Leibundgut}
  B.,  {Kirshner} R.~P.,   {Foltz} C.~B.,  1992, \mn@doi [\aj]
  {10.1086/116177}, \href
  {https://ui.adsabs.harvard.edu/abs/1992AJ....103.1632P} {103, 1632}

\bibitem[\protect\citeauthoryear{{Phillips}, {Lira}, {Suntzeff}, {Schommer},
  {Hamuy}  \& {Maza}}{{Phillips} et~al.}{1999}]{1999AJ....118.1766P}
{Phillips} M.~M.,  {Lira} P.,  {Suntzeff} N.~B.,  {Schommer} R.~A.,  {Hamuy}
  M.,   {Maza} J.,  1999, \mn@doi [\aj] {10.1086/301032}, \href
  {https://ui.adsabs.harvard.edu/abs/1999AJ....118.1766P} {118, 1766}

\bibitem[\protect\citeauthoryear{{Ponder}, {Wood-Vasey}, {Weyant}, {Barton},
  {Galbany}, {Garnavich}  \& {Matheson}}{{Ponder}
  et~al.}{2020}]{2020arXiv200613803P}
{Ponder} K.~A.,  {Wood-Vasey} W.~M.,  {Weyant} A.,  {Barton} N.~T.,  {Galbany}
  L.,  {Garnavich} P.,   {Matheson} T.,  2020, preprint, \href
  {https://ui.adsabs.harvard.edu/abs/2020arXiv200613803P} {} (\mn@eprint
  {arXiv} {2006.13803})

\bibitem[\protect\citeauthoryear{{Postnov} \& {Yungelson}}{{Postnov} \&
  {Yungelson}}{2014}]{2014LRR....17....3P}
{Postnov} K.~A.,  {Yungelson} L.~R.,  2014, \mn@doi [Living Rev. Relativ.]
  {10.12942/lrr-2014-3}, \href
  {https://ui.adsabs.harvard.edu/abs/2014LRR....17....3P} {17, 3}

\bibitem[\protect\citeauthoryear{{Pruzhinskaya}, {Novinskaya}, {Pauna}  \&
  {Rosnet}}{{Pruzhinskaya} et~al.}{2020}]{2020arXiv200609433P}
{Pruzhinskaya} M.,  {Novinskaya} A.,  {Pauna} N.,   {Rosnet} P.,  2020,
  preprint, \href {https://ui.adsabs.harvard.edu/abs/2020arXiv200609433P} {}
  (\mn@eprint {arXiv} {2006.09433})

\bibitem[\protect\citeauthoryear{{Riess} et~al.,}{{Riess}
  et~al.}{1998}]{1998AJ....116.1009R}
{Riess} A.~G.,  et~al., 1998, \mn@doi [\aj] {10.1086/300499}, \href
  {https://ui.adsabs.harvard.edu/abs/1998AJ....116.1009R} {116, 1009}

\bibitem[\protect\citeauthoryear{{Rigault} et~al.,}{{Rigault}
  et~al.}{2013}]{2013A&A...560A..66R}
{Rigault} M.,  et~al., 2013, \mn@doi [\aap] {10.1051/0004-6361/201322104},
  \href {https://ui.adsabs.harvard.edu/abs/2013A&A...560A..66R} {560, A66}

\bibitem[\protect\citeauthoryear{{Roman} et~al.,}{{Roman}
  et~al.}{2018}]{2018A&A...615A..68R}
{Roman} M.,  et~al., 2018, \mn@doi [\aap] {10.1051/0004-6361/201731425}, \href
  {https://ui.adsabs.harvard.edu/abs/2018A&A...615A..68R} {615, A68}

\bibitem[\protect\citeauthoryear{{Rose}, {Garnavich}  \& {Berg}}{{Rose}
  et~al.}{2019}]{2019ApJ...874...32R}
{Rose} B.~M.,  {Garnavich} P.~M.,   {Berg} M.~A.,  2019, \mn@doi [\apj]
  {10.3847/1538-4357/ab0704}, \href
  {https://ui.adsabs.harvard.edu/abs/2019ApJ...874...32R} {874, 32}

\bibitem[\protect\citeauthoryear{{Ruiter}}{{Ruiter}}{2020}]{2020arXiv200102947R}
{Ruiter} A.~J.,  2020, preprint, \href
  {https://ui.adsabs.harvard.edu/abs/2020arXiv200102947R} {} (\mn@eprint
  {arXiv} {2001.02947})

\bibitem[\protect\citeauthoryear{{Ruiter}, {Belczynski}  \& {Fryer}}{{Ruiter}
  et~al.}{2009}]{2009ApJ...699.2026R}
{Ruiter} A.~J.,  {Belczynski} K.,   {Fryer} C.,  2009, \mn@doi [\apj]
  {10.1088/0004-637X/699/2/2026}, \href
  {https://ui.adsabs.harvard.edu/abs/2009ApJ...699.2026R} {699, 2026}

\bibitem[\protect\citeauthoryear{{Ruiter} et~al.,}{{Ruiter}
  et~al.}{2013}]{2013MNRAS.429.1425R}
{Ruiter} A.~J.,  et~al., 2013, \mn@doi [\mnras] {10.1093/mnras/sts423}, \href
  {https://ui.adsabs.harvard.edu/abs/2013MNRAS.429.1425R} {429, 1425}

\bibitem[\protect\citeauthoryear{{Schawinski}}{{Schawinski}}{2009}]{2009MNRAS.397..717S}
{Schawinski} K.,  2009, \mn@doi [\mnras] {10.1111/j.1365-2966.2009.15065.x},
  \href {https://ui.adsabs.harvard.edu/abs/2009MNRAS.397..717S} {397, 717}

\bibitem[\protect\citeauthoryear{{Schawinski} et~al.,}{{Schawinski}
  et~al.}{2014}]{2014MNRAS.440..889S}
{Schawinski} K.,  et~al., 2014, \mn@doi [\mnras] {10.1093/mnras/stu327}, \href
  {https://ui.adsabs.harvard.edu/abs/2014MNRAS.440..889S} {440, 889}

\bibitem[\protect\citeauthoryear{{Schlafly} \& {Finkbeiner}}{{Schlafly} \&
  {Finkbeiner}}{2011}]{2011ApJ...737..103S}
{Schlafly} E.~F.,  {Finkbeiner} D.~P.,  2011, \mn@doi [\apj]
  {10.1088/0004-637X/737/2/103}, \href
  {https://ui.adsabs.harvard.edu/abs/2011ApJ...737..103S} {737, 103}

\bibitem[\protect\citeauthoryear{{Schlegel}, {Finkbeiner}  \&
  {Davis}}{{Schlegel} et~al.}{1998}]{1998ApJ...500..525S}
{Schlegel} D.~J.,  {Finkbeiner} D.~P.,   {Davis} M.,  1998, \mn@doi [\apj]
  {10.1086/305772}, \href
  {https://ui.adsabs.harvard.edu/abs/1998ApJ...500..525S} {500, 525}

\bibitem[\protect\citeauthoryear{{Shen}, {Toonen}  \& {Graur}}{{Shen}
  et~al.}{2017}]{2017ApJ...851L..50S}
{Shen} K.~J.,  {Toonen} S.,   {Graur} O.,  2017, \mn@doi [\apjl]
  {10.3847/2041-8213/aaa015}, \href
  {https://ui.adsabs.harvard.edu/abs/2017ApJ...851L..50S} {851, L50}

\bibitem[\protect\citeauthoryear{{Siebert} et~al.,}{{Siebert}
  et~al.}{2019}]{2019MNRAS.486.5785S}
{Siebert} M.~R.,  et~al., 2019, \mn@doi [\mnras] {10.1093/mnras/stz1209}, \href
  {https://ui.adsabs.harvard.edu/abs/2019MNRAS.486.5785S} {486, 5785}

\bibitem[\protect\citeauthoryear{{Silverman} et~al.,}{{Silverman}
  et~al.}{2012}]{2012MNRAS.425.1789S}
{Silverman} J.~M.,  et~al., 2012, \mn@doi [\mnras]
  {10.1111/j.1365-2966.2012.21270.x}, \href
  {https://ui.adsabs.harvard.edu/abs/2012MNRAS.425.1789S} {425, 1789}

\bibitem[\protect\citeauthoryear{{Stahl} et~al.,}{{Stahl}
  et~al.}{2019}]{2019MNRAS.490.3882S}
{Stahl} B.~E.,  et~al., 2019, \mn@doi [\mnras] {10.1093/mnras/stz2742}, \href
  {https://ui.adsabs.harvard.edu/abs/2019MNRAS.490.3882S} {490, 3882}

\bibitem[\protect\citeauthoryear{{Stritzinger} et~al.,}{{Stritzinger}
  et~al.}{2015}]{2015A&A...573A...2S}
{Stritzinger} M.~D.,  et~al., 2015, \mn@doi [\aap]
  {10.1051/0004-6361/201424168}, \href
  {https://ui.adsabs.harvard.edu/abs/2015A&A...573A...2S} {573, A2}

\bibitem[\protect\citeauthoryear{{Suh}, {Yoon}, {Jeong}  \& {Yi}}{{Suh}
  et~al.}{2011}]{2011ApJ...730..110S}
{Suh} H.,  {Yoon} S.-c.,  {Jeong} H.,   {Yi} S.~K.,  2011, \mn@doi [\apj]
  {10.1088/0004-637X/730/2/110}, \href
  {https://ui.adsabs.harvard.edu/abs/2011ApJ...730..110S} {730, 110}

\bibitem[\protect\citeauthoryear{{Sullivan} et~al.,}{{Sullivan}
  et~al.}{2010}]{2010MNRAS.406..782S}
{Sullivan} M.,  et~al., 2010, \mn@doi [\mnras]
  {10.1111/j.1365-2966.2010.16731.x}, \href
  {https://ui.adsabs.harvard.edu/abs/2010MNRAS.406..782S} {406, 782}

\bibitem[\protect\citeauthoryear{{Taubenberger}}{{Taubenberger}}{2017}]{2017hsn..book..317T}
{Taubenberger} S.,  2017, {in Alsabti~A.~W., Murdin~P., eds, The Extremes of
  Thermonuclear Supernovae, Handbook of Supernovae. Springer, Cham, p.~317}

\bibitem[\protect\citeauthoryear{{Taubenberger} et~al.,}{{Taubenberger}
  et~al.}{2011}]{2011MNRAS.412.2735T}
{Taubenberger} S.,  et~al., 2011, \mn@doi [\mnras]
  {10.1111/j.1365-2966.2010.18107.x}, \href
  {https://ui.adsabs.harvard.edu/abs/2011MNRAS.412.2735T} {412, 2735}

\bibitem[\protect\citeauthoryear{{Taylor} et~al.,}{{Taylor}
  et~al.}{2011}]{2011MNRAS.418.1587T}
{Taylor} E.~N.,  et~al., 2011, \mn@doi [\mnras]
  {10.1111/j.1365-2966.2011.19536.x}, \href
  {https://ui.adsabs.harvard.edu/abs/2011MNRAS.418.1587T} {418, 1587}

\bibitem[\protect\citeauthoryear{{Turatto}, {Benetti}, {Cappellaro},
  {Danziger}, {Della Valle}, {Gouiffes}, {Mazzali}  \& {Patat}}{{Turatto}
  et~al.}{1996}]{1996MNRAS.283....1T}
{Turatto} M.,  {Benetti} S.,  {Cappellaro} E.,  {Danziger} I.~J.,  {Della
  Valle} M.,  {Gouiffes} C.,  {Mazzali} P.~A.,   {Patat} F.,  1996, \mn@doi
  [\mnras] {10.1093/mnras/283.1.1}, \href
  {https://ui.adsabs.harvard.edu/abs/1996MNRAS.283....1T} {283, 1}

\bibitem[\protect\citeauthoryear{{Uddin}, {Mould}, {Lidman}, {Ruhlmann-Kleider}
   \& {Zhang}}{{Uddin} et~al.}{2017}]{2017ApJ...848...56U}
{Uddin} S.~A.,  {Mould} J.,  {Lidman} C.,  {Ruhlmann-Kleider} V.,   {Zhang}
  B.~R.,  2017, \mn@doi [\apj] {10.3847/1538-4357/aa8df7}, \href
  {https://ui.adsabs.harvard.edu/abs/2017ApJ...848...56U} {848, 56}

\bibitem[\protect\citeauthoryear{{Uddin} et~al.,}{{Uddin}
  et~al.}{2020}]{2020arXiv200615164U}
{Uddin} S.~A.,  et~al., 2020, preprint, \href
  {https://ui.adsabs.harvard.edu/abs/2020arXiv200615164U} {} (\mn@eprint
  {arXiv} {2006.15164})

\bibitem[\protect\citeauthoryear{{Valenti} et~al.,}{{Valenti}
  et~al.}{2009}]{2009Natur.459..674V}
{Valenti} S.,  et~al., 2009, \mn@doi [\nat] {10.1038/nature08023}, \href
  {https://ui.adsabs.harvard.edu/abs/2009Natur.459..674V} {459, 674}

\bibitem[\protect\citeauthoryear{{Whelan} \& {Iben}}{{Whelan} \&
  {Iben}}{1973}]{1973ApJ...186.1007W}
{Whelan} J.,  {Iben} I. J.,  1973, \mn@doi [\apj] {10.1086/152565}, \href
  {https://ui.adsabs.harvard.edu/abs/1973ApJ...186.1007W} {186, 1007}

\bibitem[\protect\citeauthoryear{{White} et~al.,}{{White}
  et~al.}{2015}]{2015ApJ...799...52W}
{White} C.~J.,  et~al., 2015, \mn@doi [\apj] {10.1088/0004-637X/799/1/52},
  \href {https://ui.adsabs.harvard.edu/abs/2015ApJ...799...52W} {799, 52}

\bibitem[\protect\citeauthoryear{{Wild}}{{Wild}}{1994}]{1994IAUC.5982....2W}
{Wild} P.,  1994, \iaucirc, \href
  {https://ui.adsabs.harvard.edu/abs/1994IAUC.5982....2W} {5982, 2}

\bibitem[\protect\citeauthoryear{{Wolf} et~al.,}{{Wolf}
  et~al.}{2018}]{2018PASA...35...10W}
{Wolf} C.,  et~al., 2018, \mn@doi [\pasa] {10.1017/pasa.2018.5}, \href
  {https://ui.adsabs.harvard.edu/abs/2018PASA...35...10W} {35, e010}

\bibitem[\protect\citeauthoryear{{Yaron} \& {Gal-Yam}}{{Yaron} \&
  {Gal-Yam}}{2012}]{2012PASP..124..668Y}
{Yaron} O.,  {Gal-Yam} A.,  2012, \mn@doi [\pasp] {10.1086/666656}, \href
  {https://ui.adsabs.harvard.edu/abs/2012PASP..124..668Y} {124, 668}

\bibitem[\protect\citeauthoryear{{Yungelson} \& {Livio}}{{Yungelson} \&
  {Livio}}{2000}]{2000ApJ...528..108Y}
{Yungelson} L.~R.,  {Livio} M.,  2000, \mn@doi [\apj] {10.1086/308174}, \href
  {https://ui.adsabs.harvard.edu/abs/2000ApJ...528..108Y} {528, 108}

\bibitem[\protect\citeauthoryear{{Zapartas} et~al.,}{{Zapartas}
  et~al.}{2017}]{2017A&A...601A..29Z}
{Zapartas} E.,  et~al., 2017, \mn@doi [\aap] {10.1051/0004-6361/201629685},
  \href {https://ui.adsabs.harvard.edu/abs/2017A&A...601A..29Z} {601, A29}

\bibitem[\protect\citeauthoryear{{Zenati}, {Toonen}  \& {Perets}}{{Zenati}
  et~al.}{2019}]{2019MNRAS.482.1135Z}
{Zenati} Y.,  {Toonen} S.,   {Perets} H.~B.,  2019, \mn@doi [\mnras]
  {10.1093/mnras/sty2723}, \href
  {https://ui.adsabs.harvard.edu/abs/2019MNRAS.482.1135Z} {482, 1135}

\makeatother
\end{thebibliography}

\section*{Supporting information}

Supplementary data are available at \emph{MNRAS} online.\\
\\
\textbf{PaperVIIonlinedata.csv}\\
\\
Please note: Oxford University Press is not responsible for the
content or functionality of any supporting materials supplied by
the authors. Any queries (other than missing material) should be
directed to the corresponding author for the article.
\\
\begin{flushleft}
{\small This paper has been typeset from a {\TeX/\LaTeX} file prepared by the author.}
\end{flushleft}

\label{lastpage}

\end{document}